%% file: paper.tex
\documentclass[reprint, aps, prd, letterpaper, noshowpacs, amsmath, %
amssymb, amsfonts, nofootinbib, floatfix, 
twoside]{revtex4-1}
\pdfoutput=1

\usepackage[usenames,dvipsnames]{xcolor}
\input{UsePackages}
\input{Preamble}
\input{Macros}
\input{Affiliations}

\usepackage{tikz} %
\usepackage{pgfplots} %
\usepackage{pgfplotstable} %
\pgfplotsset{width=246pt} %
\usetikzlibrary{pgfplots.groupplots} %
\pgfplotsset{axis line style={line width=1pt, black}} %
\pgfplotsset{tick style={line width=1pt, black}} %
\pgfplotsset{minor tick style={black}} %

\usetikzlibrary{pgfplots.external}
\hypersetup{urlcolor=DarkUrlColor}

\begin{document}


\graphicspath{%
  {Plots/}%
}

\title[Angular velocity of gravitational radiation and the corotating
frame] {Angular velocity of gravitational radiation from precessing
  binaries and the corotating frame}

\surnames{Michael Boyle} \author{Michael Boyle} \Cornell

\date{\today}

\begin{abstract}
  This paper defines an angular velocity for time-dependent functions
  on the sphere, and applies it to gravitational waveforms from
  compact binaries.  Because it is geometrically meaningful and has a
  clear physical motivation, the angular velocity is uniquely useful
  in helping to solve an important---and largely ignored---problem in
  models of compact binaries: the inverse problem of deducing the
  physical parameters of a system from the gravitational waves alone.
  It is also used to define the corotating frame of the waveform.
  When decomposed in this frame, the waveform has no \emph{rotational}
  dynamics and is therefore as slowly evolving as possible.  The
  resulting simplifications lead to straightforward methods for
  accurately comparing waveforms and constructing hybrids.  As
  formulated in this paper, the methods can be applied robustly to
  both precessing and nonprecessing waveforms, providing a clear,
  comprehensive, and consistent framework for waveform analysis.
  Explicit implementations of all these methods are provided in
  accompanying computer code.
\end{abstract}



\maketitle


\section{Introduction%
  \label{sec:Introduction}}%
Gravitational-wave astronomy stands on the brink of delivering
numerous observations of merging compact binaries~\cite{LIGO:2009,
  Shoemaker:2010, WaldmanEtAl:2011, Virgo:2012, KAGRA:2012,
  WeinsteinEtAl:2012}.  Though the uncertainties are large, black-hole
binaries involving large spins are expected to constitute a
significant fraction of observable events~\cite{Kalogera:2000,
  OShaughnessyEtAl:2005, GrandclementEtAl:2004, LIGO:2010}.  If these
spins are misaligned with the orbital angular velocity, the system
will precess, imprinting the gravitational radiation with strong
variations~\cite{Kidder:1995, ApostolatosEtAl:1994, WillWiseman:1996,
  Apostolatos:1996}.  While it is not clear how common such
misalignment will actually be, it is entirely clear that we will need
good models of the precessing waveforms if we hope to accurately
measure them.

We can describe the motion of a precessing binary on very short
timescales as a simple orbit in a plane; on longer timescales, that
plane rotates.  Now, we know that the gravitational-wave field of a
nonprecessing system can be decomposed into relatively simple modes
when the orbital plane is orthogonal to the $z$
axis~\cite{Einstein:1918, Thorne:1980, BlanchetEtAl:2008}.  But
precession moves the orbital plane out of alignment, causing the modes
to mix and leading to complex behaviors which complicate analysis of
the waveforms~\cite{BuonannoEtAl:2003, Blanchet:2006b, Blanchet:2007,
  ArunEtAl:2009, Blanchet:2010, MarsatEtAl:2012, BoheEtAl:2012,
  BuonannoEtAl:2003, Ajith:2011, SchmidtEtAl:2012, BrownEtAl:2012,
  BrownEtAl:2012b, OShaughnessyEtAl:2012}.  In particular, none of the
methods developed to analyze nonprecessing systems will work correctly
with precessing systems.

In the context of post-Newtonian models, Buonanno, Chen, and
Vallisneri~\cite{BuonannoEtAl:2003} proposed a convention whereby
effects of precession can be isolated from orbital motion.
Specifically, the system is analyzed at each instant in a frame with
its $z$ axis orthogonal to the orbital plane; from moment to moment,
the frame is made to rotate to follow the precession.  This method was
later rediscovered in the context of numerical relativity, and
techniques were developed for finding such a frame from the waveform
itself in a geometrically meaningful way~\cite{SchmidtEtAl:2011,
  OShaughnessyEtAl:2011, BoyleEtAl:2011}.

This paper extends previous work by developing a frame in which
\emph{all} rotational behavior is eliminated, simplifying the waveform
as much as possible, and allowing direct generalizations of methods
for analyzing nonprecessing systems.  In the process, the angular
velocity of a waveform is introduced, which also has important uses,
such as supplying a partial solution to an important inverse problem.

\subsection{The modeler's inverse problem%
  \label{sec:InverseProblem}}%
We might distinguish two significant inverse problems related to
gravitational waves: the \emph{modeler's} inverse problem and the
equally important \emph{astronomer's} inverse problem.  The
gravitational-wave astronomer's task is to deduce the parameters
(masses, spins, \etc) of a system from observations at a single point
over an extended time.  In practice, it is greatly complicated by the
presence of noise in the data.  Usually referred to as parameter
estimation, this problem has been extensively
studied~\cite{JaranowskiEtAl:1996, vanderSluysEtAl:2008a,
  vanderSluysEtAl:2008b, RaymondEtAl:2009, NissankeEtAl:2010,
  WenChen:2010, Fairhurst:2011, KlimenkoEtAl:2011, NissankeEtAl:2011,
  Vallisneri:2011, VitaleZanolin:2011, VeitchEtAl:2012}.  The
modeler's task, on the other hand, is to deduce the parameters given
observations of the entire sphere at infinity over a brief (possibly
infinitesimal) interval of time.  It is---in some sense---prior to the
astronomer's problem, because it addresses the \emph{meaning} of the
parameters in models astronomers use.\footnote{Intriguingly,
  understanding the modeler's inverse problem may help to inform the
  astronomer's inverse problem more
  directly~\cite{OShaughnessyEtAl:2012}.}  This paper concerns itself
exclusively with the modeler's inverse problem.

Various methods exist for producing gravitational
waveforms---numerical-relativity, phenomenological, post-Newtonian,
and effective-one-body models, for example.  But no one of these is
capable of producing an accurate and complete waveform on its own.
Numerical-relativity (NR) simulations are too expensive to simulate
more than a short portion of the waveform near merger.
Phenomenological models use NR data as inputs.  Post-Newtonian (\pN)
approximations break down before the merger.  Even terms generating
the effective-one-body (EOB) inspiral and the \textsl{ad hoc} method
of attaching a ringdown must be ``calibrated'' by comparison to
numerical results.  Therefore, we need more than one model to generate
a complete waveform, which means that we need to understand precisely
how the different models relate to each other.  This leads directly to
the inverse problem.

The numbers we plug in to a computation of initial data for an NR
simulation bear no clear relation to the numbers we insert into a \pN
or an EOB computation---or even to other NR simulations using
different formulations.  For example, the direction of a black-hole
spin measured in the arbitrary coordinates of NR initial data need not
correspond in any meaningful way to the direction measured in \pN
coordinates.  Even if the gauge condition used for the numerical
simulation were the same as the one used to derive the \pN formulas,
the initial data would not be the same, so the gauge itself would be
different.  For nonprecessing systems, symmetries reduce the ambiguity
to one of simple time and phase offsets, and numerous simple methods
have been suggested to resolve those ambiguities~\cite{NINJA2:2012}.
But in the precessing case, we need to be much more careful.  Simply
using the same numbers in two different models leads to comparing
systems with inherently different physics.

Fundamentally, we need to establish a mapping between the input
parameters of different models such that they produce the same physics
(as nearly as possible) during some span of time for which both models
are valid.  Because we have no access to any invariant physical
meaning behind our parameters, we need to take a different approach.
For example, given some particular set of parameters, we can run a
numerical simulation.  Then, we can work backwards from the resulting
waveform and try to find the parameters needed to generate the same
waveform with a \pN system---which is the inverse problem.

The issue of ascribing meaningful physical interpretations to
geometric quantities measured on $\scriplus$ has been investigated to
some extent~\cite{* [{}] [{ and references therein.}]
  KozamehEtAl:2008, Moreschi:2004, Helfer:2010}, but it is not clear
that these methods are useful for the immediate problem of analyzing
gravitational waveforms.  The angular-velocity vector introduced by
this paper and a related vector introduced by O'Shaughnessy
\etal~\cite{OShaughnessyEtAl:2011} provide geometrically meaningful
physical quantities which can be measured directly from the waveforms
alone, and are thus prime candidates for use in solving the inverse
problem.  Indeed, we will see in
Sec.~\ref{sec:SolvingTheInverseProblem} that these two vectors are
very closely related to input parameters for the precessing \pN
system.  This provides a partial solution to the inverse problem,
leaving three remaining degrees of freedom.  Several possibilities
will also be suggested for completing the solution of the inverse
problem, though they are beyond the scope of this paper.

\subsection{Overview of this paper%
  \label{sec:Overview}}%
Section~\ref{sec:AngularVelocityOfAWaveform} introduces the angular
velocity $\FieldAngularVelocity$ of a waveform, finding a
straightforward formula and a more intuitive interpretation of the
mathematics.  This and the related vector $\LLDominantEigenvector$
suggested in Ref.~\cite{OShaughnessyEtAl:2011} are then used in
Sec.~\ref{sec:SolvingTheInverseProblem} to find a partial solution to
the inverse problem.  It is shown that $\FieldAngularVelocity$ and
$\LLDominantEigenvector$---which are measured from the waveform
alone---can be combined to give expressions for the corresponding \pN
orbital elements.  This can then be used to deduce the parameters of
the system.  A \pN waveform is used as a test case, showing excellent
agreement between the original parameters and the parameters deduced
from the waveform alone.  In Sec.~\ref{sec:TheFrameItself}, the
angular velocity is used to determine a frame with that velocity.  The
same \pN waveform used in the previous section is decomposed in this
frame, showing that the amplitudes of the waveform modes become very
simple, and their phases become nearly constant.  Because this frame
reduces the complexity of the waveform, it is ideally suited to
practical manipulation of waveforms, which is discussed in detail in
Sec.~\ref{sec:Extrapolationcomparisonandhybridization}.  It is also
worth noting that the partial solution to the inverse problem
completely establishes all extrinsic parameters, giving us a solid
foundation for comparisons between waveforms.  Finally, the results
are summarized and suggestions for future work are collected in
Sec.~\ref{sec:Conclusions}.

The appendices provide deeper background information which may be
useful for implementing these methods or comparing to other methods.
Appendix~\ref{sec:QuaternionsAndRotations} presents a fairly
comprehensive discussion of quaternions and various related details,
including several new results.  In
Appendix~\ref{sec:RotatingSpinWeightedFunctions}, formulas are derived
for the rotation of arbitrary spin-weighted functions.  While
equivalent formulas have been derived
previously~\cite{GoldbergEtAl:1967, GualtieriEtAl:2008,
  BoyleEtAl:2011}, this derivation uses a somewhat different
technique, and carefully develops conventions for consistency
throughout this paper.  In any case, the upshot is that modes of
spin-weighted fields transform exactly as do modes of spin-weight-zero
fields.  Lastly, Appendix~\ref{sec:PreviousWork} discusses related
previous work in the same formalism used in this paper, allowing for
more direct comparisons.

\href{http://arxiv.org/src/1302.2919/anc}{Ancillary files} included
with this paper (available on the paper's arXiv page) contain computer
code implementing all of the concepts introduced here, among others.
The core functions are written in C++~\cite{Cplusplus} for speed,
using several functions from the GNU Scientific
Library~\cite{GSL:2009}.  While this code could be incorporated
directly into other C/C++ codes, an additional user interface is
provided in Python~\cite{Python, IPython:2007} code as the
\code{GWFrames} module, which simply exposes all the C++ functions
through Python.  Documentation and examples can be found among the
ancillary files.  Relevant functions or classes are mentioned where
appropriate throughout this paper.

\subsection{Quaternion notation%
  \label{sec:QuaternionNotation}}%
The techniques of this paper necessarily involve rotations, which are
best implemented in terms of the group of unit quaternions because of
the numerous advantages over direct manipulation of rotation matrices
or Euler-angle coordinates.  By using quaternions, we obtain robust
methods that can be blindly applied to general systems, including
nonprecessing ones---which simplifies the processing of large numbers
of waveforms.  Moreover, once the basics are understood, quaternion
rotations are more intuitive than either of those inferior
descriptions.  In fact, quaternions are essentially the axis-angle
description of rotations, in a more practical guise.  Therefore,
quaternion notation will be used throughout.  As mentioned above,
Appendix~\ref{sec:QuaternionsAndRotations} provides a thorough
introduction to quaternions, while computer code included in
\href{http://arxiv.org/src/1302.2919/anc}{ancillary files} with this
paper gives practical implementations of the necessary functionality
through \code{GWFrames.Quaternion}.  However, such details are not
necessary for a good understanding of this paper; the following
paragraph should provide sufficient background.

Quaternions can be thought of as generalizing the familiar complex
numbers, where the imaginary part is generalized to a
three-dimensional vector part.\footnote{In fact, both complex numbers
  and quaternions are special cases of geometric
  algebra~\cite{DoranLasenby:2003}, done in two and three dimensions,
  respectively.  Much of our intuition from complex algebra transfers
  easily to quaternion algebra when $\i=\sqrt{-1}$ is replaced by a
  unit vector.  The notable exception to this correspondence is
  noncommutativity of the quaternion product.}  We can write a
quaternion as the sum of a scalar and a vector: $\quat{Q} =
\quatcomponent{Q}{0} + \quatvec{Q}$.  The conjugate of the quaternion
is $\co{\quat{Q}} = \quatcomponent{Q}{0} - \quatvec{Q}$.  We can
multiply quaternions together [see
Eq.~\eqref{eq:QuaternionMultiplication}], the product being
associative but not commutative in general.  The norm of a quaternion
is defined according to $\abs{\quat{Q}}^{2} = \quat{Q}\, \co{\quat{Q}}
= \quatcomponent{Q}{0}^{2} + \quatvec{Q} \cdot \quatvec{Q}$.  Unit
quaternions, having norm $\abs{\quat{R}} = 1$, are especially
important, as they describe rotations.  To see this, we can consider a
vector to be a quaternion with scalar component equal to zero, in
which case it makes sense to multiply a vector by a quaternion.  Then
we can define the transformation
\begin{equation}
  \label{eq:QuaternionRotation}
  \vector{v} \mapsto \vector{v}' \defined \quat{R}\, \vector{v}\
  \co{\quat{R}}~.
\end{equation}
A simple exercise shows that this transformation is linear, and
preserves lengths and orientations, so it is just a rotation.
Ultimately, the best reason to use quaternions is the existence of
simple formulas [Eqs.~\eqref{eq:RotationAxisAngle}
and~\eqref{eq:QuaternionLogarithm}] for the exponential and logarithm,
which prove to be endlessly useful.  In particular, we can express an
arbitrary unit quaternion as $\quat{R} = \e^{\theta\, \unitvec{u}/2} =
\cos \frac{\theta}{2} + \unitvec{u}\, \sin \frac{\theta}{2}$, where
exponentiation of a quaternion is defined by the usual power
series.\footnote{Note the striking---and not coincidental---similarity
  to Euler's formula with $\unitvec{u}$ in place of the unit imaginary
  $\i$.  This results from the fact that, under quaternion
  multiplication, $\unitvec{u}\, \unitvec{u} = -1$.}  It turns out
that this $\quat{R}$ produces a rotation through the angle $\theta$
about the axis $\unitvec{u}$.  Because \emph{any} rotation may be
expressed in this form, we will use unit quaternions as our only
representation of rotations, and refer to them as \textit{rotors}.
Conversely, given a rotor $\quat{R}$, we can find the corresponding
axis and angle according to $\theta\, \unitvec{u} = 2\, \log
\quat{R}$, where the logarithm is given by
Eq.~\eqref{eq:QuaternionLogarithm}.  A frame will be described by the
rotor that generates it by rotating some standard basis frame.

\section{Angular velocity of a waveform%
  \label{sec:AngularVelocityOfAWaveform}}%
We can define the angular velocity of a gravitational waveform---or
any field on a sphere---as the opposite of the velocity of the
counter-rotation needed to keep the field as constant as possible.  In
the first part of this section, this definition will be formulated
more precisely, resulting in a surprisingly simple formula for the
angular velocity.  The formula can be interpreted as a projection of
the familiar operator equation $-\i\, \FieldAngularVelocity \cdot
\AngMomOpVec = \partial_{t}$ onto the ``rotational parts'' of the
waveform---a notion which can be made surprisingly rigorous using the
language of Hilbert spaces, as discussed in the second part of this
section.

\subsection{Finding the angular velocity%
  \label{sec:SolvingForTheRotation}}%
The essential idea here is to remove the rotational behavior of the
waveform by imposing a rotation that eliminates as much of the time
dependence as possible.  Suppose that $\RotationOperator_{j}(t)$ is a
time-dependent rotation operator acting on the wave field $\wavefield$
(usually representing $\Psi_{4}$ or $\h$) such that
$\RotationOperator_{j}(t_{j}) = 1$.  We wish to find the rotation
operator that---in some sense---minimizes the quantity
\begin{equation}
  \label{eq:RateOfRotation}
  \left. \frac{\partial}{\partial t} \left[ \RotationOperator_{j}(t)\,
      \wavefield(t; \vartheta, \varphi) \right] \right|_{t=t_{j}}~.
\end{equation}
Clearly, this is a complex function of position on the sphere.  To
reduce it to a single real number, we take its squared magnitude and
integrate over the sphere:
\begin{equation}
  \label{eq:CorotationMinimizer}
  \Xi(\RotationOperator_{j}) \defined \int_{S^{2}} \abs{
    \left. \frac{\partial}{\partial t}
      \left[\RotationOperator_{j}(t)\, \wavefield(t; \vartheta,
        \varphi) \right] \right|_{t=t_{j}}}^{2}\, \d \Omega~.
\end{equation}
If we expand the field $\wavefield$ in spin-weighted spherical
harmonics (SWSHs, discussed in
Appendix~\ref{sec:RotatingSpinWeightedFunctions}), the natural way to
express the rotation operator is in its usual form
$\RotationOperator_{j} = \exp[ - \i\, \CorotVector_{j} \cdot
\AngMomOpVec ]$, where $\AngMomOpVec$ is the standard angular-momentum
operator and $\CorotVector_{j}$ is the time-dependent axis-angle
description of the rotation.  Note that we must have
$\CorotVector_{j}(t_{j}) = 0$ because we have assumed that
$\RotationOperator_{j}(t_{j}) = 1$.  This is absolutely crucial
because it makes the differentiation in
Eq.~\eqref{eq:CorotationMinimizer} tractable.  We also define the
angular velocity\footnote{Subscripts are necessary on the rotation
  operator $\RotationOperator_{j}$ and the associated vector
  $\CorotVector_{j}$ because these have certain properties depending
  on which instant of time $t_{j}$ we are looking at.  We have
  implicitly assumed that the $\CorotVector_{j}(t)$ are all related by
  simple constant offsets, as necessary to satisfy the conditions
  $\CorotVector_{j}(t_{j})=0$.  Because the offsets are
  \emph{constant}, the angular-velocity vector $\FieldAngularVelocity$
  does not have such a dependence, and so does not need the
  subscript.}
\begin{equation}
  \label{eq:DefineAngularVelocity}
  \FieldAngularVelocity \defined - \partial_{t} \CorotVector_{j}~,
\end{equation}
where the negative sign arises because $\CorotVector_{j}$ corresponds
to the rotation needed to keep the field fixed in a moving frame,
whereas $\FieldAngularVelocity$ is intended to describe the motion of
the field relative to the initial static frame.  We have
\begin{equation}
  \label{eq:CorotationMinimizer2}
  \Xi(\FieldAngularVelocity) = \int_{S^{2}} \abs{\i\,
    \FieldAngularVelocity \cdot \AngMomOpVec\, \wavefield +
    \partial_{t} \wavefield}^{2}\, \d \Omega~.
\end{equation}

Now, we can write the integral in terms of a sum over standard matrix
elements of the angular-momentum operator and the problem simplifies
nicely.  We obtain
\begin{equation}
  \label{eq:CorotationMinimizerSimplified}
  \Xi = \FieldAngularVelocity \cdot \LLQuantity \cdot
  \FieldAngularVelocity + 2\, \FieldAngularVelocity \cdot \LdtVector +
  \sum_{\ell,m} \abs{\partial_{t} \wavefield^{\ell,m}}^{2}~,
\end{equation}
\begin{subequations}
  \label{eq:AngularVelocity}
  where we have defined the matrix\footnote{Here, the $\ket{\ell, m}$
    represent the \emph{spin-weighted} eigenfunctions, but the
    angular-momentum operator acts on these just as in the
    non-spin-weighted case~\cite{GoldbergEtAl:1967}, making this
    notation particularly familiar.  The matrix denoted here as
    $\LLQuantity^{ab}$ is precisely the quantity $\langle L_{(a}\,
    L_{b)} \rangle_{t}$ defined by O'Shaughnessy
    \etal~\cite{OShaughnessyEtAl:2011}, except that the latter is
    normalized by $\sum_{\ell,m} \lvert \wavefield^{\ell,m}
    \rvert^{2}$.}
  \begin{equation}
    \label{eq:AngularMomentumMatrix}
    \LLQuantity^{a b} \defined \sum_{\ell, m, m'}\,
    \co{\wavefield}^{\ell,m'} \braket{\ell, m'| \AngMomOp^{(a}\,
      \AngMomOp^{b)} |\ell, m}\, \wavefield^{\ell, m}~,
  \end{equation}
  and the vector
  \begin{equation}
    \label{eq:AngularMomentumVector}
    \LdtVector^{a} \defined \sum_{\ell, m, m'}\, \Im \left[
      \co{\wavefield}^{\ell, m'} \braket{\ell, m'| \AngMomOp^{a}
        |\ell, m}\, \partial_{t} \wavefield^{\ell,m} \right]~.
  \end{equation}
  Noting that the last term in
  Eq.~\eqref{eq:CorotationMinimizerSimplified} is independent of
  $\FieldAngularVelocity$, we can find the minimum\footnote{We can
    show that it is a true minimum rather than a more general
    stationary point by looking at the Hessian matrix of $\Xi$, which
    is just $2 \LLQuantity$.  We are free to rotate this matrix into a
    frame in which its dominant principal axis is along $\unitvec{z}$.
    Then, we can calculate its eigenvalues as
    $\braket{\AngMomOp_{z}^{2}}$ and $\braket{\AngMomOp^{2} -
      \AngMomOp_{z}^{2}} \pm \abs{\braket{\AngMomOp_{+}^{2}}}$.  As
    long as some mode with $m \neq 0$ is nonzero, these are always
    (strictly) positive.  Hence, $\LLQuantity$ is positive definite,
    and we have a true minimum.  Furthermore, we can calculate that
    the determinant is actually the product of these eigenvalues, and
    thus is also nonzero whenever the field is nonzero, allowing us to
    invert the matrix in Eq.~\eqref{eq:CorotationMinimized}.  Since
    $\LLQuantity$ is a geometric object, and eigenvalues and
    determinants are invariant under rotations, these conclusions hold
    in all frames.} of $\Xi$ analytically:
  \begin{equation}
    \label{eq:CorotationMinimized}
    \FieldAngularVelocity = -\LLQuantity^{-1} \cdot \LdtVector~.
  \end{equation}
\end{subequations}
The effects of $\AngMomOp^{a}$ are familiar, so this may be directly
computed from knowledge of $\wavefield^{\ell,m}(t)$, with no
optimization or solution of the eigensystem necessary.  There is no
ambiguity in the direction of the angular velocity, and we obtain a
meaningful magnitude.

In the computer code included among this paper's
\href{http://arxiv.org/src/1302.2919/anc}{ancillary files}, a waveform
object may be constructed with \code{GWFrames.Waveform}.  The angular
velocity may then be found using the \code{AngularVelocityVector}
method on such an object.

\subsection{Interpreting the mathematics%
  \label{sec:InterpretingTheMathematics}}%
Equation~\eqref{eq:AngularVelocity} gives a formula for the
angular-velocity vector of the field $\wavefield$.  Though it takes a
relatively simple form, the reason it takes this particular form may
seem somewhat opaque.  In fact, it really has quite a simple
interpretation, which may be instructive.  In fact, we can start off
with a simple observation and re-derive Eq.~\eqref{eq:AngularVelocity}
in a very different way.

The angular-momentum operator $\AngMomOpVec$ generates rotations, as
is well known.  So, for example, $-\i\, \FieldAngularVelocity \cdot
\AngMomOpVec\, \wavefield$ gives the time rate of change of the field
under a simple rotation given by $\FieldAngularVelocity$.  More
generally, the three components of $-\i\, \AngMomOpVec\, \wavefield$
form a basis generating the Hilbert subspace $\Lambda$, consisting of
functions describing possible rates of change for $\wavefield$ under
(complex) rotations.  On the other hand, we also have a second
operator $\partial_{t}$, which gives the actual time rate of change of
the field, whether that change is a simple rotation or a change in
amplitude---or a more complicated behavior.  But we can extract the
part of $\partial_{t} \wavefield$ caused by (real) rotation alone by
projecting onto the basis vectors of $\Lambda$ and taking the real
part:
\begin{equation}
  \label{eq:Projection}
  \Re \left[ \int_{S^{2}} \overline{-\i\, \AngMomOpVec\,
      \wavefield}\, \partial_{t} \wavefield\, \d \Omega \right]~.
\end{equation}
The three components of this expression completely describe the
rotational part of $\partial_{t} \wavefield$.  We take the real part
because we ordinarily take the dot product of $-\i\, \AngMomOpVec$
with a real-valued vector, so if we expect to find such terms in
$\partial_{t} \wavefield$, they must have real
components~\cite{Szekeres:2006}.

Now, the crucial point: if $\FieldAngularVelocity$ correctly describes
the rotation, the same projection of $-\i\, \FieldAngularVelocity
\cdot \AngMomOpVec\, \wavefield$ must give the same result:
\begin{equation}
  \label{eq:ProjectedEquivalence}
  \Re \left[ \int_{S^{2}} \overline{-\i\, \AngMomOpVec\, \wavefield}\,
    (-\i\, \FieldAngularVelocity \cdot \AngMomOpVec\, \wavefield)\, \d
    \Omega \right] = \Re \left[ \int_{S^{2}} \overline{-\i\,
      \AngMomOpVec\, \wavefield}\, \partial_{t} \wavefield\, \d \Omega
  \right]~.
\end{equation}
If we expand $\wavefield$ in spin-weighted spherical harmonics, it
turns out\footnote{As usual, we get integrals of the form $\int \ldots
  \ket{\vartheta, \varphi} \bra{\vartheta, \varphi} \ldots \d \Omega$,
  which are just resolutions of the identity.  Then, taking the real
  part on the left-hand side is equivalent to symmetrizing over the
  indices of the two $\AngMomOpVec$ vectors before contracting with
  $\FieldAngularVelocity$.  On the right-hand side, taking the real
  part of $\i$ times a quantity is the same as taking the negative
  imaginary part, so this is precisely the definition of
  $-\LdtVector$.} that this equation reduces to precisely
\begin{equation}
  \label{eq:CorotationRederivation}
  \LLQuantity \cdot \FieldAngularVelocity = -\LdtVector~,
\end{equation}
which is, of course, equivalent to Eq.~\eqref{eq:CorotationMinimized}.

Thus, we see the interpretation clearly.  In the case of a pure
rotation, we have $-\i\, \FieldAngularVelocity \cdot \AngMomOpVec\,
\wavefield = \partial_{t} \wavefield$.  In general, however, we have
to project onto the rotational parts of the waveform for equality to
hold, which is just what Eqs.~\eqref{eq:AngularVelocity}
and~\eqref{eq:CorotationRederivation} do.  It is also worth noting
that in the purely rotational case, we can use $-\i\,
\FieldAngularVelocity \cdot \AngMomOpVec = \partial_{t}$ directly and
calculate $\Xi \identically 0$.  Recalling the definition of $\Xi$ in
Eq.~\eqref{eq:CorotationMinimizer}, this says that the time variation
is completely eliminated.

Interestingly, we can see this projection working directly by showing
that $\LLQuantity$ and $\LdtVector$ are insensitive to changes in the
amplitudes of the modes.  Clearly, $\LLQuantity$ does not depend on
any derivatives with respect to time.  To see that $\LdtVector$ is
insensitive to changing amplitude, we first note that it is a
geometric object so we can evaluate it in any frame we choose---we
choose a frame in which it is aligned with the $z$ axis.  Next, we
decompose the field into (logarithmic) amplitude and phase parts:
\begin{equation}
  \label{eq:MagArgDefined}
  \wavefield^{\ell,m}(t) = \exp \left[ \chi^{\ell,m}(t) + \i\,
    \phi^{\ell,m}(t) \right]~.
\end{equation}
A pure rotation about the $z$ axis leads to $\dot{\chi}^{\ell,m} = 0$
and $\dot{\phi}^{\ell,m} = -m \abs{\FieldAngularVelocity}$, so we
expect that a projection onto the rotational part will eliminate
$\dot{\chi}^{\ell,m}$ but must not eliminate $\dot{\phi}^{\ell,m}$.
In fact, we can explicitly calculate
\begin{subequations}
  \label{eq:LdtVectorCalculated}
  \begin{align}
    \LdtVector &= \unitvec{z}\, \sum_{\ell, m}\, \Im \left[
      \co{\wavefield}^{\ell,m} \braket{\ell,m | \AngMomOp^{z} |
        \ell,m} \partial_{t} \wavefield^{\ell, m} \right] \\ &=
    \unitvec{z}\, \sum_{\ell, m}\, \Im \left[
      \left(\dot{\chi}^{\ell,m} +\i\, \dot{\phi}^{\ell,m}\right)\, m\,
      \abs{\wavefield^{\ell, m}}^{2} \right] \\
    &= \label{eq:LdtVectorResult} \unitvec{z}\, \sum_{\ell, m}\, m\,
    \dot{\phi}^{\ell,m}\, \abs{\wavefield^{\ell, m}}^{2}~,
  \end{align}
\end{subequations}
Here, taking the imaginary part has caused $\dot{\chi}^{\ell,m}$ to
drop out entirely, leaving only $\dot{\phi}^{\ell,m}$, supporting the
claim that we have removed non-rotational parts of the waveform.  Note
that this formula is entirely general; we have not assumed any
particular behavior of $\chi$ or $\phi$, for example.

\section{Solving the inverse problem%
  \label{sec:SolvingTheInverseProblem}}%
Section~\ref{sec:InverseProblem} established the need to solve the
inverse problem.  Essentially, in order to create a complete
gravitational waveform, we need to be able to take a finite or even
infinitesimal portion of a waveform and infer the parameters of the
\pN (or similar) system that result in that waveform.  Because it is
the most extensively developed system, we will discuss the
quasicircular \pN model as a concrete example.  In this section, we
will first describe the parameters that need to be established.  This
will involve reviewing the basic elements of the \pN model.  We will
then see how to solve part of the inverse problem using the
angular-velocity vector $\FieldAngularVelocity$ and the dominant
eigenvector of $\LLQuantity$ (denoted
$\LLDominantEigenvector$)~\cite{OShaughnessyEtAl:2011}, showing the
effectiveness of this method with an example.

\subsection{The required parameters%
  \label{sec:ThePNSystem}}%
To the extent that different formulations of the \pN model are
correct, they predict the same physics, and so we are free to choose
between them as we wish.  Certain formulations may be better than
others with regard to solving the inverse problem.  Here, we follow
Refs.~\cite{FayeEtAl:2006, BoheEtAl:2012}.  First, we assume a pair of
particles with masses $M_{1}$ and $M_{2}$, and spins $\vector{S}_{1}$
and $\vector{S}_{2}$.  The unit vector pointing from the second to the
first is $\unitvec{n}$.  The orbital angular velocity is defined as
\begin{subequations}
  \label{eq:PNAngularVelocities}
  \begin{equation}
    \label{eq:OrbitalAngularVelocity}
    \OrbitalAngularVelocity \defined \unitvec{n} \times
    \unitvecdot{n}~.
  \end{equation}
  The direction of this vector is frequently expressed in the
  literature as $\LNhat (\identically \OrbitalAngularVelocityHat)$.
  There is (in general) an additional rotation of the system due to
  precession, denoted $\PrecessionAngularVelocity$.  Now, if this
  vector were to have any component orthogonal to $\unitvec{n}$, that
  would contradict the definition of $\OrbitalAngularVelocity$, so it
  must simply be proportional to $\unitvec{n}$:\footnote{Note that for
    other formulations of the \pN model, this equation may not be
    true.  See Refs.~\cite{ApostolatosEtAl:1994, Kidder:1995,
      BuonannoEtAl:2003, Blanchet:2006b, Blanchet:2007, ArunEtAl:2009,
      Blanchet:2010, MarsatEtAl:2012, BoheEtAl:2012, Ajith:2011} for
    more details.}
  \begin{equation}
    \label{eq:PrecessionalAngularVelocity}
    \PrecessionAngularVelocity \defined \PrecessionalFrequency\,
    \unitvec{n}~.
  \end{equation}
  We also define the sum of these:
  \begin{equation}
    \label{eq:TotalAngularVelocity}
    \TotalAngularVelocity \defined \OrbitalAngularVelocity +
    \PrecessionAngularVelocity~.
  \end{equation}
\end{subequations}
During the evolution we must record the minimal-rotation
frame\footnote{See Sec.~\ref{sec:ImposingTheMinimalRotationCondition}
  and Ref.~\cite{BoyleEtAl:2011}.}  aligned with
$\OrbitalAngularVelocity$ and the accumulated orbital phase
$\OrbitalPhase$ measured relative to $\unitvec{n}$.  Then, the frame
of the binary will given by rotating the minimal-rotation frame by
$\OrbitalPhase$ about its $z$ axis.  These are the orbital elements of
the system.  Their evolution is not of particular concern here, as the
details have no effect on our conclusions.  The waveform can be
calculated in this frame using standard formulas, and transformed to
an inertial frame if needed to complete the construction of the
waveform.

The initial data we need to begin a \pN calculation, then, are the
values of $(M_{1}, M_{2}, \vector{S}_{1}, \vector{S}_{2},
\OrbitalAngularVelocity, \unitvec{n})$ at some initial time.  These
might be termed the \textit{intrinsic} parameters of the
system~\cite{Owen:1996, BuonannoEtAl:2003, PanEtAl:2004}.  They are
geometrically meaningful, and covariant under certain symmetries
assumed for our system---namely time translation and rotation of
coordinates.  But this brings up a subtlety.  We can think of two more
classes of parameters: the \textit{extrinsic} and the
\textit{fiducial}.  Extrinsic parameters depend on the observer, and
can be thought of generally in terms of degrees of gauge freedom like
the time offset or an overall rotation.  Fiducial parameters are
selected values of intrinsic quantities that depend on extrinsic
parameters.  By solving for the intrinsic parameters relative to a
particular time function and a particular basis for the vectors, we
will be tacitly setting the extrinsic parameters.  Then, when
comparing two waveforms, we must choose fiducial parameters and ensure
that the extrinsic parameters are the same.  We will find that it is a
simple matter to ascertain the intrinsic parameters except for three
degrees of freedom in the directions of the spin vectors.  It will
also be straightforward to completely establish the extrinsic
parameters.

\subsection{Deducing the parameters%
  \label{sec:DeducingParameters}}%
We might only expect quantities to be meaningful if they are covariant
objects measured at infinity---\eg, waveforms or ADM-type quantities.
Coordinate locations of black holes in a simulation, for example,
depend too much on details of gauge conditions and vagaries of initial
data and junk radiation to be of any real use.  On the other hand,
some quantities are also reasonably well defined when the black holes
are very widely separated.  Therefore, if we find that certain
quantities change slowly during the early part of the NR simulation,
and are not expected to have changed much previously, then we might
also be able to use those quantities in our analysis.  This is
typically true of the masses and spin magnitudes when measured
appropriately~\cite{Owen:2007, CookWhiting:2007, [{Appendix~A of }]
  [{.}] LovelaceEtAl:2008}, except to the extent that they are
expected to change~\cite{Alvi:2001, LovelaceEtAl:2012}.  Therefore, we
assume that $M_{1}$, $M_{2}$, $\lvert \vector{S}_{1} \rvert$, and
$\lvert \vector{S}_{2} \rvert$ can be measured in the simulation and
used directly.  The rest of our intrinsic parameters will come from
the waveforms (or possibly other measurements on $\scriplus$).

To see how we can derive orbital elements from quantities observable
from the waveform, we need to see how orbital elements give rise to
the waveform.  Familiar calculations~\cite{Einstein:1918, Thorne:1980}
tell us that the \pN waveform is created by motion of the binary.  The
complete motion is described by $\TotalAngularVelocity$, so we expect
that $\FieldAngularVelocity$ should be the same.  On the other hand,
the component along $\unitvec{n}$ does not lead to changing multipole
moments (to our level of approximation).  So only the component of
$\TotalAngularVelocity$ orthogonal to $\unitvec{n}$ is involved---but
that is precisely $\OrbitalAngularVelocity$.  We can therefore expect
that the waveform is oriented along this vector, in some sense.  Now,
the vector $\vector{z}$ happens to be the dominant eigenvector of
$\LLQuantity$ for an individual spin-weighted spherical harmonic
(though not necessarily for a combination of them).  It also happens
to be the dominant eigenvector when the field $\wavefield$ is
symmetric under reflection through the $x$--$y$
plane~\cite{OchsnerOShaughnessy:2012}---as the \pN waveform is in the
frame aligned with $\OrbitalAngularVelocity$.  Therefore, we should
expect $\OrbitalAngularVelocity$ to be parallel to
$\LLDominantEigenvector$.

Putting these considerations together, we can expect the following
approximate equalities:
\begin{subequations}
  \label{eq:OrbitalElementsFromWaveform}
  \begin{align}
  \label{eq:TotalAngularVelocityFromWaveform}
    \TotalAngularVelocity &\asymptoticallyequal
    \FieldAngularVelocity~, \\
  \label{eq:OrbitalAngularVelocityFromWaveform}
    \OrbitalAngularVelocity &\asymptoticallyequal \left(
      \LLDominantEigenvector \cdot \FieldAngularVelocity \right)\,
    \LLDominantEigenvector~, \\
  \label{eq:PrecessionAngularVelocityFromWaveform}
    \PrecessionAngularVelocity &\asymptoticallyequal
    \FieldAngularVelocity - \left( \LLDominantEigenvector \cdot
      \FieldAngularVelocity \right)\, \LLDominantEigenvector~.
  \end{align}
\end{subequations}
Inspection of the \pN model suggests that these expressions should
become more exactly true in the asymptotic limit of low orbital
velocities.  Note that Eq.~\eqref{eq:PrecessionalAngularVelocity}
shows that $\unitvec{n}$ is along $\PrecessionAngularVelocity$, so we
effectively obtain that quantity as well, whenever the precession is
nonzero.

\begin{figure*}
  \includegraphics{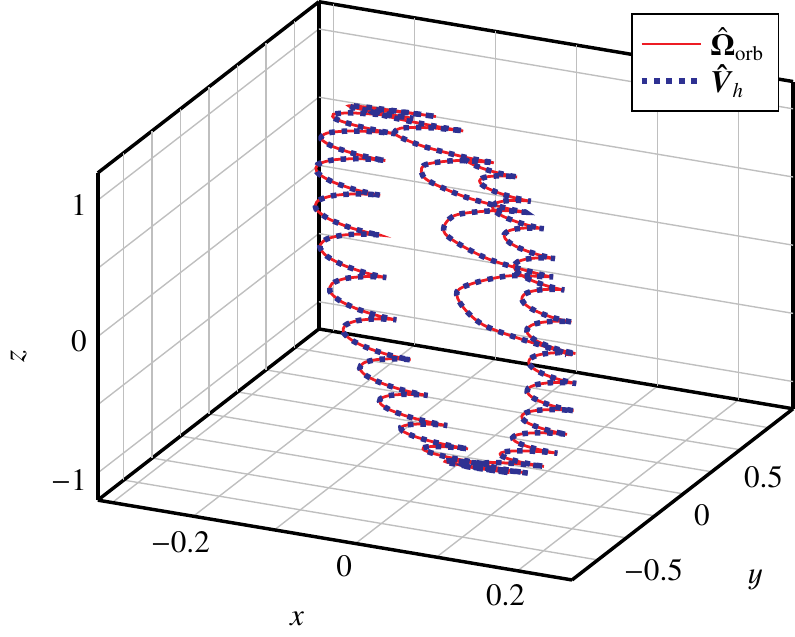} %
  \hfill %
  \includegraphics{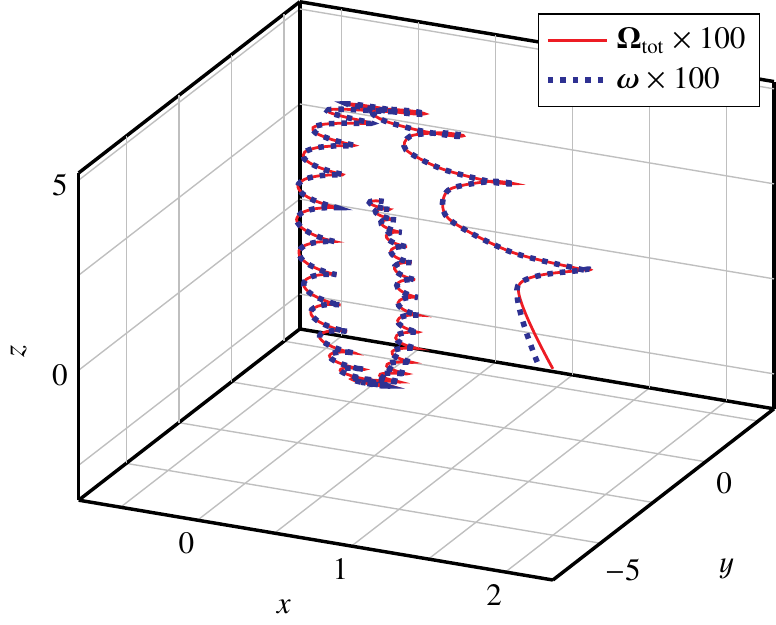} %
  \caption{ \label{fig:OrbitalElementsFromWaveform} %
    \CapName{Orbital elements compared to waveform quantities} These
    plots show the \pN orbital elements $\OrbitalAngularVelocityHat$
    (left) and $\TotalAngularVelocity$ (right), compared to the
    ``\pN-equivalent'' quantities derived from the waveforms alone
    given in Eqs.~\eqref{eq:OrbitalElementsFromWaveform}.  The binary
    has a $6:1$ mass ratio.  Initially, the larger black hole has a
    spin of $S_{1}/M_{1}^{2} = 0.9$ in the $(\vartheta, \varphi) =
    (2.00, 0.25)$ direction; the smaller black hole has a spin of
    $S_{2}/M_{2}^{2} = 0.3$ in the $(\vartheta, \varphi) = (2.4, 2.9)$
    direction.  These parameters were chosen because the resulting
    orbital velocity happens to execute a complete flip, passing very
    close to $-\unitvec{z}$, which is a rigorous test of these
    methods.  In both plots, tighter oscillations correspond to
    earlier times; the last $3600\, M$ before merger are shown.  The
    directions $\OrbitalAngularVelocityHat$ and
    $\LLDominantEigenvector[h]$ are identical to within the numerical
    accuracy throughout the inspiral.  The vectors
    $\TotalAngularVelocity$ and $\FieldAngularVelocity$ are the same
    to within a few parts in $10^{5}$ early in the inspiral, though
    differences grow somewhat as the system approaches merger (roughly
    the end of the data shown here). %
  }
\end{figure*}

Figure~\ref{fig:OrbitalElementsFromWaveform} compares the orbital
elements to the related waveform expressions, for a \pN system with
significant precession.  The direction of $\OrbitalAngularVelocity$
coincides extremely well with $\LLDominantEigenvector$---they agree to
within the numerical precision throughout the inspiral.
$\TotalAngularVelocity$ and $\FieldAngularVelocity$ agree to within a
few parts in $10^{5}$ early in the inspiral, though the disagreement
grows near merger.  However, it may be possible to remove even this
disagreement through more careful treatment of the distinction between
the orbital phase and the phase of a waveform in \pN theory.

Now, in each case of Eqs.~\eqref{eq:OrbitalElementsFromWaveform}, the
quantities on the right-hand side are measured directly from the
waveform.  Thus, if we have a numerical waveform, we can simply
measure the right-hand sides and \emph{define} the ``\pN-equivalent''
orbital elements according to these equations.  A \pN system given
those parameters as initial conditions will necessarily be as similar
to the numerical system as possible---at least by the measures of
$\FieldAngularVelocity$ and $\LLDominantEigenvector$.

This gives us a partial solution to the inverse problem.  We are
lacking four degrees of freedom corresponding to the directions of the
two spin vectors.  One additional piece of information is also
available from the foregoing considerations.  The magnitude
$\PrecessionalFrequency$ is given in \pN theory as a bilinear function
of $\vector{S}_{1} \cdot \unitvec{n}$ and $\vector{S}_{2} \cdot
\unitvec{n}$, where the coefficients depend on the \pN-expansion
parameter $v \defined (M\, \OrbitalFrequency)^{1/3}$, and are
therefore already known.  Thus, we can solve for $\vector{S}_{1} \cdot
\unitvec{n}$, for example.  This suggests other possible methods to
find the remaining three components of spin.  For example, \pN
expressions for the orbital angular \emph{momentum} $\vector{L}$ are
available in terms of the orbital elements and the various projections
of the spin vectors.  If it is practical to measure the total angular
momentum of the spacetime $\vector{J}$ in the numerical
solution~\cite{Moreschi:1986, AdamoEtAl:2009}, we could then use the
\pN expression for $\vector{L} + \vector{S}$ to solve for the
\pN-equivalent components of spin.  This would complete the solution
of the inverse problem.

Alternatively, we might measure various modes of the waveform and
equate them to the \pN expressions for those modes.  Again, these
expressions contain various known quantities, as well as bilinear
combinations of $\vector{S}_{1} \cdot \OrbitalAngularVelocityHat$ and
$\vector{S}_{2} \cdot
\OrbitalAngularVelocityHat$~\cite{WillWiseman:1996,
  BuonannoEtAl:2012}.  Therefore, we could solve for these
combinations of the spin components.  Seemingly, this would rely on
the accuracy of the \pN expressions, which is not very high for spin
terms.  On the other hand, the influence of any errors that result
would be similarly diminished.  One final degree of freedom would
remain in this example, and would have to be fixed by other means.  In
any case, we leave these considerations to future work.

In the computer code included among this paper's
\href{http://arxiv.org/src/1302.2919/anc}{ancillary files}, a waveform
object may be constructed with \code{GWFrames.Waveform}, or a \pN
waveform may be constructed with \code{GWFrames.PNWaveform}.  The
\pN-equivalent orbital and precessional angular velocities may then be
calculated using the \code{PNEquivalentOrbitalAV} and
\code{PNEquivalentPrecessionalAV} methods.

\section{The corotating frame%
  \label{sec:TheFrameItself}}%
So far, we have calculated only the instantaneous angular velocity of
the waveform, $\FieldAngularVelocity$
[Eq.~\eqref{eq:CorotationMinimized}].  While this has already proven
useful in the previous section, it can also be advantageous in
determining a frame in which to decompose the waveform.  Specifically,
we seek a frame whose angular velocity is just
$\FieldAngularVelocity$.  When decomposed in this frame, the waveform
will have no rotation, and will be as constant as possible.  The frame
compares favorably to other frames introduced
previously~\cite{SchmidtEtAl:2011, OShaughnessyEtAl:2011,
  BoyleEtAl:2011} (see Appendix~\ref{sec:PreviousWork}).  It has
practical benefits and suggests simple techniques for measuring,
comparing, and processing waveforms from numerical simulations.

\subsection{Finding the corotating frame}
\label{sec:FindingTheCorotatingFrame}
Our task here is to find the rotor $\quat{R}(t)$ describing a frame
whose angular velocity is $\FieldAngularVelocity$.  We can relate the
two by a simple equation:
\begin{equation}
  \label{eq:FrameAngularVelocity}
  \FieldAngularVelocity = 2\, \dot{\quat{R}}\, \co{\quat{R}}~.
\end{equation}
(See Ref.~\cite{BoyleEtAl:2011} and
Sec.~\ref{sec:IntegratingAngularVelocity}.)  Unfortunately, the
solution we might naively write down is wrong:
\begin{equation}
  \label{eq:BadCorotatingFrame}
  \quat{R}(t) \neq \exp \left[ \frac{1}{2}\, \int^{t}
    \FieldAngularVelocity(t')\, \d t' \right]~,
\end{equation}
except when the system is nonprecessing.  Ultimately, the reason for
the failure of this formula in general is that $\FieldAngularVelocity$
is not parallel to its derivative (or integral).  In the language of
quaternions,\footnote{Note that the failure to commute is by no means
  specific to the quaternion description of rotations; it is a feature
  of rotations themselves.  Quaternions do, however, provide a very
  effective means of solving the problem.} $\FieldAngularVelocity$
fails to commute with its derivative (or integral); to find the
correct version of Eq.~\eqref{eq:BadCorotatingFrame}, we need to
account for that noncommutativity.

Given $\FieldAngularVelocity$, we could solve
Eq.~\eqref{eq:FrameAngularVelocity} for $\dot{\quat{R}}$ and integrate
as we would a vector equation.  But in practice this would quickly
violate the constraint that $\quat{R}$ should be a unit quaternion.
Instead, we will need an expression in terms of the logarithm of this
rotor:
\begin{equation}
  \label{eq:RotorDerivativeAsLogarithms}
  \dot{\quat{R}}\, \co{\quat{R}} = \dot{\quatlog{R}} +
  \frac{\sin^{2}\abs{\quatlog{R}}} {2 \abs{\quatlog{R}}^{2}}\,
  [\quatlog{R}, \dot{\quatlog{R}}] + \frac{\abs{\quatlog{R}} -
    \sin\abs{\quatlog{R}}\, \cos\abs{\quatlog{R}}} {4\,
    \abs{\quatlog{R}}^{3}}\, \big[\quatlog{R}, [\quatlog{R},
  \dot{\quatlog{R}}] \big]~,
\end{equation}
with $\quatlog{R}(t) \defined \log \quat{R}(t)$.  (See
Sec.~\ref{sec:IntegratingAngularVelocity} for the derivation.)  The
second and third terms on the right-hand side of this expression
account for the noncommutativity as needed.  Setting the right-hand
side equal to $\FieldAngularVelocity/2$, we can solve for
$\dot{\quatlog{R}}$ to find
\begin{subequations}
  \label{eq:CorotatingFrame}
  \begin{equation}
    \label{eq:FrameVectorDerivative}
    \dot{\quatlog{R}} = \left( \FieldAngularVelocity -
      \frac{\quatlog{R}\, (\quatlog{R} \cdot \FieldAngularVelocity)}
      {\lvert \quatlog{R} \rvert^{2}} \right)\, \frac{\lvert
      \quatlog{R} \rvert\, \cot \lvert \quatlog{R} \rvert} {2} +
    \frac{\quatlog{R}\, (\quatlog{R} \cdot \FieldAngularVelocity)}
    {2\, \lvert \quatlog{R} \rvert^{2}} + \frac{1}{2}
    \FieldAngularVelocity \times \quatlog{R}~.
  \end{equation}
  This is just an explicit first-order ordinary differential equation,
  so we can integrate numerically using standard techniques to arrive
  at the appropriate $\quatlog{R}(t)$ and find the corotating frame
  \begin{equation}
    \label{eq:CorotatingFrameRotor}
    \Rcorot(t) = \exp \left[ \quatlog{R}(t) \right]~.
  \end{equation}
\end{subequations}
Using the fact that $\quat{R} = \exp\quatlog{R}$ and $-\quat{R} = \exp
\left[ \frac{\lvert \quatlog{R} \rvert -\pi} {\lvert \quatlog{R}
    \rvert}\, \quatlog{R} \right]$ describe the same frame, we can
reset the value of $\quatlog{R}$ between steps of the numerical
integration to keep its magnitude small.  This improves the quality of
the numerical integration, though it may then be useful to go back and
flip the signs of rotors as necessary to keep $\quat{R}(t)$ as
continuous as possible.  The procedure is described in more detail in
Appendix~\ref{sec:IntegratingAngularVelocity}.

The advantage of this method over direct integration of
Eq.~\eqref{eq:FrameAngularVelocity} is that it ensures that the
resulting quaternion truly does have norm $1$.  When integrated
directly, the quaternion in Eq.~\eqref{eq:FrameAngularVelocity} has
four degrees of freedom, whereas a unit quaternion has only three.  By
integrating Eq.~\eqref{eq:FrameVectorDerivative} instead, we eliminate
the extra degree of freedom, reducing this to a truly
three-dimensional problem while automatically satisfying the
constraint on the norm.  In general, transforming equations in such a
way improves the accuracy of numerical results significantly---as is
certainly the case with this system when tested.

Naturally, imposing a condition on the angular velocity of a frame
leaves its overall orientation free.  Assuming $\Rcorot(t)$ describes
a frame whose angular velocity is $\FieldAngularVelocity$, then the
frame $\Rcorot(t)\, \quat{R}_{\text{c}}$ will have the same angular
velocity for any constant $\quat{R}_{\text{c}}$.  Alternatively, the
frame $\quat{R}_{\text{c}}\, \Rcorot(t)$ would have angular velocity
$\FieldAngularVelocity$ rotated by $\quat{R}_{\text{c}}$.  In the
interests of simplifying the waveform, it is best to choose some
particular time during the inspiral at which to align the $z$ axis of
the frame with $\LLDominantEigenvector$, as suggested by O'Shaughnessy
\etal~\cite{OShaughnessyEtAl:2011}.  Once this is done at one instant
of time, $z$ and $\LLDominantEigenvector$ should be aligned at all
other times to very high accuracy.  We are still free to rotate about
the $z$ axis, so we can set the phase of the $(\ell,m)=(2,2)$ mode to
$0$ at this instant, for example.  We will see below that the phase is
very slowly varying in the corotating frame, so this will not be a
delicate operation.  Alternatively, if the waveform is precessing, we
can align the $x$ axis with the \pN-equivalent $\unitvec{n}$, which
should be roughly equivalent to setting the $(2,2)$ phase to $0$.
When comparing two waveforms, the only requirement is that these
instants of time be comparable, which is assured by choosing a common
fiducial quantity.  These issues are discussed further in
Sec.~\ref{sec:Extrapolationcomparisonandhybridization}.

In the computer code included among this paper's
\href{http://arxiv.org/src/1302.2919/anc}{ancillary files}, the
function \code{GWFrames.FrameFromAngularVelocity} returns the
corotating frame, given an array of quaternions representing the
angular-velocity vector as a function of time.  Alternatively, a
waveform object may be constructed with \code{GWFrames.Waveform} and
transformed to the corotating frame with the
\code{TransformToCorotatingFrame} method.

\subsection{Gravitational waveforms in the corotating frame%
  \label{sec:GWsInCorotatingFrame}}%
\begin{figure*}
  \tikzsetfigurename{Figure2_} %
  \includegraphics{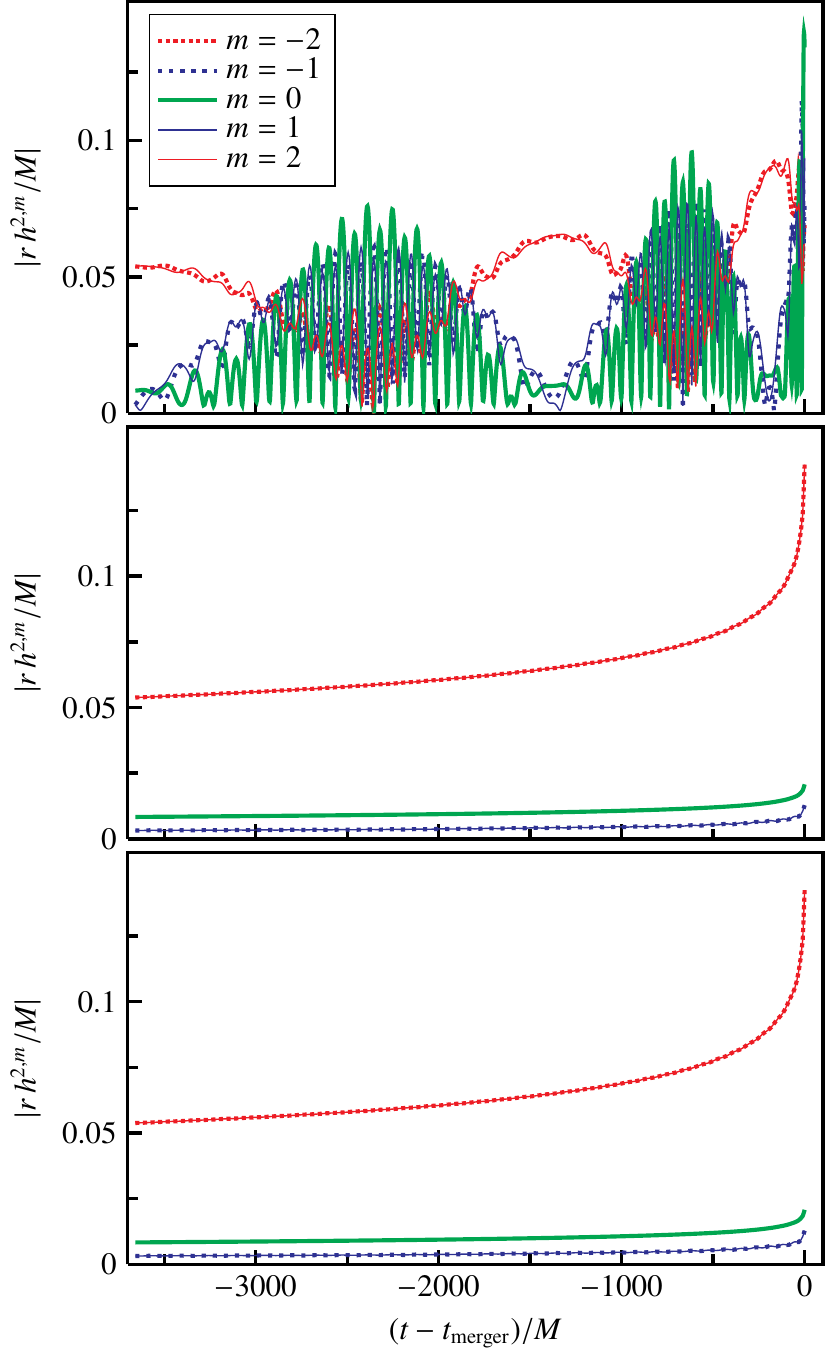} %
  \hfill %
  \includegraphics{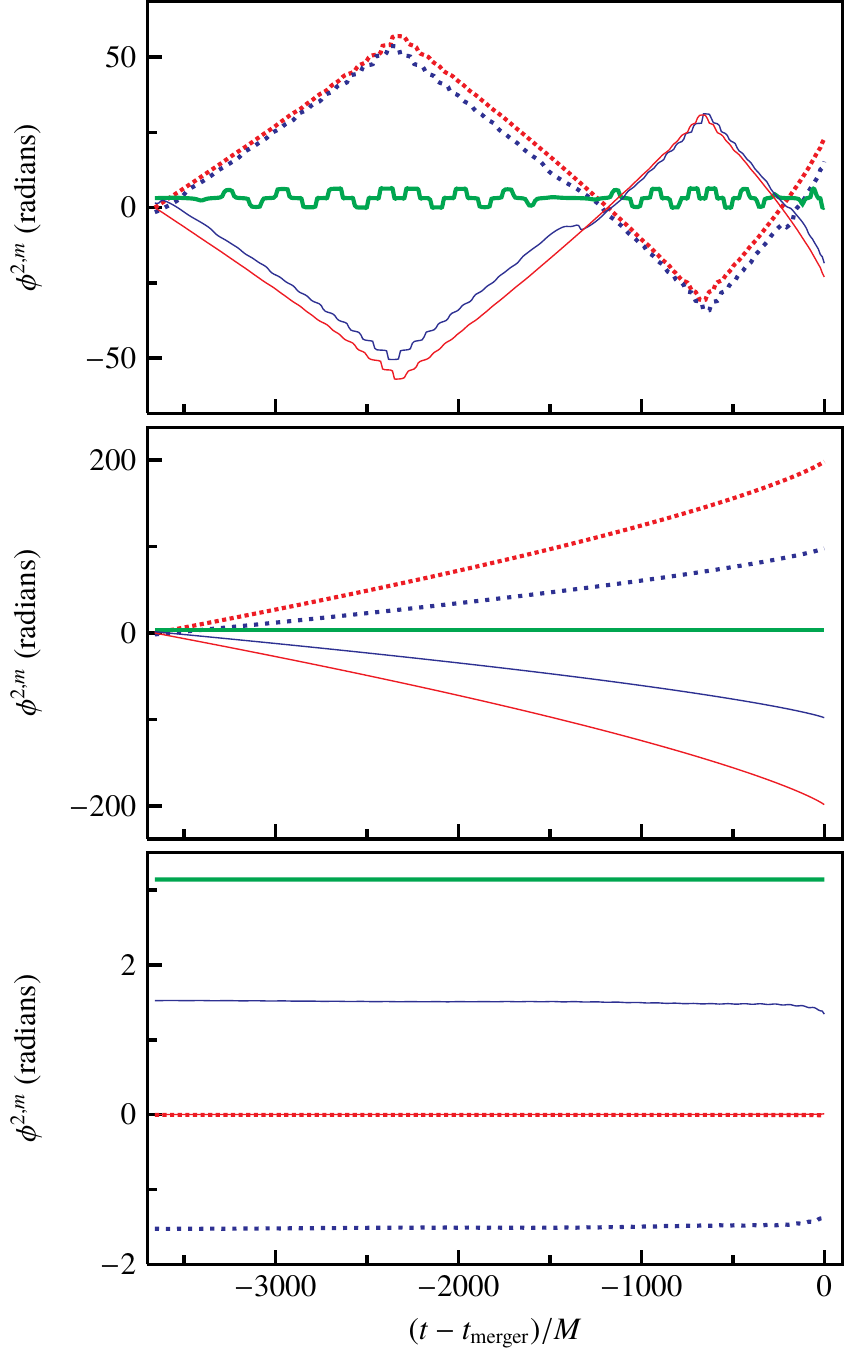} %
  \caption{ \label{fig:Waveforms} %
    \CapName{Precessing waveform in various frames} These plots show
    the modulus (left) and phase (right) of the $\ell=2$ modes of a
    post-Newtonian waveform in the inertial (top), waveform-aligned
    minimal-rotation (middle), and corotating (bottom) frames.  The
    phase is defined as usual~\cite{IUPAPRedBookSymbols:1987,
      ISO80000-2:2009, NINJA2:2011} so that $h^{\ell,m} = \lvert
    h^{\ell,m} \rvert\, \exp[\i\, \phi^{\ell,m}]$, with branch-cut
    discontinuities removed.    The system
    is the same as the one shown in
    Fig.~\ref{fig:OrbitalElementsFromWaveform}.
    Going from the inertial frame to the aligned frame drastically
    simplifies the waveform amplitudes and significantly simplifies
    the phase.  In the aligned frame, the waveform looks very much
    like a nonprecessing waveform~\cite{SchmidtEtAl:2012}.  Expressing
    the waveform in the corotating frame retains the smoothness in
    amplitude seen in the aligned frame, but makes the phases of the
    modes nearly constant, with values of roughly $0$, $\pm \pi/2$,
    and $\pi$.  Similar results can also be seen for modes with
    $\ell>2$. %
  }
\end{figure*}

Figure~\ref{fig:Waveforms} demonstrates the effects of decomposing the
\pN waveform described in Fig.~\ref{fig:OrbitalElementsFromWaveform}
in various frames.  
First is the usual inertial frame, where a stationary observer at
infinity has constant coordinate position.  In this frame, the moduli
of the modes oscillate wildly, as power shifts between them (upper
left panel of Fig.~\ref{fig:Waveforms}).  Similarly, the phase (upper
right panel) shows strange features.  The $\ell=\pm 2$ and $\ell=\pm
1$ modes have roughly the same frequency, as power from the dominant
modes leaks into and overwhelms the $\ell = \pm 1$ modes.  Those
phases also change direction each time the rotation axis passes
through the $x$--$y$ plane (at times of roughly $-2400$ and $-600$).
Both the modulus and phase are very complicated functions in this
frame.  They would be hard to model directly, and their rapid
variations are not conducive to accurate numerics.

The next pair of panels shows the waveform in the waveform-aligned
frame suggested by O'Shaughnessy \etal~\cite{OShaughnessyEtAl:2011}
supplemented with the minimal-rotation condition (see
Sec.~\ref{sec:ImposingTheMinimalRotationCondition} and
Ref.~\cite{BoyleEtAl:2011}).  Here, $\LLQuantity$ is evaluated using
all modes up to $\ell=8$, the dominant eigenvector is found, and the
fame is rotated so that its $z$ axis coincides with that eigenvector
while obeying the minimal-rotation condition.  This drastically
simplifies both the modulus and phase, as seen in the middle panels of
Fig.~\ref{fig:Waveforms}.  The modulus is very smoothly sweeping up as
the binary spirals in toward merger.  The phases are now separated as
usual, with slopes more nearly equal to $-m\, \OrbitalFrequency$.

Decomposing the waveform in the waveform-aligned frame also requires
recording the orientation of the frame.  In that case, there is no
additional overhead in going to the corotating frame, which is shown
in the lower panels of Fig.~\ref{fig:Waveforms}.  The modulus plot is
identical to the one in the previous case.  However, there is further
improvement in the phase, with each mode having nearly constant phase
throughout.  Similar behavior is seen in other modes with $\ell>2$.
Naturally, such a waveform is particularly well suited to
interpolation and hybridization~\cite{CampanelliEtAl:2009,
  BoyleEtAl:2011, SchmidtEtAl:2012}.

A minor feature to note in the phase is the non-constancy of the $(2,
\pm 1)$ modes.  Considered on their own, these variations could be
removed by a rotation because the curves change in opposite
directions.  On the other hand, this would cause the phases of other
modes to vary.  Equation~\eqref{eq:CorotationMinimized} automatically
balances these concerns; the amplitudes of the $(2, \pm 1)$ modes are
so small that they do not carry much weight.  By transforming to the
corotating frame, we isolate the waveform's intrinsic dynamics---seen
here in $\phi^{2, \pm 1}$---from the rotational dynamics of the
system, allowing for separate analyses.  This is important because
they are separately modeled, so it is useful to be able to inspect
each effect on its own.

\subsection{Extrapolation, comparison, alignment, and hybridization%
  \label{sec:Extrapolationcomparisonandhybridization}}%
As a practical matter, we need to manipulate numerical waveforms in
various ways.  We must eliminate physical and gauge effects associated
with extraction of data at finite radius, usually by
extrapolation~\cite{BoyleEtAl:2007, Boyle:2008,
  BoyleMroue:2009}.\footnote{Cauchy-characteristic extraction is
  another method of finding the correct waveform at
  $\scriplus$~\cite{BishopEtAl:1996, Winicour:2012,
    ReisswigEtAl:2012}.  This can proceed as usual, and the final
  waveform can be transformed to the corotating frame.  A recent
  implementation~\cite{ReisswigEtAl:2012} has improved the efficiency
  to make this a more attractive alternative to extrapolation.}  To
compare numerical waveforms to each other or to analytical waveforms,
we need to determine the extrinsic parameters corresponding to freedom
in choosing the zero of time and the overall orientation of our axes.
To construct a complete waveform, we may occasionally need to
hybridize waveforms from different systems~\cite{CampanelliEtAl:2009,
  BoyleEtAl:2009, Boyle:2011, BoyleEtAl:2011, MacDonaldEtAl:2011,
  MacDonaldEtAl:2013, SchmidtEtAl:2012}.  Various approaches to these
problems have been introduced, but most are designed exclusively for
nonprecessing systems, or are otherwise incomplete or fragile.  The
angular velocity and corotating frame provide excellent tools for
addressing these issues more generally, and are especially robust when
implemented by means of quaternions.

Practical extrapolation relies on smoothness of the extrapolated
functions~\cite{BoyleMroue:2009}.  In the corotating frame, the modes
of the waveform are essentially constant during inspiral---except for
the overall growth in modulus---suggesting that this is the ideal
frame for extrapolation.  We can implement such a procedure in the
usual way, with one minor addition.  We first impose a
time-retardation offset to all the data, as usual.  Then, we add the
step of finding the corotating frame of the outermost extracted data,
and transforming the data at all radii to that frame.  (We cannot
rotate data at each radius into its own corotating frame, as that
would require extrapolation of rotors, which is not well understood.)
The waveforms at the other radii will not be precisely in their own
corotating frames, but should be close enough that the data are quite
smooth.  The extrapolation may then proceed as usual, resulting in an
extrapolated waveform in the corotating frame of the outermost data.
Again, the extrapolated result will not be precisely its own
corotating frame, but the transformation is routine.

In the nonprecessing case, a multitude of methods have been suggested
to compare and align waveforms (fixing the extrinsic parameters) and
to construct hybrids of NR and \PN waveforms~\cite{NINJA2:2012}.
These all require generalization to use in the case of significant
precession.  By using the corotating frame, we simplify the modes of
the waveform decomposition sufficiently that ordinary methods can
still be used.  (Comparing phase differences or relative differences
in modulus, for example.)  But each waveform now comes with its own
frame, $\quat{R}_{A}(t)$ and $\quat{R}_{B}(t)$, encoding most of the
phase dynamics, so we will also need to compare the difference
between the frames themselves.  Fortunately, quaternions provide us
with \emph{geometrically meaningful} measures of the difference.

\begin{subequations}
  \label{eq:RotorDifferences}
  The difference itself is given in quaternion form as\footnote{The
    inverse of an arbitrary nonzero quaternion $\quat{Q}$ is just
    $\co{\quat{Q}} / \lvert \quat{Q} \rvert^{2}$.  Since rotors have
    norm $1$, the inverse of a rotor $\quat{R}$ is just
    $\co{\quat{R}}$.  Therefore, this formula is analogous to
    subtraction, but applied to rotation operators.}
  \begin{equation}
    \label{eq:DifferenceRotation}
    \quat{R}_{\Delta}(t) \defined \quat{R}_{A}(t)\,
    \co{\quat{R}}_{B}(t)~.
  \end{equation}
  This is the rotation taking frame $B$ into frame $A$, and is
  independent of the basis with respect to which $A$ and $B$ are
  defined.  Now, we might want to know how ``big'' this difference
  rotation is.  As noted in Sec.~\ref{sec:QuaternionNotation}, we can
  write any rotation, including $\quat{R}_{\Delta}$, in axis-angle
  form:
  \begin{equation}
    \label{eq:DifferenceRotationAxisAngle}
    \quat{R}_{\Delta}(t) = \exp \left[ \vector{\Phi}_{\Delta}/2
    \right]~.
  \end{equation}
  We can easily solve this equation for $\vector{\Phi}_{\Delta}$ by
  taking the logarithm.  In particular, its magnitude is the angle
  through which the system must be rotated:
  \begin{equation}
    \label{eq:DifferenceRotationAngle}
    \Phi_{\Delta}(t) = 2 \abs{\log \quat{R}_{\Delta}(t)}~.
  \end{equation}
  This can be used as a simple but complete description of the phase
  difference between two systems.\footnote{It is crucial to note that
    $\log (\quat{R}_{A}\, \co{\quat{R}}_{B}) \neq \log \quat{R}_{A} -
    \log \quat{R}_{B}$ because rotations do not commute.  In
    particular, the latter depends on the basis frame, and is
    therefore not a useful measure of the difference between frames.}
\end{subequations}

We can understand this better and make contact with previous work by
recognizing $\vector{\Phi}_{\Delta}$ as a more general version of a
common measure of the difference between waveforms common in analysis
of nonprecessing systems.  That measure is $\Delta \phi^{\ell,m}$, the
difference between the phases of the modes as measured in the static
frame.  Because the angular velocity is conventionally chosen to be
along the $z$ axis, we can usually relate the orbital phase
$\OrbitalPhase$ to the waveform phase as $\phi^{\ell,m} \approx -m\,
\OrbitalPhase$.  Here, we have the similar expression
\begin{equation}
  \label{eq:RelatingMeasuresOfPhaseDifference}
  \Delta \phi^{\ell,m} \approx -m\, \Phi_{\Delta}~.
\end{equation}
However, as we saw in the previous section, $\Delta \phi^{\ell,m}$ is
a less useful measure for precessing systems.  In contrast,
$\vector{\Phi}_{\Delta}$ encapsulates the differences in both the
orbital and the precessional dynamics\footnote{If needed, the orbital
  evolution can be further isolated from the precessional dynamics
  using the minimal-rotation frame.} in one convenient function while
leaving the waveform dynamics separate, and is equally relevant in
both precessing and nonprecessing systems.

The quantity $\Phi_{\Delta}$ gives us a compact description of the
difference between two waveforms as measured in their corotating
frames.  But it depends on the extrinsic parameters discussed in
Sec.~\ref{sec:ThePNSystem}: the overall time offset and orientation of
the static basis frame.  Alignment of waveforms consists of minimizing
differences between the waveforms at some instant or over some span of
time by adjusting the extrinsic parameters as needed.  This can be
seen as a restricted version of the inverse problem, where we simply
assume that the intrinsic parameters are identical---as when we wish
to compare waveforms evolved the same initial data with different
numerical resolution.  Section~\ref{sec:DeducingParameters} used
implicit assumptions that the time coordinate and basis frame would be
defined to be the same in both waveforms, and the physical parameters
are expected to be the same in that frame.  Here, we are simply given
two waveforms, which have arbitrary time and orientation offsets.
They must be aligned more actively.

We can separate this into two steps: first align the time, then align
the frames.  To align the time, we will need some measure of the
waveform that is independent of orientation.  For example, we can use
the magnitude of the angular velocity $\lvert \FieldAngularVelocity
\rvert$.  We then choose some fiducial time $\tfid$ and find the value
of $\delta t$ such that
\begin{equation}
  \label{eq:AlignTime}
  \abs{ \FieldAngularVelocity_{A}(\tfid) } = \abs{
    \FieldAngularVelocity_{B}(\tfid + \delta t) }~.
\end{equation}
The time coordinates of waveform $B$ may then be shifted as $t \mapsto
t + \delta t$.  The main limitation with this method is that the
magnitude $\lvert \FieldAngularVelocity \rvert$ is not always strictly
monotonic for highly precessing systems.  Usually it is possible to
find a time for which it is monotonic.  Alternatively, we can find
$\delta t$ by minimizing the squared difference between the two sides
of Eq.~\eqref{eq:AlignTime} integrated over some significant span of
time.  There may also be quantities other than $\lvert
\FieldAngularVelocity \rvert$ that may be slightly more robust against
this non-monotonicity or numerical noise---quantities such as flux or
the total power in the waveform.

Now, once the time coordinates have been properly aligned, it is a
simple matter to align the frames.  We simply apply the transformation
$\quat{R}_{B}(t) \mapsto \quat{R}_{\Delta}(\tfid)\, \quat{R}_{B}(t)$,
using $\quat{R}_{\Delta}$ as given in
Eq.~\eqref{eq:DifferenceRotation}.  Then, at $\tfid$, there will be
precisely no difference between the frames; in particular,
$\Phi_{\Delta}(\tfid)=0$.  Reference~\cite{BoyleEtAl:2011} suggested
essentially this same transformation, but included an additional
rotation about the $z$ axis because there was still one degree of
rotational freedom in that paper.  Here, we have assumed that the
orientation of each waveform has been completely fixed at $\tfid$, as
discussed in Sec.~\ref{sec:FindingTheCorotatingFrame}, though we may
require a fixed rotation to the physical system of one.  In
particular, following the discussion at the end of
Sec.~\ref{sec:FindingTheCorotatingFrame}, we see that this
transformation actually rotates $\FieldAngularVelocity_{B}$ by
$\quat{R}_{\Delta}$.  Note that no rotation of the waveform modes is
to be done here; we are only changing how we think of the frame in
which those modes are decomposed.

An alternative approach involves simultaneously fixing the time and
frame.  In previous work with nonprecessing systems, this was done by
minimizing the squared difference in some quantity ($\Delta
\phi^{2,2}$, for example) integrated over some span of time.  This was
used in an effort to nullify spurious effects such as junk radiation
or residual eccentricity~\cite{MacDonaldEtAl:2011,
  MacDonaldEtAl:2013}.  A similar program can certainly be applied to
$\Phi_{\Delta}$ by minimizing
\begin{equation}
  \label{eq:AlignmentMinimization}
  \Upsilon(\delta t, \quat{R}_{\delta}) \defined
  \int_{t_{1}}^{t_{2}}\, 4\, \abs{\log \left[ \quat{R}_{A}(t)\,
      \co{\quat{R}}_{B}(t + \delta t)\, \co{\quat{R}}_{\delta} \right]
  }^{2}\, \d t~.
\end{equation}
This requires simultaneously optimizing over the time offset and all
three degrees of rotational freedom.  In particular, a simplification
that occurs in the nonprecessing case and allows the problem to be
reduced to one dimension~\cite{BoyleEtAl:2008} will not work in the
precessing case due to noncommutativity of rotations; the problem must
remain truly four-dimensional.  Nonetheless, the minimization is
straightforward, and the frame is adjusted as $\quat{R}_{B}(t) \mapsto
\quat{R}_{\delta}\, \quat{R}_{B}(t + \delta t)$.  Again, the waveform
modes themselves are not to be rotated; just the frame information.

Once the waveforms are aligned, it is a simple matter to hybridize
them with a slight generalization of the standard method.  Typically,
we use only information from waveform $A$ before some time $t_{1}$,
and only information from waveform $B$ after some $t_{2}$, with a
transition in between.  The waveforms are assumed to have been aligned
somewhere in the range between $t_{1}$ and $t_{2}$.  The transition
may be accomplished with some (usually smooth) monotonic function
$\tau(t)$ that equals $1$ before $t_{1}$ and $0$ after $t_{2}$.  Then,
the hybrid version of any mode $\wavefield^{\ell,m}$ may be defined as
a simple linear interpolation between the two waveforms:\footnote{In
  previous work, this formula is usually applied separately to the
  phase and modulus of the mode.  With this new description, there
  seems to be no advantage in decomposing the waveform in this way.
  The formula given here is written assuming complex mode data.}
\begin{subequations}
  \label{eq:Hybridization}
  \begin{equation}
    \label{eq:HybridWavefield}
    \wavefield^{\ell,m}_{\text{hybrid}} \defined
    \wavefield^{\ell,m}_{A}(t)\, \tau(t) +
    \wavefield^{\ell,m}_{B}(t)\, [1-\tau(t)]~.
  \end{equation}
  Again, however, each waveform comes with its own frame, and these
  frames have to be hybridized.  As suggested in
  Ref.~\cite{BoyleEtAl:2011}, this can be accomplished with a form of
  linear interpolation defined for rotors:
  \begin{equation}
    \label{eq:HybridRotation}
    \quat{R}_{\text{hybrid}} \defined L\Big(\tau(t); \quat{R}_{A}(t),
    \quat{R}_{B}(t) \Big)~,
  \end{equation}
\end{subequations}
where the interpolant $L$ is given by Eq.~\eqref{eq:Slerp}.  As
discussed in Appendix~\ref{sec:Interpolation}, it is critically
important to use the correct interpolant.  Finally, we must note we
only have the frames $\quat{R}_{A}$ and $\quat{R}_{B}$ sampled at
discrete (essentially arbitrary) points, so we will need to
interpolate between those points to the desired $t$.  For this,
smoother interpolation is required.  Appendix~\ref{sec:Interpolation}
discusses a method using cubic splines reinterpreted for rotors.

In the computer code included among this paper's
\href{http://arxiv.org/src/1302.2919/anc}{ancillary files}, waveform
objects may be constructed with \code{GWFrames.Waveform}.  They may be
aligned, compared, and hybridized with methods such as
\code{AlignTime}, \code{AlignFrame}, \code{AlignTimeAndFrame},
\code{Compare}, and \code{Hybridize}.

\section{Conclusions%
  \label{sec:Conclusions}}%
The angular velocity of a waveform was defined in
Sec.~\ref{sec:AngularVelocityOfAWaveform} by the rotation which
minimizes the time dependence of the waveform.  This fairly nebulous
criterion was reformulated precisely, and led to a simple formula,
providing us with a geometrically meaningful description of the motion
of a waveform.  We also saw that $\FieldAngularVelocity$ can be
considered to be that vector which makes the action of the operator
$-\i\, \FieldAngularVelocity \cdot \AngMomOpVec$ as equal as possible
to the action of $\partial_{t}$, in a sense that can also be made
surprisingly precise.

The angular-velocity vector and the dominant eigenvector
$\LLDominantEigenvector$ of $\LLQuantity$ proposed by O'Shaughnessy
\etal~\cite{OShaughnessyEtAl:2011} provide us with powerful tools to
understand and manipulate waveforms, with no reference to meaningless
gauge quantities.  Section~\ref{sec:SolvingTheInverseProblem} showed
that these two vectors can be used very effectively and accurately to
find at least part of the solution to the important inverse problem.
Determining the three remaining degrees of freedom is left for future
work, though some suggestions were made for how to do this.

Beyond this fundamental benefit, $\FieldAngularVelocity$ also provides
key practical advantages.  We can readily calculate the corotating
frame, which also has angular velocity $\FieldAngularVelocity$.
Transformed to this frame, the waveform is literally as constant as
possible.  When functions are slowly varying, they are easily
approximated by low-order functions; they can be numerically
interpolated and differentiated quite accurately; and fewer data
points are needed to record their values than for quickly varying
functions.  This type of technique has already seen great success in
numerical simulations themselves~\cite{ScheelEtAl:2006,
  HembergerEtAl:2012}.

Putting these together, we can also perform all the standard
manipulations needed for waveform analysis.  Data collected from a
simulation at different radii can be extrapolated nicely.  Two
waveforms (\eg, different resolutions of a numerical simulation, or an
NR and a \pN waveform) can be aligned, compared, and hybridized
readily.  The only additional steps necessary are comparison and
hybridization of the frames, but these are easily achieved using
formulas given by Eqs.~\eqref{eq:DifferenceRotation}
and~\eqref{eq:HybridRotation}.  Notably, the use of quaternions vastly
improves numerics and allows us to access the geometrically meaningful
elements of rotations.

The code included with this paper implements all the techniques
discussed above, showing that they are ready to use in waveform
analysis.  There are, however, issues that may benefit from further
investigation.  As mentioned, more work is needed to complete the
solution of the inverse problem.  Also, it is certainly possible that
different techniques could further simplify the ringdown, for example.
While preliminary results show that reasonable, smooth results for
$\FieldAngularVelocity$ and $\LLDominantEigenvector$ are obtained
throughout the inspiral, merger, and ringdown, the waveform in any
frame still has very complicated structure during ringdown, presumably
stemming from the difference between spin-weighted \emph{spheroidal}
harmonics and \emph{spherical} ones~\cite{PanEtAl:2011,
  KellyBaker:2012}.  More specific methods~\cite{YangEtAl:2012,
  OShaughnessyEtAl:2012, KellyBaker:2012} will likely be needed to
adequately capture features of general ringdowns with simple models.

Nonetheless, we can conclude that $\FieldAngularVelocity$ and
$\LLDominantEigenvector$ already deliver a complete system for
waveform analysis.  When implemented with quaternion methods, the
system is robust enough to be applied blindly to both precessing and
nonprecessing systems.  This consistency simplifies the production and
analysis of both types of waveform.

\begin{acknowledgments}
  It is my great pleasure to thank Larry Kidder, Abdul Mrou{\'{e}},
  Evan Ochsner, Richard O'Shaughnessy, Sergei Ossokine, Robert Owen,
  Harald Pfeiffer, Christian Reisswig, and Saul Teukolsky for useful
  conversations and comments on earlier drafts of this paper.  I also
  appreciate the hospitality of the Kavli Institute for Theoretical
  Physics at UC Santa Barbara during the early stages of this work.
  This project was supported in part by a grant from the Sherman
  Fairchild Foundation; by NSF Grants No.\ PHY-0969111, No.\
  PHY-1005426, No.\ PHY11-25915, and No.\ PHY11-25915; and by NASA
  Grant No.\ NNX09AF96G.
\end{acknowledgments}
\appendix 

\section{Quaternions and rotations%
  \label{sec:QuaternionsAndRotations}}%
Unit quaternions clearly constitute the representation of choice when
computing with spatial rotations.  Quaternions have become the
dominant technique in fields as diverse as computer graphics,
robotics, molecular dynamics, navigation, and orbital mechanics.  They
are closely related to the axis--angle formalism, which gives us clear
geometric intuition and avoids the problem of gimbal lock associated
with singularities of the Euler angles.  But there is also a clear
notion of them as operators, giving us all the advantages of the
matrix representation of rotations.  They are trivially inverted and
easily composed, and the logarithm and exponential functions are easy
to evaluate, presenting further advantages over all other
representations.

For all these reasons, this paper and the accompanying code use the
notation of quaternions.  Here, the basic elements of quaternion math
are summarized, Wigner's $\D$ matrices and the spin-weighted spherical
harmonics are expressed directly in terms of quaternions, and formulas
for the linear interpolants and splines of rotors are given.  The
computer code included among this paper's
\href{http://arxiv.org/src/1302.2919/anc}{ancillary files} contains
all of the quaternion functions discussed here.  The fundamental
object is the \code{GWFrames.Quaternion}, which has numerous methods.
See the documentation for more details.  Also, note that
\code{Mathematica}~\cite{Mathematica8} returns incorrect results for
logarithms of general quaternions, and is thus not a reliable tool for
most of the calculations in this paper.

\subsection{Elements of quaternion mathematics%
  \label{sec:ElementsOfQuaternionMath}}%
\begin{table}
  \caption{\label{tab:QuaternionNotation} Quaternion notation}
  \begin{ruledtabular}
    \begin{tabular}{rcll}
      &$\quat{Q}$ & Quaternion &\\
      &$\quatcomponent{Q}{\alpha}$ & Component $\alpha$ of the
      quaternion &\\
      &$\co{\quat{Q}}$ & Conjugate: $(\quatcomponent{Q}{0},
      -\quatcomponent{Q}{1}, -\quatcomponent{Q}{2},
      -\quatcomponent{Q}{3})$ &\\
      &$\quatabs{Q}$ & Norm: $\sqrt{\smash[b]{\quatcomponent{Q}{0}^{2}
          + \quatcomponent{Q}{1}^{2} + \quatcomponent{Q}{2}^{2} +
          \quatcomponent{Q}{3}^{2}}}$ &\\
      &$\quatvec{Q}$ & Vector part: $( \quatcomponent{Q}{1},
      \quatcomponent{Q}{2}, \quatcomponent{Q}{3} )$ &\\
      &$\quatvecmag{Q}$ & Magnitude of vector part:
      $\sqrt{\smash[b]{\quatcomponent{Q}{1}^{2} +
          \quatcomponent{Q}{2}^{2} + \quatcomponent{Q}{3}^{2}}}$ &\\
      &$\quatunitvec{Q}$ & Normalized vector part: $\quatvec{Q} /
      \quatvecmag{Q}$ &\\
      &$\quatlog{Q}$ & Logarithm: $\log \quat{Q}$ \\
      &$\quatlogmag{Q}$ & Magnitude of the logarithm:
      $\abs{\quatlog{Q}}$ \\
      &$\quatangle{R}$ & Angle of a rotation: $2\, \quatlogmag{R} =
      2\, \abs{ \log \quat{R} }$
    \end{tabular}
  \end{ruledtabular}
\end{table}

A quaternion is a set of four numbers, usually denoted as
\begin{equation}
  \label{eq:Quaternion}
  \quat{Q} = (\quatcomponent{Q}{0}, \quatcomponent{Q}{1},
  \quatcomponent{Q}{2}, \quatcomponent{Q}{3}) = \quatcomponent{Q}{0} +
  \quatvec{Q}~.
\end{equation}
We summarize the notation in Table~\ref{tab:QuaternionNotation}.  The
quaternions form an algebra, meaning that the quaternions form a
vector space (over the real numbers), as well as a group where the
product is defined by
\begin{equation}
  \label{eq:QuaternionMultiplication}
  \quat{P}\, \quat{Q} = (\quatcomponent{P}{0}\, \quatcomponent{Q}{0} -
  \quatvec{P} \cdot \quatvec{Q}) + (\quatcomponent{P}{0}\, \quatvec{Q}
  + \quatcomponent{Q}{0}\, \quatvec{P} + \quatvec{P} \times
  \quatvec{Q})~.
\end{equation}
Here, the dot product and cross product of vectors take their usual
meanings.  Note that this product is neither commutative nor
anti-commutative in general.  The conjugate of a quaternion is defined
as
\begin{equation}
  \label{eq:QuaternionConjugate}
  \co{\quat{Q}} \defined (\quatcomponent{Q}{0}, -\quatcomponent{Q}{1},
  -\quatcomponent{Q}{2}, -\quatcomponent{Q}{3}) = \quatcomponent{Q}{0}
  - \quatvec{Q}~,
\end{equation}
and the norm of a quaternion according to
\begin{equation}
  \label{eq:QuaternionNorm}
  \quatabs{Q}^{2} = \quat{Q}\, \co{\quat{Q}} =
  \quatcomponent{Q}{0}^{2} + \quatcomponent{Q}{1}^{2} +
  \quatcomponent{Q}{2}^{2} + \quatcomponent{Q}{3}^{2} =
  \quatcomponent{Q}{0}^{2} + \quatvec{Q} \cdot \quatvec{Q}~.
\end{equation}
A unit quaternion is simply a quaternion with unit norm.  Because
quaternion multiplication is associative, we can find a useful inverse
of a quaternion by taking the conjugate and dividing by the squared
norm:
\begin{equation}
  \label{eq:ConjugateInverse}
  \quat{Q}^{-1} = \frac{\co{\quat{Q}}} {\quatabs{Q}^{2}}~.
\end{equation}
In particular, the inverse of a unit quaternion is just its conjugate.
Note, however, that while a unit quaternion has \emph{norm}
$\quatabs{R} = 1$, its \emph{square} is not $1$ in general.  For
example, if $\quat{R} = \unitvec{u}$ is some unit vector, we have
$\quat{R}^{2} = -1$.

Now, given any vector $\vector{v}$, we can define the transformation
law
\begin{equation}
  \label{eq:VectorRotation}
  \vector{v}\,' = \quat{R}\, \vector{v}\, \co{\quat{R}}~,
\end{equation}
where the right-hand side involves quaternion multiplication with
$\vector{v}$ interpreted as a quaternion with scalar part
$\quatcomponent{V}{0}=0$.  It is not hard to check that if $\quat{R}$
has unit magnitude, then this transformation law preserves
orientation, angles, and lengths---and is therefore a rotation.  These
rotations compose in the natural way, and we will see below that we
can construct a unit quaternion representing any desired rotation,
which means that the unit quaternions form a representation of the
rotation group.\footnote{In fact, because of the double-sided rotation
  law, Eq.~\eqref{eq:VectorRotation}, $\quat{R}$ and $-\quat{R}$
  represent the same rotation, so the unit quaternions provide a
  double cover of the rotation group $\SO3$; the group of unit
  quaternions is actually isomorphic to $\SU2$.  Unsurprisingly, the
  logarithms [defined in Eq.~\eqref{eq:QuaternionLogarithm}] of unit
  quaternions form a group isomorphic to $\su2$.  Hence the notation
  $\quatlog{R} = \log \quat{R}$.}

Using the product law for quaternions, we can define the exponential
of a quaternion according to the standard power series:
\begin{equation}
  \label{eq:QuaternionExponential}
  \exp \quat{Q} \defined \sum_{n=0}^{\infty} \frac{\quat{Q}^{n}}
  {n!}~.
\end{equation}
Note that, because of the non-commutativity of quaternion
multiplication, the usual rules of exponents do not apply.  In
particular, $\exp [ \quat{P} + \quat{Q} ] \neq \exp \quat{P}\,
\exp\quat{Q}$ unless $\quat{P}$ and $\quat{Q}$ commute---which happens
precisely when their vector parts are parallel.  Given some angle
$\theta$ and some unit vector $\unitvec{u}$, we can show that the unit
quaternion\footnote{More generally, the exponential of any quaternion
  is $\exp \quat{Q} = \exp \quatabs{Q}\, \exp(\quat{Q}/\quatabs{Q})$,
  where the second factor can be evaluated according to
  Eq.~\eqref{eq:RotationAxisAngle}.}
\begin{equation}
  \label{eq:RotationAxisAngle}
  \quat{R} = \exp \left[ \frac{\theta}{2}\, \unitvec{u} \right] = \cos
  \frac{\theta}{2} + \unitvec{u}\, \sin \frac{\theta}{2}
\end{equation}
represents a rotation through the angle $\theta$ about the axis
$\unitvec{u}$ (in the positive sense, using the right-hand rule).
This illustrates the connection between the axis--angle and the
unit-quaternion representations of rotation.  The factor of $1/2$
needed in the exponential is a result of the double-sided rotation
law, Eq.~\eqref{eq:VectorRotation}.

By inspection of Eq.~\eqref{eq:RotationAxisAngle}, we see that we can
also define a reasonable logarithm of nonzero
quaternions:\footnote{Again, this expression generalizes the more
  familiar complex relation $\log z = \log\, \lvert z \rvert + \i\,
  \arctan \Im z / \Re z$, where $\i$ is replaced by a general
  three-vector.}
\begin{equation}
  \label{eq:QuaternionLogarithm}
  \log \quat{Q} \defined \log \quatabs{Q} + \frac{\quatvec{Q}}
  {\quatvecmag{Q}}\, \arctan \frac{\quatvecmag{Q}}
  {\quatcomponent{Q}{0}}~.
\end{equation}
Note that the logarithm of a unit quaternion will be a pure
vector---$\log \quatabs{Q}=0$.  For compactness, we define the
notation $\quatlog{Q} \defined \log \quat{Q}$ and $\quatlogmag{Q}
\defined \abs{\quatlog{Q}}$.  As with the usual complex logarithm and
the real arctangent function, this function is multivalued; the
magnitude of the vector part is ambiguous up to integer multiples of
$2\,\pi$.  We typically choose the principal value so that the norm of
the vector part is in $[0, \pi]$, as with the complex logarithm.
Choosing the branch must be done carefully, in order to obtain correct
geometric results and reasonably continuous functions of time.  When
differentiating the logarithm (as in
Sec.~\ref{sec:IntegratingAngularVelocity} for example), we will treat
the function as being continuous.  On the other hand, sometimes in the
very same formula, we will assume the logarithm takes on its principal
value.

The principal value of the quaternion logarithm can actually be
restricted further if our purpose is only to cover $\SO3$.  Because of
the double-sided rotation law of Eq.~\eqref{eq:VectorRotation}, the
final vector is invariant under $\quat{R} \mapsto -\quat{R}$.  This is
equivalent to
\begin{equation}
  \label{eq:RestrictedPrincipalValue}
  \quatlog{R} \mapsto \frac{\quatlogmag{R}-\pi} {\quatlogmag{R}}\,
  \quatlog{R}~.
\end{equation}
In particular, we can apply this formula when $\quatlogmag{R} >\pi/2$,
ensuring that $\quatlogmag{R} \in [0,\pi/2]$.  The transformation
gives rise to different \emph{rotors}, but the same
\emph{rotation}---at least for objects of integral spin weight, like
vectors.  We will use this fact when integrating the angular velocity
in Sec.~\ref{sec:IntegratingAngularVelocity} to eliminate an edge case
and improve numerical behavior.

As with exponents of real numbers, we can define
\begin{equation}
  \label{eq:QuaternionPower}
  \quat{Q}^{\quat{P}} \defined \exp[ \quat{P}\, \log \quat{Q}]~.
\end{equation}
This formula will be usually be applied in cases where $\quat{P}$ is a
pure real number, though other formulas may be advantageous in such
cases---as illustrated in the case of the square-root below.

The square root of a quaternion is particularly useful in constructing
rotations taking one vector into another as directly as possible.  We
can find a formula for it with an elegant geometric interpretation and
important numerical advantages over Eq.~\eqref{eq:QuaternionPower}
with $\quat{P}=1/2$.  The product of two unit vectors $-\unitvec{u}\,
\unitvec{w}$ is a rotation in the $\unitvec{u}$--$\unitvec{w}$ plane
of twice the angle between those vectors, in the sense from
$\unitvec{w}$ to $\unitvec{u}$.  The square root of this product is
the same rotation through only half that angle---in particular,
$\sqrt{-\unitvec{u}\, \unitvec{w}}$ is the most direct rotation taking
$\unitvec{w}$ into $\unitvec{u}$.  We need to bisect the angle between
them, and a familiar geometric construction that achieves this is the
diagonal of the rhombus having $\unitvec{w}$ and $\unitvec{u}$ as
sides:
\begin{equation}
  \label{eq:AverageVector}
  \unitvec{v} = \frac{\unitvec{u}+\unitvec{w}}
  {\abs{\unitvec{u}+\unitvec{w}}}~.
\end{equation}
Then the rotation we want is
\begin{equation}
  \label{eq:SquareRootOfVectors}
  \sqrt{-\unitvec{u}\, \unitvec{w}} = \pm\, \unitvec{v}\, \unitvec{w}
  = \pm \frac{\unitvec{u}\, \unitvec{w}-1}
  {\abs{\unitvec{u}+\unitvec{w}}} = \pm \frac{1-\unitvec{u}\,
    \unitvec{w}} {\sqrt{2 [1 - (\unitvec{u} \, \unitvec{w})_{0}]}}~.
\end{equation}
Computing the square root using this expression is easier than using
Eq.~\eqref{eq:QuaternionPower}, in the sense that no transcendental
functions are required and fewer singularities are encountered.  This
expression is very robust and deals well with finite numerical
precision.  This expression is ill defined whenever
$\unitvec{u}+\unitvec{w}=0$---which is not surprising, as there are
infinitely many ``shortest'' ways to rotate a vector into its
opposite.  These are two ways of expressing the fact that there are
infinitely many square roots of $-1$ among the unit quaternions.

\subsection{Formulas for rotations and SWSHs%
  \label{sec:RotatingSWSHs}}%
We now express Wigner's $\D$ matrices and the spin-weighted spherical
harmonics (SWSHs) directly in terms of quaternions, so that no
conversion to or from the more usual Euler-angle representation is
necessary.  In the following, we will treat the general case in which
the spin weight $s$ is arbitrary; the formulas given here do not
assume $s = \pm 2$.

The SWSHs form a basis for spin-weighted functions on the
sphere~\cite{NewmanPenrose:1965, JanisNewman:1965,
  NewmanPenrose:1966}.  Goldberg \etal~\cite{GoldbergEtAl:1967} showed
that the SWSHs can be expressed as special cases of Wigner's $\D$
matrices, so that by constructing $\D^{(\ell)}_{m', m}$, we will
obtain $\sYlm{\ell,m}$.  Defining the parts of the quaternion
$\quat{Q}$ as
\begin{equation}
  \label{eq:QuaternionParts}
  \quatpart{Q}{a} \defined \quatcomponent{Q}{0} + \i\,
  \quatcomponent{Q}{3} %
  \quad \text{and} \quad %
  \quatpart{Q}{b} \defined \quatcomponent{Q}{2} + \i\,
  \quatcomponent{Q}{1}~,
\end{equation}
we can express quaternion multiplication as
\begin{subequations}
  \begin{gather}
    \quatpart{(P Q)}{a} = \quatpart{P}{a}\, \quatpart{Q}{a} -
    \quatpartco{P}{b}\,
    \quatpart{Q}{b}~, \\
    \quatpart{(P Q)}{b} = \quatpart{P}{b}\, \quatpart{Q}{a} +
    \quatpartco{P}{a}\, \quatpart{Q}{b}~.
  \end{gather}
\end{subequations}
Quaternions are isomorphic to (Pauli) spinors, and the two parts of
the quaternion defined here are essentially the two components of the
spinor.  The choices of signs in Eq.~\eqref{eq:QuaternionParts}
are---to some extent---arbitrary conventions.  However, care must be
taken to ensure that the resulting $\D$ matrices form a representation
of the rotation group rather than an anti-representation, and to
ensure that the handedness of space is preserved.  Our purpose in
choosing these particular signs is to reproduce the standard SWSHs as
special cases.  In particular, note that the presence of
$\quatcomponent{Q}{3}$ in the definition of $\quatpart{Q}{a}$ is what
picks out the $z$ axis as the point of reference on the sphere, so
that the polar angle is measured with respect to it, rather than the
$x$ or $y$ axes.

Now, following the standard derivation~\cite{[{Chapter~15 of
  }] [{.  Note that Wigner defines the Euler angles differently,
    interchanging the first and last: $\alpha \leftrightarrow
    \gamma$.}]  Wigner:1959}, we obtain
\begin{widetext}
  \begin{equation}
    \label{eq:WignerDFullExpression}
    \D^{(\ell)}_{m',m} (\quat{R}) =
    \begin{cases}
      \delta_{m',-m}\, \quatpart{R}{b}^{2m}\, (-1)^{\ell+m} &
      \text{when $\quatpart{R}{a} = 0$,} \\ %
      \delta_{m',m}\, \quatpart{R}{a}^{2m} & \text{when
        $\quatpart{R}{b} = 0$,} \\ %
      \sqrt{ \frac{ (\ell+m)!\, (\ell-m)! } { (\ell+m')!\, (\ell-m')!
        } }\, \abs{\quatpart{R}{a}}^{2\ell-2m}\,
      \quatpart{R}{a}^{m+m'}\, \quatpart{R}{b}^{m-m'}\, \sum_{\rho}
      (-1)^{\rho}\, \binom{\ell+m'} {\rho}\, \binom{\ell-m'}
      {\ell-\rho-m}\, \left( \frac{\abs{\quatpart{R}{b}}}
        {\abs{\quatpart{R}{a}}} \right)^{2\rho} & \text{otherwise.}
    \end{cases}
  \end{equation}
\end{widetext}
This expression is valid for all integral and half-integral values of
$\ell \geq 0$; naturally, we only need integral values $\ell \geq 2$
for the $s=-2$ fields discussed in this paper.  Note in particular
that $\D^{(\ell)}_{m',m}(-\quat{R}) = (-1)^{2m}\,
\D^{(\ell)}_{m',m}(\quat{R})$, which may be nontrivial when $m$ can
take half-integral values.  This is another statement of the fact that
rotation through $2\pi$ may not return fields of half-integral spin
weight to their original values, but will return fields of integral
spin weight.  This fact will be useful below when we integrate the
angular velocity.

To recover the usual expressions for $\D$ in terms of Euler angles, we
use $\quat{R} = \e^{\alpha\, \unitvec{z}/2}\, \e^{\beta\,
  \unitvec{y}/2}\, \e^{\gamma\, \unitvec{z}/2}$, from which we can
easily find
\begin{subequations}
  \label{eq:PartsOfEulerRotation}
  \begin{align}
    \quatpart{R}{a} = \cos \frac{\beta}{2}\, \e^{\i\,
      \frac{\gamma+\alpha} {2} } \qquad \quatpart{R}{b} = \sin
    \frac{\beta}{2}\, \e^{\i\, \frac{\gamma-\alpha} {2} }~.
  \end{align}
\end{subequations}
It must be emphasized, of course, that evaluating
Eq.~\eqref{eq:WignerDFullExpression} directly is faster and deals with
numerical-precision issues better than using the form with sines and
cosines.

Now, to express the SWSHs in terms of these $\D$ matrices, we adopt
conventions to agree with Ref.~\cite{AjithEtAl:2007}, which attempts
to establish uniform conventions for use in numerical relativity.  We
have
\begin{equation}
  \label{eq:SWSHs}
  \sYlm{\ell,m}(\vartheta, \varphi) = (-1)^{s}\, \sqrt{ \frac{2\, \ell
      + 1} {4\, \pi} }\, \D^{(\ell)}_{m,-s}\left( \e^{\varphi\,
      \unitvec{z}/2}\, \e^{\vartheta\, \unitvec{y}/2} \right)~.
\end{equation}
These functions are implemented in the
\href{http://arxiv.org/src/1302.2919/anc}{ancillary files} as
\code{GWFrames.WignerDMatrix} and \code{GWFrames.SWSH}.

\subsection{Integrating the angular velocity%
  \label{sec:IntegratingAngularVelocity}}%
In many contexts, quaternion-valued functions of time turn up.  These
may be differentiated or integrated with respect to time, much as
vector-valued functions may be.  However, noncommutativity leads to
certain problems.  In the next section, we will see how interpolation
can be handled sensibly.  Here, we prove a vital formula used in the
main text of this paper to integrate the angular velocity vector to
find the rotor describing the frame with that angular velocity.

First, we need formula for the derivative of the inverse, which can be
obtained by differentiating $\quat{Q}\, \quat{Q}^{-1} = 1$:
\begin{equation}
  \label{eq:DerivativeOfInverse}
  \frac{\d}{\d t} \quat{Q}^{-1} = - \quat{Q}^{-1}\, \frac{\d \quat{Q}}
  {\d t}\, \quat{Q}^{-1}~.
\end{equation}
This is the crucial relation that allows us to calculate the angular
velocity $\FrameAngularVelocity$ of a frame described by the rotor
$\quat{R}(t)$~\cite{BoyleEtAl:2011}.  Suppose that a vector
$\vector{v}_{0}$ is stationary in the rotating frame.  Then, that
vector is given in the inertial frame as $\vector{v}(t) =
\quat{R}(t)\, \vector{v}_{0}\, \co{\quat{R}}(t)$.  We also know that
$\d\vector{v}/\d t = \FrameAngularVelocity \times \vector{v}$.  Using
the definition of quaternion multiplication and the usual commutator
(Lie product), we can calculate $\FrameAngularVelocity \times
\vector{v} = \frac{1}{2} [\FrameAngularVelocity, \vector{v}]$.
Another way of writing this is
\begin{subequations}
  \label{eq:StationaryToRotatingVectorDeriv_Quaternions}
  \begin{align}
    \label{eq:StationaryToRotatingVectorDeriv_Quaternions_a}
    \frac{\d}{\d t} \big( \quat{R}\,\vector{v}_{0}\,\co{\quat{R}}
    \big) &= \frac{1}{2}\, \big[ \FrameAngularVelocity, \quat{R}\,
    \vector{v}_{0}\,
    \co{\quat{R}} \big] \\
    \label{eq:StationaryToRotatingVectorDeriv_Quaternions_b}
    &= \big[ \dot{\quat{R}}\, \co{\quat{R}}, \quat{R}\,
    \vector{v}_{0}\, \co{\quat{R}} \big]~,
  \end{align}
\end{subequations}
where the second line comes from simply evaluating the left-hand side
and using Eq.~\eqref{eq:DerivativeOfInverse}.  It is not hard to show
that $\dot{\quat{R}}\, \co{\quat{R}}$ is a pure vector, using
Eq.~\eqref{eq:DerivativeOfInverse} and the fact that any arbitrary
quaternion $\quat{Q}$ is a pure vector if and only if $\quat{Q} = -
\co{\quat{Q}}$.  Then, if
Eq.~\eqref{eq:StationaryToRotatingVectorDeriv_Quaternions} is to be
true for all vectors $\vector{v}_{0}$, we must have
\begin{equation}
  \label{eq:FrameAngularVelocity2}
  \FrameAngularVelocity = 2\, \dot{\quat{R}}\, \co{\quat{R}}~.
\end{equation}
The factor of $2$ appears here because we are using quaternions; this
factor does not appear in the equivalent result for rotation
operators.

As explained in Sec.~\ref{sec:TheFrameItself}, we need an expression
for the right-hand side in terms of logarithms.  To borrow notation
from the theory of Lie groups, we define the adjoint operator using
the familiar commutator:
\begin{equation}
  \label{eq:AdjointOperator}
  \ad{\quat{P}}\, \quat{Q} \defined [\quat{P}, \quat{Q}] = \quat{P}\,
  \quat{Q} - \quat{Q}\, \quat{P}~.
\end{equation}
This notation is convenient because we will need repeated applications
of the commutators.  For example, $\ad{\quat{P}}^{2} \quat{Q} =
\big[\quat{P}, [\quat{P}, \quat{Q}] \big]$.  Now, if $\quat{P}$ and
$\quat{Q}$ are unit quaternions, their logarithms will be pure
vectors: $\log{\quat{P}} = \quatlog{P}$ and $\log{\quat{Q}} =
\quatlog{Q}$.  We will also use the notation $\quatlogmag{P} \defined
\abs{\quatlog{P}}$, \etc.  Again, we can use the definition of
quaternion multiplication in Eq.~\eqref{eq:QuaternionMultiplication}
to see that $[\quatlog{P}, \quatlog{Q}] = 2\, \quatlog{P} \times
\quatlog{Q}$, which allows us to use familiar properties of the cross
product to calculate
\begin{equation}
  \label{eq:AdjointActionSO3}
  \ad{\quatlog{P}}^n\, \quatlog{Q} = \begin{cases}
    \quatlog{Q} & \text{$n=0$;} \\
    (-1)^{(n-1)/2}\, [\quatlog{P},\quatlog{Q}]\,
    (2 \quatlogmag{P})^{n-1} & \text{$n$ odd;} \\
    (-1)^{(n-2)/2}\, \big[ \quatlog{P}, [\quatlog{P}, \quatlog{Q}]
    \big]\, (2 \quatlogmag{P})^{n-2} & \text{$n>0$ even.}
  \end{cases}
\end{equation}
The proof is a simple induction.  A standard formula~\cite{*[{Lemma
    5.3 of }] [{.}] Miller:1972} says
\begin{equation}
  \label{eq:ConjugateOfLogarithm}
  \e^{\quatlog{P}}\, \quatlog{Q}\, \e^{-\quatlog{P}} =
  \sum_{n=0}^{\infty}\, \frac{1}{n!}\, \ad{\quatlog{P}}^{n}
  \quatlog{Q}~,
\end{equation}
while a somewhat less-standard formula~\cite{*[{Derived from
    Eq.~(B.10) of }] [{.}]  DuistermaatKolk:1999} gives us
\begin{equation}
  \label{eq:DerivativeOfExponential}
  \dot{\quat{P}} = \frac{\d \e^{\quatlog{P}}}{\d t} = \int_{0}^{1}
  \e^{s\, \quatlog{P}}\, \frac{\d \quatlog{P}} {\d t}\, \e^{(1-s)\,
    \quatlog{P}}\, \d s
\end{equation}
for $\quatlogmag{P} \leq \pi$.  We can multiply this formula on the
right by $\e^{-\quatlog{P}}$, and substitute using
Eq.~\eqref{eq:ConjugateOfLogarithm}.  We then separate the resulting
sum into three parts, corresponding to the three cases in
Eq.~\eqref{eq:AdjointActionSO3}.  These can be readily evaluated,
yielding simple trigonometric functions, which can then be
integrated:\footnote{Again, note the assumption that $\quatlogmag{R}
  \leq \pi$, which is essential to the correctness of
  Eq.~\eqref{eq:DerivativeOfRotorDerivation}, where the actual
  magnitude $\quatlogmag{R}$ is used.  Nonetheless, we also assume
  that the derivative $\dot{\quatlog{R}}$ exists and is continuous
  everywhere, which must be enforced by removing branch-cut
  discontinuities before differentiating.}
\begin{subequations}
  \label{eq:DerivativeOfRotorDerivation}
  \begin{align}
    \label{eq:DerivativeOfRotorDerivation1}
    \dot{\quat{R}}\, \co{\quat{R}} &= \int_{0}^{1} \e^{s\,
      \quatlog{R}}\, \frac{\d \quatlog{R}} {\d t}\, \e^{-s\,
      \quatlog{R}}\, \d s \\
    &= \int_{0}^{1} \left( \sum_{n=0}^{\infty} \frac{1}{n!}\, \ad{s
        \quatlog{R}}^{n} \dot{\quatlog{R}} \right)\, \d s \\
    &= \int_{0}^{1} \left( \dot{\quatlog{R}} + \frac{\sin(2\, s\,
        \quatlogmag{R})} {2\, \quatlogmag{R}}\, [\quatlog{R},
      \dot{\quatlog{R}}] + \frac{\sin^{2}(s\, \quatlogmag{R})} {2\,
        \quatlogmag{R}^{2}}\, \big[\quatlog{R}, [\quatlog{R},
      \dot{\quatlog{R}}] \big]
    \right)\, \d s \\
    &= \dot{\quatlog{R}} + \frac{\sin^{2} \quatlogmag{R}} {2\,
      \quatlogmag{R}^{2}}\, [\quatlog{R}, \dot{\quatlog{R}}] +
    \frac{\quatlogmag{R} - \sin\quatlogmag{R}\, \cos\quatlogmag{R}}
    {4\, \quatlogmag{R}^{3}}\, \big[\quatlog{R}, [\quatlog{R},
    \dot{\quatlog{R}}] \big]~.
  \end{align}
\end{subequations}
As discussed in Sec.~\ref{sec:ElementsOfQuaternionMath}, we evaluate
the derivative $\dot{\quatlog{R}}$ by treating $\quatlog{R}$ as a
continuous function, removing any branch cuts.  On the other hand,
when used without differentiating, we have assumed that
$\quatlogmag{R} \leq \pi$.

We can re-express this relation as a matrix equation by defining
\begin{equation}
  \label{eq:DerivativeOfRotorMatrix}
  \begin{split}
    \DerivativeOfRotorMatrix &\defined \left\{
      \begin{pmatrix}
        1 & 0 & 0 \\
        0 & 1 & 0 \\
        0 & 0 & 1
      \end{pmatrix}
      + \frac{\sin^{2}\quatlogmag{R}} {\quatlogmag{R}^{2}}\,
      \begin{pmatrix}
        0 & -\quatlog{R}_{3} & \quatlog{R}_{2} \\
        \quatlog{R}_{3} & 0 & -\quatlog{R}_{1} \\
        -\quatlog{R}_{2} & \quatlog{R}_{1} & 0
      \end{pmatrix}
    \right. \\ & \quad \left. \hphantom{abcd} - \frac{\quatlogmag{R} -
        \sin\quatlogmag{R}\, \cos\quatlogmag{R}}
      {\quatlogmag{R}^{3}}\,
      \begin{pmatrix}
        \quatlog{R}_{2}^{2} + \quatlog{R}_{3}^{2} & -\quatlog{R}_{1}\,
        \quatlog{R}_{2} & -\quatlog{R}_{1}\, \quatlog{R}_{3} \\
        -\quatlog{R}_{1}\, \quatlog{R}_{2} & \quatlog{R}_{1}^{2} +
        \quatlog{R}_{3}^{2} & -\quatlog{R}_{2}\, \quatlog{R}_{3} \\
        -\quatlog{R}_{1}\, \quatlog{R}_{3} & -\quatlog{R}_{2}\,
        \quatlog{R}_{3} & \quatlog{R}_{1}^{2} + \quatlog{R}_{2}^{2}
      \end{pmatrix}
    \right\}~,
  \end{split}
\end{equation}
in which case we have the much more compact formula
\begin{equation}
  \label{eq:DerivativeOfRotorUnsimplified}
  \dot{\quat{R}}\, \co{\quat{R}} = \frac{\FrameAngularVelocity}{2} =
  \DerivativeOfRotorMatrix\, \dot{\quatlog{R}}~.
\end{equation}
The determinant of the matrix simplifies to $\sin^{2} \quatlogmag{R} /
\quatlogmag{R}^{2}$, and is thus invertible for $\quatlogmag{R} <
\pi$.  For the rare edge case with exactly $\quatlogmag{R} = \pi$, the
rotation $\exp \quatlog{R}=-1$, which corresponds to the identity
rotation, so we should have $\dot{\quatlog{R}} =
\FrameAngularVelocity/2$.  For all other cases, we can invert the
matrix explicitly to find\footnote{An equivalent formula was found
  through a very different derivation by Grassia~\cite{Grassia:1998}.}
\begin{equation}
  \label{eq:DerivativeOfRotor}
  \dot{\quatlog{R}} = \left( \FrameAngularVelocity -
    \frac{\quatlog{R}\, (\quatlog{R} \cdot \FrameAngularVelocity)}
    {\quatlogmag{R}^{2}} \right)\, \frac{\quatlogmag{R}\, \cot
    \quatlogmag{R}} {2} + \frac{\quatlog{R}\, (\quatlog{R} \cdot
    \FrameAngularVelocity)} {2\, \quatlogmag{R}^{2}} + \frac{1}{2}
  \FrameAngularVelocity \times \quatlog{R}~.
\end{equation}
Thus, we are left with an ordinary differential equation to solve for
$\quatlog{R}$, as discussed in Sec.~\ref{sec:TheFrameItself}.
Finally, we obtain (up to the constant of integration) $\quat{R} =
\exp \quatlog{R}$.  We can improve the numerics and avoid the edge
case with $\quatlogmag{R} = \pi$ by using the mapping of
Eq.~\eqref{eq:RestrictedPrincipalValue} between time steps whenever
$\quatlogmag{R} > \pi/2$.  If the resulting rotor function is to be
interpolated (or used for any other purpose for which $\quat{R}$ and
$-\quat{R}$ are not equivalent), it may be useful to go back and make
$\quat{R}(t)$ as continuous as possible by flipping the sign of
$\quat{R}(t_{i})$ whenever $\abs{\quat{R}(t_{i})-\quat{R}(t_{i-1})} >
\sqrt{2}$, for example.  The complete algorithm is implemented in the
\href{http://arxiv.org/src/1302.2919/anc}{ancillary files} as
\code{GWFrames.FrameFromAngularVelocity}.

\subsection{Interpolation%
  \label{sec:Interpolation}}%
When comparing waveforms, one of the most basic requirements is the
ability to interpolate.  The description of a gravitational waveform
has now expanded to include both the SWSH modes of the waveform and
the rotor describing the frame of that decomposition.  So we need a
way to interpolate rotors.  But interpolation of rotors is complicated
by the fact that the interpolant needs to remain normalized to unity
at all times.  While it is possible to simply interpolate the
quaternions in $\mathbb{R}^{4}$ and normalize the result, the
interpolant will generally exhibit unnatural accelerations between the
interpolated points, even in the simplest case of uniform rotation.
Interpolation of rotation matrices is just as bad.  It goes without
saying, of course, that interpolation of Euler angles leads to
complete nonsense---the result is highly sensitive to the orientation
of the coordinate basis, and depends very strongly on the conventions
for which directions the successive Euler rotations take.  A
reasonable suggestion might be to interpolate the logarithms of the
rotors and exponentiate the interpolant.  However, this also leads to
unnatural behaviors in fairly simple cases, whenever the logarithms of
the rotors are not parallel.  Fortunately, there are well-motivated
solutions to the problem of quaternion interpolation that can give
reasonable results in very general cases.

Recognizing that the unit quaternions can also be regarded as points
on the unit sphere $S^{3}$, we might further expect an interpolant to
follow the geodesic between two points on the sphere.  In fact,
achieving this property is actually quite simple, using the fact that
the quaternions operate as a (Lie) group.  A simple interpolation
between unit quaternions $\quat{R}_{0}$ and $\quat{R}_{1}$ that
preserves the normalization is given by~\cite{Shoemake:1985}
\begin{equation}
  \label{eq:Slerp}
  L(\tau; \quat{R}_{0}, \quat{R}_{1}) = \left( \quat{R}_{1}\,
    \co{\quat{R}}_{0} \right)^{\tau}\, \quat{R}_{0} = \quat{R}_{0}\,
  \left( \co{\quat{R}}_{0}\, \quat{R}_{1} \right)^{\tau}~.
\end{equation}
Obviously, $L(0; \quat{R}_{0}, \quat{R}_{1}) = \quat{R}_{0}$ and $L(1;
\quat{R}_{0}, \quat{R}_{1}) = \quat{R}_{1}$, and the norm of $L(\tau;
\quat{R}_{0}, \quat{R}_{1})$ is always 1.  This formula is strongly
analogous to the formula for standard linear interpolation, except
that multiplication by $\tau$ becomes exponentiation and addition
becomes multiplication.\footnote{This analogy should not be carried
  too far because quaternion multiplication is noncommutative.  In
  particular, it is crucial to note that the right-hand side of
  Eq.~\eqref{eq:Slerp} is not equal to $\quat{R}_{1}\,
  \quat{R}_{0}^{1-\tau}$, for example, whenever $\quat{R}_{0}$ and
  $\quat{R}_{1}$ do not commute; such a formula actually gives very
  poor interpolation in many cases.  Equation~\eqref{eq:Slerp} is
  preferable because the path it describes is a geodesic in the space
  of unit quaternions.} This interpolation is referred to as ``slerp''
for \textit{s}pherical \textit{l}inear int\textit{erp}olation.  It
will be useful to note that
\begin{subequations}
  \label{eq:SlerpVelocity}
  \begin{align}
    \label{eq:SlerpVelocityLeft} %
    \frac{\d}{\d \tau}\, L(\tau; \quat{R}_{0}, \quat{R}_{1}) &= \log
    \left( \quat{R}_{1}\, \co{\quat{R}}_{0} \right)\, L(\tau;
    \quat{R}_{0}, \quat{R}_{1})~, \\
    \label{eq:SlerpVelocityRight} %
    &= L(\tau; \quat{R}_{0}, \quat{R}_{1})\, \log \left(
      \co{\quat{R}}_{0}\, \quat{R}_{1} \right)~.
  \end{align}
\end{subequations}
This formula shows us that the speed along the slerp path is constant,
as it must be for a geodesic.

Many problems only call for a linear interpolation like slerp.  In
particular, when blending \pN and NR waveforms, each waveform
possesses its own frame.  To transition between the two waveforms, we
must transition between the frames, which is just a simple linear
interpolation at each instant of time where the extent of the
interpolation (the $\tau$ argument to the function above) depends on
the time.  However, we also need to be able to interpolate each
individual waveform as a function of time.\footnote{For example, the
  \pN waveform and the NR waveform will generally be calculated at
  different instants of time.  To compare them, we need to be able to
  interpolate the values of one waveform onto the time steps of the
  other.}  For this, we cannot use linear interpolation for the motion
of the frame of either waveform.  If we did, we would see the frame
abruptly change rotation speed as it goes through each original data
point, just as a linearly interpolated graph changes slope abruptly as
it passes through each original data point.  Instead, we would prefer
some higher-order technique.

We can approach this problem in analogy with the construction of
curves in space, which suggests various approaches such as the
de~Casteljau algorithm for constructing \Bezier curves.
Unfortunately, the various methods---while being equivalent for real
numbers---are not equivalent when using quaternions because of
noncommutativity~\cite{KimEtAl:1995}.  It is not clear that any
particular formulation will give better results than any other, so we
may take the pragmatic approach of simply choosing one which is easily
implemented.  The result will be a \textit{s}pherical interpolation
based on the \textit{quad}rilateral of a standard spline, referred to
as \textit{squad}.

\begin{widetext}
  In that spirit, we will define the cubic-spline interpolant in terms
  of the linear interpolant:
  \begin{equation}
    \label{eq:Squad}
    C(t; \quat{R}_{i}, \quat{A}_{i}, \quat{B}_{i+1}, \quat{R}_{i+1}) =
    L \Big( 2\tau_{i}(1-\tau_{i}); L(\tau_{i}; \quat{R}_{i},
    \quat{R}_{i+1}), L(\tau_{i}; \quat{A}_{i}, \quat{B}_{i+1}) \Big)~,
  \end{equation}
  where $\quat{A}_{i}$ and $\quat{B}_{i+1}$ are ``control points'' to
  be solved for.  We have also defined $\tau_{i}(t) = (t - t_{i}) /
  (t_{i+1} - t_{i})$, where $i$ is assumed to be the index of the
  nearest time sample such that $t_{i} \leq t$.  We can evaluate the
  derivative
  \begin{subequations}
    \label{eq:dSquaddtau}
    \begin{align}
      \frac{\d}{\d\tau_{i}} C %
      &= \frac{\d}{\d\tau_{i}} \bigg\{ \exp \Big[
      2\tau_{i}(1-\tau_{i})\, \log \Big( L(\tau_{i}; \quat{A}_{i},
      \quat{B}_{i+1})\, L(\tau_{i}; \quat{R}_{i}, \quat{R}_{i+1})^{-1}
      \Big) \Big]\, L(\tau_{i}; \quat{R}_{i}, \quat{R}_{i+1}) \bigg\}
      \\%
      &= (2-4\tau_{i})\, \log \Big( L(\tau_{i}; \quat{A}_{i},
      \quat{B}_{i+1})\, L(\tau_{i}; \quat{R}_{i}, \quat{R}_{i+1})^{-1}
      \Big)\, C + 2 \tau_{i} (1-\tau_{i})\, G + C\,
      \log(\co{\quat{R}}_{i}\, \quat{R}_{i+1})~,
    \end{align}
  \end{subequations}
  where $G$ is a complicated expression, which is easy to compute, but
  messy to write; fortunately do not need to evaluate it because that
  term drops out when we evaluate at $\tau_{i}=0$ or $\tau_{i}=1$, as
  the factor in front of $G$ goes to zero.  We wish to ensure that the
  time derivatives are equal at the end of one segment and the
  beginning of the next:
  \begin{equation}
    \label{eq:SquadEqualityCondition1}
    \left. \frac{\d}{\d t} C(\tau_{i-1}, \quat{R}_{i-1},
      \quat{A}_{i-1}, \quat{B}_{i}, \quat{R}_{i})
    \right|_{\tau_{i-1}=1} = \left. \frac{\d}{\d t} C(\tau_{i},
      \quat{R}_{i}, \quat{A}_{i}, \quat{B}_{i+1},
      \quat{R}_{i+1})\right|_{\tau_{i}=0}~.
  \end{equation}
  Note that we differentiate with respect to $t$, rather than
  $\tau_{i}$, to account for differences in the time steps of the
  given data.
  Plugging in the result of Eq.~\eqref{eq:dSquaddtau} and
  simplifying, we get
  \begin{gather}
    \label{eq:SquadEqualityConditionDerivation}
    \frac{1} {\Delta t_{i-1}}\, \left\{ -2\, \log \Big( \quat{B}_{i}\,
      \co{\quat{R}}_{i} \Big)\, \quat{R}_{i} + \quat{R}_{i}\,
      \log(\co{\quat{R}}_{i-1}\, \quat{R}_{i}) \right\} = \frac{1}
    {\Delta t_{i}}\, \left\{ 2\, \log \Big( \quat{A}_{i}\,
      \co{\quat{R}}_{i} \Big)\, \quat{R}_{i} + \quat{R}_{i}\,
      \log(\co{\quat{R}}_{i}\, \quat{R}_{i+1}) \right\}~, \\ %
    \intertext{or equivalently} %
    \frac{1} {\Delta t_{i-1}}\, \quat{R}_{i}\, \left\{ -2\, \log \Big(
      \co{\quat{R}}_{i}\, \quat{B}_{i}\, \Big) +
      \log(\co{\quat{R}}_{i-1}\, \quat{R}_{i}) \right\} = \frac{1}
    {\Delta t_{i}}\, \quat{R}_{i}\, \left\{ 2\, \log \Big(
      \co{\quat{R}}_{i}\, \quat{A}_{i} \Big) +
      \log(\co{\quat{R}}_{i}\, \quat{R}_{i+1}) \right\}~.
  \end{gather}
  We need one more condition to solve for both variables
  $\quat{A}_{i}$ and $\quat{B}_{i}$.  We may choose\footnote{This
    choice has the nice property of agreeing with our intuition in the
    case of ``straight-line'' motion.  To be precise: if the
    transformation from $\quat{R}_{i-1}$ to $\quat{R}_{i}$ is written
    as multiplication by $\quat{R}_{i}\, \co{\quat{R}}_{i-1}$, then
    ``straight-line'' motion occurs when $\quat{R}_{i+1} =
    \quat{R}_{i}\, \co{\quat{R}}_{i-1}\, \quat{R}_{i}$ (and for
    simplicity, we assume $\Delta t_{i-1} = \Delta t_{i}$).  Then this
    average velocity is $\quat{R}_{i}\, \log (\co{\quat{R}}_{i-1}\,
    \quat{R}_{i}) / \Delta t$, which is precisely the velocity of a
    linear interpolation at that point.} to set the velocity at either
  side equal to the average velocity of linear interpolations on those
  two sides, giving us the following two equations:
  \begin{subequations}
    \label{eq:AverageVelocityConditions}
    \begin{gather}
      \quat{R}_{i}\, \frac{\log(\co{\quat{R}}_{i}\, \quat{R}_{i+1}) /
        \Delta t_{i} + \log(\co{\quat{R}}_{i-1}\, \quat{R}_{i}) /
        \Delta t_{i-1}} {2} = \frac{1} {\Delta t_{i-1}}\,
      \quat{R}_{i}\, \left\{ -2\, \log \Big( \co{\quat{R}}_{i}\,
        \quat{B}_{i}\, \Big) + \log(\co{\quat{R}}_{i-1}\,
        \quat{R}_{i}) \right\} \\ %
      \quat{R}_{i}\, \frac{\log(\co{\quat{R}}_{i}\, \quat{R}_{i+1}) /
        \Delta t_{i} + \log(\co{\quat{R}}_{i-1}\, \quat{R}_{i}) /
        \Delta t_{i-1}} {2} = \frac{1} {\Delta t_{i}}\, \quat{R}_{i}\,
      \left\{ 2\, \log \Big( \co{\quat{R}}_{i}\, \quat{A}_{i} \Big) +
        \log(\co{\quat{R}}_{i}\, \quat{R}_{i+1}) \right\}~.
    \end{gather}
  \end{subequations}
  We can now solve for the control points:
  \begin{subequations}
    \label{eq:ControlPoints}
    \begin{gather}
      \quat{A}_{i} = \quat{R}_{i}\, \exp \left[
        \frac{\log(\co{\quat{R}}_{i}\, \quat{R}_{i+1}) +
          \log(\co{\quat{R}}_{i-1}\, \quat{R}_{i}) \, \Delta t_{i} /
          \Delta t_{i-1} - 2\, \log(\co{\quat{R}}_{i}\,
          \quat{R}_{i+1})} {4} \right] \\ %
      \quat{B}_{i} = \quat{R}_{i}\, \exp \left[
        -\frac{\log(\co{\quat{R}}_{i}\, \quat{R}_{i+1})\, \Delta
          t_{i-1} / \Delta t_{i} + \log(\co{\quat{R}}_{i-1}\,
          \quat{R}_{i}) - 2\, \log(\co{\quat{R}}_{i-1}\,
          \quat{R}_{i})} {4} \right]~.
    \end{gather}
  \end{subequations}
  To apply these formulas to the edge cases of $i=0$ and $i=N-1$, we
  also define the quantities $\quat{R}_{-1} = \quat{R}_{0}\,
  \co{\quat{R}}_{1}\, \quat{R}_{0}$ and $\quat{R}_{N} =
  \quat{R}_{N-1}\, \co{\quat{R}}_{N-2}\, \quat{R}_{N-1}$, which
  roughly represent straight-line motion.
\end{widetext}

In the computer code included among this paper's
\href{http://arxiv.org/src/1302.2919/anc}{ancillary files}, the
functions \code{GWFrames.Slerp} and \code{GWFrames.Squad} implement
linear and cubic interpolations of rotors.

\section{Rotating spin-weighted functions%
  \label{sec:RotatingSpinWeightedFunctions}}%
Gravitational radiation is a complex field of nonzero spin weight,
meaning that it picks up a position-dependent phase under
rotation~\cite{NewmanPenrose:1966}.  The reason for this is its
definition with respect to a dyadic which is itself defined in terms
of a \emph{coordinate} basis; when the coordinates rotate, the dyadic
rotates.  Depending on details of the definition of the
gravitational-wave field, the spin weight may be $s=2$ or $s=-2$---the
most common choice being the latter.  Throughout the rest of this
paper, we have assumed $s=-2$; in order to discuss the properties of
general spin-weighted fields, this appendix will apply to general
values of $s$.

Suppose we have a field $\wavefield$ of spin weight $s$ on the sphere.
To measure this field, we first need some standard basis for our
space, $(\unitvec{x}, \unitvec{y}, \unitvec{z})$.  We can define the
usual spherical coordinates relative to this basis, and write the
field as a function of the coordinates, so that in some particular
direction $\Point$, we have $\wavefield(\Point) =
\wavefield(\vartheta, \varphi)$.  We define the rotor
\begin{equation}
  \label{eq:SphericalCoordinateQuaternion}
  \quat{R}_{(\vartheta, \varphi)} \defined \e^{\varphi\, \unitvec{z} /
    2}\, \e^{\vartheta\, \unitvec{y} / 2}
\end{equation}
and note that
\begin{equation}
  \label{eq:DirectionAndRotor}
  \Point = \quat{R}_{(\vartheta, \varphi)}\, \unitvec{z}\,
  \co{\quat{R}}_{(\vartheta, \varphi)}~.
\end{equation}
Now, suppose we rotate the physical system by some
$\PhysicalRotation$.  Then, the corresponding direction in the rotated
field is $\Point' = \PhysicalRotation\, \Point\, \PhysicalRotationCo$.
We know that there must be some angles $(\Rotated{\vartheta},
\Rotated{\varphi})$ such that
\begin{equation}
  \label{eq:RotatedDirectionAndRotor}
  \Point' = \quat{R}_{(\Rotated{\vartheta}, \Rotated{\varphi})}\,
  \unitvec{z}\, \co{\quat{R}}_{(\Rotated{\vartheta},
    \Rotated{\varphi})}~.
\end{equation}
But these conditions are not enough to fully restrict our rotations.
There is some other angle\footnote{This angle is required to account
  for the full three-dimensional freedom in choosing
  $\PhysicalRotation$.  It is always possible to find such an angle.
  However, this angle need not be unique for certain orientations;
  $\gamma$ may be degenerate with $\varphi$.  Similarly, because of
  the familiar singularities of the spherical coordinates, there may
  not be a unique choice of $(\vartheta, \varphi)$ or
  $(\Rotated{\vartheta}, \Rotated{\varphi})$ for certain positions.
  Nonetheless, the rotations $\quat{R}_{(\vartheta, \varphi)}\,
  \e^{\gamma \unitvec{z} / 2}$ and $\quat{R}_{(\Rotated{\vartheta},
    \Rotated{\varphi})}$ generated by these angles \emph{will} be
  uniquely determined, much as the North and South Poles are uniquely
  determined despite the ill-defined longitude at those points.}
$\gamma$ such that
\begin{equation}
  \label{eq:AngleRelationship}
  \quat{R}_{(\Rotated{\vartheta}, \Rotated{\varphi})} =
  \PhysicalRotation\, \quat{R}_{(\vartheta, \varphi)}\, \e^{\gamma\,
    \unitvec{z}/2}~.
\end{equation}
The term involving $\gamma$ represents an \emph{initial} rotation
through that angle in the positive sense about the direction
$\unitvec{z}$, which is equivalent to a \emph{final} rotation about
the direction $\Point'$.  For spin-weighted functions, this
corresponds~\cite{[{}] [{.  Note that the sense of the rotation used
    to define spin weights is somewhat counterintuitive.}]
  NewmanPenrose:1966} to multiplication of the function value by
$\e^{-\i\, s\, \gamma}$.  Thus, in this basis, we measure a different
field $\Rotated{\wavefield}$, related to the field $\wavefield$
measured in the first basis by
\begin{equation}
  \label{eq:RotatedWavefield}
  \Rotated{\wavefield}(\Rotated{\vartheta}, \Rotated{\varphi}) =
  \wavefield(\vartheta, \varphi)\, \e^{-\i\, s\, \gamma}~.
\end{equation}
For $s=0$, we recover the familiar result that a scalar field does not
depend on the frame in which it is measured.

The spin-weighted spherical harmonics (SWSHs) form a basis for
spin-weighted functions on the sphere~\cite{NewmanPenrose:1965,
  JanisNewman:1965, NewmanPenrose:1966}, just as standard spherical
harmonics form a basis for spin-zero functions.  We can write
\begin{subequations}
  \label{eq:SWSHDecomposition}
  \begin{gather}
    \label{eq:SWSHDecompositionUnrotated}
    \wavefield(\vartheta, \varphi) = \sum_{\ell,m}
    \wavefield^{\ell,m}\,
    \sYlm{\ell, m} (\vartheta, \varphi)~, \\
    \label{eq:SWSHDecompositionRotated}
    \Rotated{\wavefield}(\Rotated{\vartheta}, \Rotated{\varphi}) =
    \sum_{\ell,m} \Rotated{\wavefield}^{\ell,m}\, \sYlm{\ell, m}
    (\Rotated{\vartheta}, \Rotated{\varphi})~.
  \end{gather}
\end{subequations}
The SWSHs themselves are just special cases of the Wigner $\D$
matrices (see Eq.~\eqref{eq:SWSHs} and Ref.~\cite{GoldbergEtAl:1967}).
We can then use the fact that the $\D$ matrices form a representation
of the rotation group to find the transformation law for SWSHs:
\begin{equation}
  \label{eq:WignerDAsRepresentation}
  \D^{(\ell)}_{m',m} (\quat{R}_{1}\, \quat{R}_{2}) = \sum_{m''}
  \D^{(\ell)}_{m',m''} (\quat{R}_{1})\, \D^{(\ell)}_{m'',m}
  (\quat{R}_{2})
\end{equation}
implies, using Eq.~\eqref{eq:AngleRelationship}, that
\begin{equation}
  \label{eq:SWSHTransformation}
  \sYlm{\ell, m}(\Rotated{\vartheta}, \Rotated{\varphi}) = \sum_{m'}
  \sYlm{\ell, m'}(\vartheta, \varphi)\,
  \D^{(\ell)}_{m,m'}(\PhysicalRotation)\, \e^{-\i\, s\, \gamma}~.
\end{equation}
The dependence of $\gamma$ on $(\vartheta, \varphi)$ means that,
strictly speaking, the SWSHs with $s \neq 0$ do \emph{not} transform
among themselves under rotations.  Naturally, when coupled to the
appropriate spin-weighted tensors, the complete object transforms as
expected~\cite{Thorne:1980}.  Similarly, the modes $\wavefield^{\ell,
  m}$ transform nicely thanks to a convenient cancellation.  Inserting
Eqs.~\eqref{eq:SWSHDecomposition} and~\eqref{eq:SWSHTransformation}
into Eq.~\eqref{eq:RotatedWavefield}, we can show that
\begin{subequations}
  \label{eq:FieldTransformations}
  \begin{equation}
    \label{eq:FieldTransformation}
    \wavefield^{\ell, m} = \sum_{m'} \Rotated{\wavefield}^{\ell, m'}\,
    \D^{(\ell)}_{m',m}(\PhysicalRotation)~,
  \end{equation}
  or equivalently
  \begin{equation}
    \label{eq:FieldTransformationInv}
    \Rotated{\wavefield}^{\ell, m} = \sum_{m'} \wavefield^{\ell, m'}\,
    \D^{(\ell)}_{m',m}(\PhysicalRotationCo)~.
  \end{equation}
\end{subequations}
These are precisely the same as the transformation laws for modes of
standard ($s=0$) spherical harmonics, and do not depend on $\gamma$.

It is worth pointing out that a rotation of the physical system by
$\PhysicalRotation$ is equivalent to a rotation of the basis with
respect to which that system is measured by $\FrameRotation =
\PhysicalRotationCo$.  But there is an important subtlety to be
observed in the context of composing rotations.  Physical rotations
compose by left multiplication, whereas rotations of the frame compose
by right multiplication.  That is, if we first perform a physical
rotation $\quat{R}_{\text{p}1}$ then rotate that system by
$\quat{R}_{\text{p}2}$, it is equivalent to rotating the original
system by $\quat{R}_{\text{p}2}\, \quat{R}_{\text{p}1}$---just as with
vectors.  On the other hand, if we first rotate the frame by
$\quat{R}_{\text{f}1}$, then by $\quat{R}_{\text{f}2}$, it is
equivalent to rotating the frame by $\quat{R}_{\text{f}1}\,
\quat{R}_{\text{f}2}$---which is opposite to the usual behavior.  We
must carefully bear in mind the type of rotation we are performing.

Expressions for the Wigner $\D$ matrices are given directly in terms
of the rotor in Eq.~\eqref{eq:WignerDFullExpression}, which avoids the
need for conversion to Euler angles.  The SWSHs are expressed as
particular components of these matrices in Eq.~\eqref{eq:SWSHs}.  In
the computer code included among this paper's
\href{http://arxiv.org/src/1302.2919/anc}{ancillary files}, the Wigner
$\D$ matrices and SWSHs are implemented as
\code{GWFrames.WignerDMatrix} and \code{GWFrames.SWSH}.  Waveform
objects may be constructed with \code{GWFrames.Waveform}, and
transformed to different frames using the methods
\code{RotatePhysicalSystem} and \code{RotateDecompositionBasis}.

\section{Other methods of choosing a frame%
  \label{sec:PreviousWork}}%

In the interests of completeness, and to facilitate direct comparisons
using common language, we now review three other methods of choosing a
frame to eliminate mode-mixing in waveforms from precessing systems.
Each of these methods constructs a new frame by ensuring that the
$\unitvec{z}'$ direction lies along some chosen axis which is roughly
the axis of rotation of the waveform.  These differ from the
corotating frame introduced in the main text of this paper, in that
the waveform is still rotating in these new frames.  The
considerations of Sec.~\ref{sec:SolvingTheInverseProblem} (and in
particular the left panel of
Fig.~\ref{fig:OrbitalElementsFromWaveform}) suggest that among these
three, the preferred method is that of O'Shaughnessy \etal
supplemented with the minimal-rotation
condition~\cite{BoyleEtAl:2011}.  In particular, when setting the
integration constant discussed near the end of
Sec.~\ref{sec:FindingTheCorotatingFrame}, that is the method of
choice.  However, over all, the corotating frame is generally still a
preferable choice.

We first describe two methods suggested by Schmidt
\etal~\cite{SchmidtEtAl:2011} and O'Shaughnessy
\etal~\cite{OShaughnessyEtAl:2011}, using a common notation which
allows a common implementation by means of explicit maximization of a
quality function.  We then describe the method of O'Shaughnessy \etal
in the way in which it was introduced, which allows a second
implementation by solution of an eigensystem.  A third possible axis
suggests itself given the results of this paper: the angular-velocity
vector $\FieldAngularVelocity$ given by
Eq.~\eqref{eq:AngularVelocity}.  All three need an additional step to
remove sharp features in the waveforms, given by the minimal-rotation
condition.  In this section, we review each
alternative in the language of quaternions, suggesting improvements
for numerical accuracy and robustness.

\subsection{Maximization%
  \label{sec:DirectMaximization}}%
In general, we can describe the process of finding the radiation axis
as a maximization over $\quat{R}$ of the quantity
\begin{subequations}
  \label{eq:RadiationAxisQuantity}
  \begin{align}
    \label{eq:RadiationAxisQuantityDefined}
    Q(\quat{R}) &= \sum_{\ell, m} w_{\ell,m}
    \abs{\Rotated{\wavefield}^{\ell,m}}^{2} \\
    \label{eq:RadiationAxisQuantityUsefulForm}
    &= \sum_{\ell, m} w_{\ell,m} \abs{\sum_{m'} \wavefield^{\ell,m'}\,
      \D^{(\ell)}_{m',m}(\co{\quat{R}})}^{2}~.
  \end{align}
\end{subequations}
Here, the $w_{\ell,m}$ are simply weighting factors.  Schmidt \etal
took these factors to be $w_{2,\pm 2} = 1$, and zero otherwise;
O'Shaughnessy \etal effectively chose $w_{\ell,m} = m^{2}$, with some
cutoff $\ell$ above which $w_{\ell,m} = 0$.

This function is actually degenerate with respect to initial rotations
about the $\unitvec{z}$ axis, because such rotations simply affect the
overall phase of the term inside the absolute value.  For numerical
efficiency, we need to restrict $Q$ to some nondegenerate domain for
efficient numerical maximization.  Whereas previous
references~\cite{SchmidtEtAl:2011, OShaughnessyEtAl:2011,
  BoyleEtAl:2011} used rotations of the form $\quat{R}_{(\vartheta,
  \varphi)}$, we choose instead to use rotations of the form
$\quat{R}_{(\vartheta, \varphi)}\, \e^{-\varphi\, \unitvec{z}/2}$.
All such rotations can be written as $\quat{R}_{\vector{v}} =
\e^{\vector{v}}$ for some vector $\vector{v}$ in the $x$--$y$ plane,
of magnitude less than or equal to $\pi/2$.  Using rotations of this
form significantly simplifies calculation of the Wigner $\D$ matrices
and eliminates the degeneracy near the identity, which substantially
improves numerical accuracy and stability for mildly precessing
systems.

Of course, the sphere cannot be covered homeomorphically by a single
coordinate chart, so an additional degeneracy remains: all vectors on
the boundary of our domain result in rotations with equal values of
$Q$.  However, this set has measure zero in the domain itself, meaning
that it is almost never encountered.  Moreover, the effect of this
degeneracy will be completely eliminated in
Sec.~\ref{sec:ImposingTheMinimalRotationCondition}.  In fact, we find
it convenient to extend the domain further.  We maximize
$Q(\e^{\vector{v}})$ for all vectors $\vector{v}$ in the entire
$x$--$y$ plane, parameterizing the function arguments by the usual
coordinates $(x, y) \in \mathbb{R}^{2}$.  There are now degeneracies
on circles of radius $n\, \pi/2$ centered on the origin for all
integers $n>0$.  Again, however, these degeneracies cause no practical
difficulties.

Given values for the modes $\wavefield^{\ell,m}$ and the weights
$w_{\ell,m}$, the right-hand side of
Eq.~\eqref{eq:RadiationAxisQuantityUsefulForm} is known analytically,
using Eq.~\eqref{eq:WignerDFullExpression}, as are its derivatives
with respect to $x$ and $y$.  These functions are ungainly, but can be
written down explicitly, plugged into a computer, and used in
efficient numerical optimization routines.  Direct maximization of
Eq.~\eqref{eq:RadiationAxisQuantity} is simple to implement, and can
be made reasonably efficient and robust.  It does have disadvantages,
however.  At each step in the minimization routine, all relevant $\D$
matrices need to be recomputed.  When the $w_{\ell,m}$ are nonzero for
many values, this can become very expensive.  In such cases, it can be
significantly more efficient to find a radiation axis using the
following method.

In the computer code included among this paper's
\href{http://arxiv.org/src/1302.2919/anc}{ancillary files}, a waveform
object may be constructed with \code{GWFrames.Waveform}.  The axis
suggested by Schmidt \etal may then be found by applying the
\code{SchmidtEtAlVector} method.

\subsection{Dominant principal axis%
  \label{sec:DominantPrincipalAxis}}%
In general, if $w_{\ell,m} = w_{\ell}\, m^{2}$ where $w_{\ell}$ only
depends on $\ell$, then this can be presented in a different form and
solved as an eigenvector problem---which is the approach O'Shaughnessy
\etal actually used when introducing their method.
Define\footnote{O'Shaughnessy \etal used $w_{2}=1$, and $0$ for all
  other weights, as well as an overall normalization which is ignored
  here for simplicity.}
\begin{equation}
  \label{eq:AngularMomentumMatrix_weighted}
  \langle L_{(a}\, L_{b)} \rangle \defined \sum_{\ell, m, m'}\,
  w_{\ell}\, \co{\wavefield}^{\ell,m'} \braket{\ell, m'| L_{(a}\,
    L_{b)} |\ell, m}\, \wavefield^{\ell, m}~,
\end{equation}
where $L_{a}$ is the usual angular-momentum operator.  The radiation
axis is chosen to be the dominant principal axis
$\LLDominantEigenvector$ of this tensor---the eigenvector with the
eigenvalue of largest magnitude, which can be found with standard
algebraic techniques.  We can find some rotation $\YawfulRot$ taking
the $z$ axis into the dominant principal axis.
Reference~\cite{BoyleEtAl:2011} showed that such a rotation maximizes
the function $Q$ of Eq.~\eqref{eq:RadiationAxisQuantity}.

Again, however, this rotation is not unique.  Moreover, the dominant
principal axis is only defined up to a sign, and numerical
implementations may choose between the two options effectively
randomly.  A naive choice of $\YawfulRot(t)$, then, may flip back and
forth discontinuously.  Fortunately, we can overcome this problem
easily by taking $\unitvec{a}_{i} \mapsto -\unitvec{a}_{i}$ whenever
$\unitvec{a}_{i} \cdot \unitvec{a}_{i-1} < 0$.  Then, we can ensure
that the appropriate $\YawfulRot(t_{i})$ is as close\footnote{The
  distance between two rotations $\quat{R}_{1}$ and $\quat{R}_{2}$ can
  be defined as $2\, \abs{ \log(\co{\quat{R}}_{1}\, \quat{R}_{2}) }$,
  which is the minimum angle needed to rotate one into the other.  See
  Appendix~\ref{sec:ElementsOfQuaternionMath} for more details.} as
possible to $\YawfulRot(t_{i-1})$ by choosing
\begin{subequations}
  \label{eq:DominantPrincipalAxisRotation}
  \begin{gather}
    \label{eq:DominantPrincipalAxisRotationVector}
    \RelativeFrame \defined \sqrt{-\unitvec{a}_{i} \,
      \unitvec{a}_{i-1}}~,
    \\
    \label{eq:DominantPrincipalAxisRotationTotal}
    \YawfulRot(t_{i}) = \RelativeFrame\, \YawfulRot(t_{i-1})~.
  \end{gather}
\end{subequations}
In Eq.~\eqref{eq:DominantPrincipalAxisRotationVector}, the vectors are
multiplied as quaternions, and the square root may be found with the
help of Eq.~\eqref{eq:SquareRootOfVectors}.  We start out with
$\unitvec{a}_{-1} = \unitvec{z}$, and build up the frame by stepping
forward in time according to
Eq.~\eqref{eq:DominantPrincipalAxisRotation}, where each
$\unitvec{a}_{i}$ is the dominant principal axis at that instant of
time.  However, for reasons of numerical stability,
$\unitvec{a}_{i-1}$ in
Eq.~\eqref{eq:DominantPrincipalAxisRotationVector} should be expressed
as $\YawfulRot(t_{i-1})\, \unitvec{z}\, \YawfulRotCo(t_{i-1})$, rather
than as the principal axis at the previous instant.

The frame found by this method has certain advantages over
maximization of Eq.~\eqref{eq:RadiationAxisQuantity}.  In particular,
the matrix in Eq.~\eqref{eq:AngularMomentumMatrix_weighted} need only
be computed once for each time step.  The dominant principal axis is
then obtained from this.  No calculations of Wigner's $\D$ matrices
are necessary, which tends to make the computation fast.  Also, some
minor care is needed to make the present method robust, mostly
involving choosing the direction of the axis to be consistent from
moment to moment.

In the computer code included among this paper's
\href{http://arxiv.org/src/1302.2919/anc}{ancillary files}, a waveform
object may be constructed with \code{GWFrames.Waveform}.  The dominant
principal axis of $\langle L_{(a}\, L_{b)} \rangle$ may then be found
by applying the \code{OShaughnessyEtAlVector} method.

\subsection{Aligned with the angular velocity%
  \label{sec:AlignedWithAngularVelocity}}%
A very similar frame can be defined, using the angular-velocity vector
$\FieldAngularVelocity$ in place of the dominant principal axis of
$\langle L_{(a}\, L_{b)} \rangle$.  The vector $\FieldAngularVelocity$
was found in Sec.~\ref{sec:SolvingForTheRotation} and is given
explicitly by Eq.~\eqref{eq:AngularVelocity}.  This can be used for
the $\unitvec{a}_{i}$ in Eq.~\eqref{eq:DominantPrincipalAxisRotation}.
Note that using the only the direction of the vector to align the axis
of the new frame throws away some information.  Specifically, the
magnitude of $\FieldAngularVelocity$ is meaningful and is used in
Sec.~\ref{sec:TheFrameItself} to derive a fully corotating frame.
Nonetheless, the waveform rotation in this frame is about the $z'$
axis at each instant, making the time dependence of the waveform in
this frame quite similar to that of a nonprecessing system in a
stationary frame.

In the computer code included among this paper's
\href{http://arxiv.org/src/1302.2919/anc}{ancillary files}, a waveform
object may be constructed with \code{GWFrames.Waveform}.  The angular
velocity may then be found using the \code{AngularVelocityVector}
method on such an object.

\subsection{The minimal-rotation condition%
  \label{sec:ImposingTheMinimalRotationCondition}}%
Each of the three methods discussed above is critically flawed when
applied to a time-series of data, unless followed by the procedure
described here.  The end result of any of the three previous methods
is some rotation $\YawfulRot(t)$ that takes the $z$ axis into the
chosen radiation axis: $\unitvec{a}(t) = \YawfulRot(t)\, \unitvec{z}\,
\YawfulRotCo(t)$.  As mentioned, however, this is by no means the only
such rotation.  Indeed, because of the invariance of $\unitvec{z}$
under rotations about the $z$ axis, any rotation of the form
\begin{equation}
  \label{eq:TotalRotation}
  \quat{R}(t) = \YawfulRot(t)\, \exp\left[ \frac{\gamma(t)}{2}\,
    \unitvec{z} \right]
\end{equation}
will do the same.  Arbitrarily setting $\gamma(t)=0$ leaves us with
large extraneous features in the phase of each mode of the waveform.
Reference~\cite{BoyleEtAl:2011} showed that it is easy to impose a
condition on $\gamma(t)$ such that the total rotation $\quat{R}$
satisfies a geometrically and physically meaningful criterion referred
to as the \textit{minimal-rotation condition}.  This section simply
reiterates the previous description in quaternion form and suggests a
more accurate way of finding $\YawfulRotDot$ in some cases.

To motivate this condition, we first define the radiation frame's
instantaneous angular-velocity vector $\FrameAngularVelocity$.  Then,
for any vector $\vector{v}$ that is stationary in the radiation frame,
its derivative in an inertial frame is given by
\begin{equation}
  \label{eq:StationaryToRotatingVectorDeriv}
  \vectordot{v} = \FrameAngularVelocity \times \vector{v}~,
\end{equation}
where a dot denotes differentiation with respect to time.  A radiation
frame is---by definition---a frame in which the radiation axis
$\unitvec{a}$ is stationary.  So, its derivative in an inertial frame
is $\unitvecdot{a} = \FrameAngularVelocity \times \unitvec{a}$.
Taking the cross product of both sides of this equation by
$\unitvec{a}$, using the standard vector triple product formula with
the fact that $\unitvec{a}$ has unit magnitude, then rearranging, we
find
\begin{equation}
  \label{eq:FrameRotationVector}
  \FrameAngularVelocity = \unitvec{a} \times \unitvecdot{a} +
  (\unitvec{a} \cdot \FrameAngularVelocity)\, \unitvec{a}~.
\end{equation}
Now, we might hope that since $\unitvec{a}(t)$ is actually measured
from the waveform, this might be enough to specify the frame.
Unfortunately, Eq.~\eqref{eq:FrameRotationVector} defines the
component of $\FrameAngularVelocity$ along $\unitvec{a}$ circularly;
it is undetermined, so we need another condition.  Of course, an
obvious solution presents itself.  When the radiation axis is
stationary, we can expect that the frame should be stationary.  To
achieve this, Eq.~\eqref{eq:FrameRotationVector} shows that we must
have $\FrameAngularVelocity \cdot \unitvec{a} = 0$.  Because
$\unitvec{a}$ is a geometric object, independent of the frame in which
it is measured, this relation is geometrically meaningful.  We
therefore require this condition even in the nonprecessing case.
This minimizes the magnitude of $\FrameAngularVelocity$, so we refer
to it as the \textit{minimal-rotation
  condition}~\cite{BuonannoEtAl:2003, BoyleEtAl:2011}.  We adopt this
condition as the criterion for selecting the radiation frame.

Of course, the frame is not given by its instantaneous rotation
vector, but by its orientation at each instant of time.  So we need to
express $\FrameAngularVelocity$ in terms of $\quat{R}(t)$ to impose
our condition.  This is conveniently calculated in
Sec.~\ref{sec:IntegratingAngularVelocity}, which shows that
$\FrameAngularVelocity = 2\, \dot{\quat{R}}\, \co{\quat{R}}$.  Because
the radiation axis is given by $\unitvec{a} = \quat{R}\, \unitvec{z}\,
\co{\quat{R}}$, the minimal-rotation condition becomes
$\FrameAngularVelocity \cdot \unitvec{a} = 2\, \dot{\quat{R}}\,
\co{\quat{R}} \cdot \quat{R}\, \unitvec{z}\, \co{\quat{R}} = 0$.
Invariance of the dot product under rotation shows that we can also
write this as $\co{\quat{R}}\, \dot{\quat{R}} \cdot \unitvec{z} = 0$.
Expanding $\quat{R}$ as given in Eq.~\eqref{eq:TotalRotation}, we see
that the minimal-rotation condition is satisfied if $\gamma(t)$
satisfies
\begin{equation}
  \label{eq:ImposingMinRot}
  \dot{\gamma}(t) = - 2\, \YawfulRotCo(t)\, \YawfulRotDot(t) \cdot
  \unitvec{z} = 2 \left( \YawfulRotCo(t)\, \YawfulRotDot(t)\,
    \unitvec{z} \right)_{0}~,
\end{equation}
where the subscript $0$ here denotes the scalar part.  Now, since
$\YawfulRot$ is assumed to be known---perhaps by one of the three
foregoing methods---we can evaluate the right-hand side, then
integrate in time, and insert the result into
Eq.~\eqref{eq:TotalRotation}.  Note that the integration constant
$\gamma(0)$ is undetermined.  This corresponds to the usual freedom in
choosing a phase, familiar from nonprecessing systems, and will have
to be fixed in a similar way.

As a practical matter, the rotor $\YawfulRot$ is typically computed
using Eq.~\eqref{eq:SquareRootOfVectors} with $\unitvec{w} =
\unitvec{z}$ and $\unitvec{u} = \unitvec{a}$.  We can easily
differentiate this, assuming $\unitvec{w}$ is constant, and arrive at
an analytical formula for $\YawfulRotDot$ in terms of
$\unitvecdot{a}$.  In the construction of \pN waveforms, the latter is
known analytically, and may be inserted into this formula for higher
accuracy:
\begin{subequations}
  \label{eq:SquareRootDotOfVectors}
  \begin{align}
    \YawfulRotDot &= \pm \partial_{t} \frac{1-\unitvec{a}\,
      \unitvec{z}} {\sqrt{2 [1 - (\unitvec{a} \, \unitvec{z})_{0}]}}
    \\ &= \pm \left( \frac{-\partial_{t} \unitvec{a}\, \unitvec{z}}
      {\sqrt{2 [1 + \unitvec{a}_{3}]}} - \frac{\partial_{t}
        \unitvec{a}_{3}} {2 [1 + \unitvec{a}_{3}]} \YawfulRot
    \right)~.
  \end{align}
\end{subequations}
Here, we have used $- (\unitvec{a} \, \unitvec{z})_{0} =
\unitvec{a}_{3}$ for simplicity.

In the computer code included among this paper's
\href{http://arxiv.org/src/1302.2919/anc}{ancillary files}, an array
of quaternions can be put into minimal-rotation form using the
\code{GWFrames.MinimalRotation} function.  A waveform object
constructed with \code{GWFrames.Waveform} can be transformed into the
frames discussed in this section using methods beginning with
\code{TransformTo}.

\vfil
\let\c\Originalcdefinition %
\let\d\Originalddefinition %
\let\i\Originalidefinition %
\hypersetup{urlcolor=UrlColor}
\bibliography{References}



\end{document}

%% file: UsePackages.tex
\usepackage[T3,T1]{fontenc}
\usepackage{mathrsfs} 
\usepackage{bm} 
\makeatletter
\@ifclassloaded{beamer}
  { 
    \typeout{UsePackages: Detected beamer}
    \usepackage{tgheros}
    
  }
  { 
    \typeout{UsePackages: Did not detect beamer}
    \ifx\asybeamer\undefined 
    \typeout{UsePackages: Detected article}
    \usepackage[varg]{txfonts} 
    \usepackage{tgtermes} 
    \else 
    \typeout{Fonts: Detected asy for beamer}
    \usepackage{tgheros}
    
    \fi
  }
\usepackage{microtype} 
\def\MT@register@subst@font{
  \MT@exp@one@n\MT@in@clist\font@name\MT@font@list
  \ifMT@inlist@\else\xdef\MT@font@list{\MT@font@list\font@name,}\fi}
\makeatother

\usepackage{graphicx} %
\usepackage{color} %
\usepackage[all]{onlyamsmath} 
\usepackage{xspace}
\usepackage{braket}
\usepackage{accents}
\usepackage{siunitx}
\usepackage{mathtools}
\DeclareSymbolFontAlphabet{\mathrm}{operators}

\definecolor{CiteColor}{rgb}{0.18039, 0.18824, 0.57255}
\definecolor{UrlColor} {rgb}{0.741, 0.173, 0.000}
\definecolor{DarkUrlColor} {rgb}{0.500, 0.110, 0.000}
\definecolor{LinkColor}{rgb}{0.25098, 0.47843, 0.04706}


%% file: Preamble.tex
\makeatletter

\def\@seccntformat#1{\csname the#1\endcsname.~}%

\def\section{%
  \@startsection {section}
  {1} {\z@} {0.55cm \@plus1ex \@minus .02ex}%
    {0.225cm} { \normalfont\bfseries \centering}%
}%
\def\subsection{%
  \@startsection {subsection}
  {2} {\z@ } {0.45cm \@plus 0.8ex \@minus 0.2ex}%
  {0.1125cm}{\normalfont \bfseries \centering }}
\def\subsubsection{%
  \@startsection {subsubsection}
  {3} {\z@ } {0.4cm \@plus 0.6ex \@minus 0.1ex}%
  {0.075cm}{\normalfont \it \centering }}

\newcommand{\surnames}[1]{\def\@surnamelist{#1}\relax}
\surnames{\ }
\ifx \@shorttitle \@empty
\else
  \def\@oddhead{\small \MakeUppercase{\@shorttitle} \hfill}
\fi
\ifx \@surnamelist \@empty
\else
  \def\@evenhead{\small \MakeUppercase{\@surnamelist} \hfill}
\fi
\def\@oddfoot{\reset@font\hfil\thepage\hfil}%
\def\@evenfoot{\reset@font\hfil\thepage\hfil}%
\g@addto@macro\maketitle{\global\@specialpagetrue\gdef\@specialstyle{plain}}

\usepackage[colorlinks, plainpages=false,
hyperfigures=true]{hyperref}
\hypersetup{linkcolor=LinkColor}
\hypersetup{citecolor=CiteColor}
\hypersetup{urlcolor=UrlColor}
\hypersetup{setpagesize=false}
\hypersetup{pdfborder=0 0 0}

\makeatother



%% file: Macros.tex
\let\Originalddefinition\d
\renewcommand{\d}{\ensuremath{\mathrm{d}}}

\newcommand{\e}{\ensuremath{\mathrm{e}}}
\let\Originalidefinition\i
\renewcommand{\i}{\ensuremath{\mathrm{i}}}

\newcommand{\scriplus}{\ensuremath{\mathscr{I}^{+}}}

\let\Originalcdefinition\c
\renewcommand{\c}{\mathrm{c}}

\newcommand{\MSun}{\ensuremath{M_\odot}\xspace}

\DeclareSIUnit{\strain}{strain}
\DeclareSIPrePower{\root}{1/2}
\DeclareSIUnit{\parsec}{pc}
\DeclareSIUnit{\yr}{yr}
\DeclareSIUnit{\year}{yr}
\DeclareSIUnit{\lightyear}{ly}
\DeclareSIUnit{\SolarMass}{\ensuremath{\MSun}}
\DeclareSIUnit{\Mass}{\ensuremath{M}}

\newcommand{\abs} [1]{\left\lvert{#1}\right\rvert}

\newcommand{\co}[1]{\ensuremath{\bar{#1}}}

\newcommand{\defined}{\coloneqq}
\newcommand{\identically}{\equiv}
\newcommand{\asymptoticallyequal}{\simeq}

\newcommand{\D}{\ensuremath{\mathfrak{D}} }

\newcommand{\sYlm}[1]{\ensuremath{\scripts{_{s}}{Y}{_{#1}}}}

\DeclareSymbolFont{tipa}{T3}{tipa}{m}{n}
\DeclareMathAccent{\ibreve}{\mathalpha}{tipa}{'020}
\newcommand{\Rotated}[1]{\ensuremath{\ibreve{#1}}}

\newcommand{\SO}[1]{\ensuremath{\mathit{SO}(#1)}}

\newcommand{\SU}[1]{\ensuremath{\mathit{SU}(#1)}}
\newcommand{\su}[1]{\ensuremath{\mathfrak{su}(#1)}}

\newcommand{\h}{\ensuremath{h} }

\newcommand{\tfid}{\ensuremath{t_{\text{fid}}}}

\makeatletter
\newcommand{\foreign}[1]{\textrm{#1}}
\newcommand{\etal}{\textit{et~al}\@ifnextchar{\relax}{.\relax}{\ifx\@let@token.\else\ifx\@let@token~.\else.\@\xspace\fi\fi}}
\newcommand{\etc}{\foreign{etc}\@ifnextchar{\relax}{.\relax}{\ifx\@let@token.\else\ifx\@let@token~.\else.\@\xspace\fi\fi}}
\newcommand{\eg}{\foreign{e.g}\@ifnextchar{\relax}{.\relax}{\ifx\@let@token.\else\ifx\@let@token~.\else.\@\xspace\fi\fi}}
\newcommand{\ie}{\foreign{i.e}\@ifnextchar{\relax}{.\relax}{\ifx\@let@token.\else\ifx\@let@token~.\else.\@\xspace\fi\fi}}
\makeatother

\newcommand{\pN}{\text{PN}\xspace}
\newcommand{\PN}{\pN\xspace}

\definecolor{NoteColor}{rgb}{0.900, 0.218, 0.000}

\definecolor{NewColor}{rgb}{0,.55,0}

\newcommand{\code}[1]{\texttt{\small{#1}}}

\makeatletter
\newcommand{\CapName}[1]{\textbf{#1}.}
\newcommand{\ShowDimensions}{%
  \typeout{The font encoding is \f@encoding}        %
  \typeout{The font family is \f@family}            %
  \typeout{The font series is \f@series}            %
  \typeout{The font shape is \f@shape}              %
  \typeout{The font size is \f@size}                %
  \typeout{The baselineskip is \f@baselineskip}     %
  \typeout{The math font size is \tf@size}          %
  \typeout{The math script size is \sf@size}        %
  \typeout{The math scriptscript size is \ssf@size} %
  \typeout{The linewidth is \the\linewidth}         %
}
\newcommand{\prefixscripts}[2]{%
  \@mathmeasure\z@\displaystyle{#2}%
  \global\setbox\@ne\vbox to\ht\z@{}\dp\@ne\dp\z@
  \setbox\tw@\box\@ne
  \@mathmeasure4\displaystyle{\copy\tw@#1}%
  \@mathmeasure6\displaystyle{#2}%
  \dimen@-\wd6 \advance\dimen@\wd4 \advance\dimen@\wd\z@
  \hbox to\dimen@{}{\kern-\dimen@\box4\box6}%
}
\newcommand{\scripts}[3]{%
  \@mathmeasure\z@\displaystyle{#2}%
  \global\setbox\@ne\vbox to\ht\z@{}\dp\@ne\dp\z@
  \setbox\tw@\box\@ne
  \@mathmeasure4\displaystyle{\copy\tw@#1}%
  \@mathmeasure6\displaystyle{#2#3}%
  \dimen@-\wd6 \advance\dimen@\wd4 \advance\dimen@\wd\z@
  \hbox to\dimen@{}{\kern-\dimen@\box4\box6}%
}
\makeatother

\makeatletter
\let\protect\relax
{\catcode`\|=\active
  \xdef\InnerProduct{\protect\expandafter\noexpand\csname InnerProduct \endcsname}
  \expandafter\gdef\csname InnerProduct \endcsname#1{%
    \begingroup
    \ifx\SavedDoubleVert\relax
    \let\SavedDoubleVert\|\let\|\IpDoubleVert
    \fi
    \mathcode`\|32768\let|\IPVert
    \left({#1}\right)
    \endgroup
  }
}
\def\IPVert{\@ifnextchar|{\|\@gobble}
     {\egroup\,\mid@vertical\,\bgroup}}
\def\IPDoubleVert{\egroup\,\mid@dblvertical\,\bgroup}
\let\SavedDoubleVert\relax
\def\midvert{\egroup\mid\bgroup}
\def\SetVert{\@ifnextchar|{\|\@gobble}
    {\egroup\;\mid@vertical\;\bgroup}}
\def\SetDoubleVert{\egroup\;\mid@dblvertical\;\bgroup}
\def\mid@vertical{\mskip1mu\vrule\mskip1mu}
\def\mid@dblvertical{\mskip1mu\vrule\mskip2.5mu\vrule\mskip1mu}
\makeatother

\renewcommand{\vector}[1]{\bm{#1}}
\newcommand{\vectordot}[1]{\bm{{\dot{#1}}}}
\newcommand{\unitvec}[1]{\vector{{\hat{#1}}}}
\newcommand{\unitvecdot}[1]{\vector{{\dot{\hat{#1}}}}}

\newcommand{\Bezier}{B{\'{e}}zier\xspace}

\newcommand{\ad}[1]{\ensuremath{\text{ad}_{#1}}}

\newcommand{\LdtVector}{\ensuremath{\braket{\vector{L}\, \partial_{t}}}}
\newcommand{\LLQuantity}{\ensuremath{\braket{\vector{L}\vector{L}}}}
\newcommand{\LLDominantEigenvector}[1][\wavefield]{\ensuremath{\unitvec{V}_{#1}}}

\newcommand{\Lvec}{\ensuremath{\vector{L}}}
\newcommand{\Lhat}{\ensuremath{\hat{\Lvec}}}

\newcommand{\LNhat}{\ensuremath{\Lhat_{\text{N}}}}

\newcommand{\AngMomOp}{\ensuremath{L}}
\newcommand{\AngMomOpVec}{\ensuremath{\vector{\AngMomOp}}}

\newcommand{\CorotVector}{\ensuremath{\vector{\theta}}}

\newcommand{\Rcorot}{\ensuremath{\quat{R}_{\text{corot}}}}

\newcommand{\PhysicalRotation}{\ensuremath{\quat{R}_{\text{phys}}}}
\newcommand{\PhysicalRotationCo}{\ensuremath{\co{\quat{R}}_{\text{phys}}}}
\newcommand{\FrameRotation}{\ensuremath{\quat{R}_{\text{frame}}}}

\newcommand{\FieldAngularVelocity}{\ensuremath{\vector{\omega}}}
\newcommand{\FrameAngularVelocity}{\ensuremath{\vector{\varpi}}}

\newcommand{\OrbitalPhase}{\ensuremath{\Phi_{\text{orb}}}}
\newcommand{\OrbitalFrequency}{\ensuremath{\Omega_{\text{orb}}}}
\newcommand{\PrecessionalFrequency}{\ensuremath{\Omega_{\text{prec}}}}
\newcommand{\OrbitalAngularVelocity}{\ensuremath{\vector{\Omega}_{\text{orb}}}}
\newcommand{\PrecessionAngularVelocity}{\ensuremath{\vector{\Omega}_{\text{prec}}}}
\newcommand{\TotalAngularVelocity}{\ensuremath{\vector{\Omega}_{\text{tot}}}}
\newcommand{\OrbitalAngularVelocityHat}{\ensuremath{\hat{\vector{\Omega}}_{\text{orb}}}}

\newcommand{\DerivativeOfRotorMatrix}{\ensuremath{A}}

\newcommand{\RotationOperator}{\ensuremath{\mathscr{R}}}

\newcommand{\wavefield}{f}

\newcommand{\YawfulRot}{\ensuremath{\quat{R}_{\text{ax}}}}
\newcommand{\YawfulRotCo}{\ensuremath{\co{\quat{R}}_{\text{ax}}}}
\newcommand{\YawfulRotDot}{\ensuremath{\dot{\quat{R}}_{\text{ax}}}}
\newcommand{\YawfulRotCoDot}%
        {\ensuremath{\dot{\co{\quat{R}}}_{\text{ax}}}}

\newcommand{\RelativeFrame}{\ensuremath{\quat{R}_{\Delta}}}

\newcommand{\quat}[1]{\ensuremath{\mathbf{\MakeUppercase{#1}}}}
\newcommand{\quatcomponent}[2]{\ensuremath{\MakeLowercase{#1}_{#2}}}
\newcommand{\quatcomponentco}[2]%
        {\ensuremath{\co{\MakeLowercase{#1}}_{#2}}}
\newcommand{\quatpart}[2]{\ensuremath{\MakeUppercase{#1}_{#2}}}
\newcommand{\quatpartco}[2]{\ensuremath{\co{\MakeUppercase{#1}}_{#2}}}
\newcommand{\quatabs}[1]{\ensuremath{\abs{\quat{#1}}}}
\newcommand{\quatvec}[1]{\ensuremath{\vector{\lowercase{#1}}}}
\newcommand{\quatvecmag}[1]{\ensuremath{\lowercase{#1}}}
\newcommand{\quatunitvec}[1]{\ensuremath{\unitvec{\lowercase{#1}}}}
\newcommand{\quatlog}[1]{\ensuremath{\bm{\mathfrak{\lowercase{#1}}}}}
\newcommand{\quatlogmag}[1]{\ensuremath{\mathfrak{\lowercase{#1}}}}
\DeclareMathOperator{\Angle}{ang}
\newcommand{\quatangle}[1]{\ensuremath{\Angle{\quat{#1}}}}

\newcommand{\Point}{\ensuremath{\unitvec{n}}}


%% file: Affiliations.tex
\newcommand{\Cornell}{\affiliation{Center for Radiophysics and
    Space Research, Cornell University, Ithaca, New York 14853, USA}}%

%% file: paper.bbl
\begin{thebibliography}{99}%
\makeatletter
\providecommand \@ifxundefined [1]{%
 \@ifx{#1\undefined}
}%
\providecommand \@ifnum [1]{%
 \ifnum #1\expandafter \@firstoftwo
 \else \expandafter \@secondoftwo
 \fi
}%
\providecommand \@ifx [1]{%
 \ifx #1\expandafter \@firstoftwo
 \else \expandafter \@secondoftwo
 \fi
}%
\providecommand \natexlab [1]{#1}%
\providecommand \enquote  [1]{``#1''}%
\providecommand \bibnamefont  [1]{#1}%
\providecommand \bibfnamefont [1]{#1}%
\providecommand \citenamefont [1]{#1}%
\providecommand \href@noop [0]{\@secondoftwo}%
\providecommand \href [0]{\begingroup \@sanitize@url \@href}%
\providecommand \@href[1]{\@@startlink{#1}\@@href}%
\providecommand \@@href[1]{\endgroup#1\@@endlink}%
\providecommand \@sanitize@url [0]{\catcode `\\12\catcode `\$12\catcode
  `\&12\catcode `\#12\catcode `\^12\catcode `\_12\catcode `\%12\relax}%
\providecommand \@@startlink[1]{}%
\providecommand \@@endlink[0]{}%
\providecommand \url  [0]{\begingroup\@sanitize@url \@url }%
\providecommand \@url [1]{\endgroup\@href {#1}{\urlprefix }}%
\providecommand \urlprefix  [0]{URL }%
\providecommand \Eprint [0]{\href }%
\providecommand \doibase [0]{http://dx.doi.org/}%
\providecommand \selectlanguage [0]{\@gobble}%
\providecommand \bibinfo  [0]{\@secondoftwo}%
\providecommand \bibfield  [0]{\@secondoftwo}%
\providecommand \translation [1]{[#1]}%
\providecommand \BibitemOpen [0]{}%
\providecommand \bibitemStop [0]{}%
\providecommand \bibitemNoStop [0]{.\EOS\space}%
\providecommand \EOS [0]{\spacefactor3000\relax}%
\providecommand \BibitemShut  [1]{\csname bibitem#1\endcsname}%
\let\auto@bib@innerbib\@empty
\bibitem [{\citenamefont {{The LIGO Scientific
  Collaboration}}(2009)}]{LIGO:2009}%
  \BibitemOpen
  \bibfield  {author} {\bibinfo {author} {\bibnamefont {{The LIGO Scientific
  Collaboration}}},\ }\href {\doibase 10.1088/0034-4885/72/7/076901} {\bibfield
   {journal} {\bibinfo  {journal} {Repts. Prog. Phys.}\ }\textbf {\bibinfo
  {volume} {72}},\ \bibinfo {pages} {076901} (\bibinfo {year}
  {2009})}\BibitemShut {NoStop}%
\bibitem [{\citenamefont {Shoemaker}(2010)}]{Shoemaker:2010}%
  \BibitemOpen
  \bibfield  {author} {\bibinfo {author} {\bibfnamefont {D.}~\bibnamefont
  {Shoemaker}},\ }\href
  {https://dcc.ligo.org/cgi-bin/DocDB/ShowDocument?docid=2974} {\emph {\bibinfo
  {title} {{LIGO-T0900288-v3}: {A}dvanced {LIGO} anticipated sensitivity
  curves}}},\ \bibinfo {type} {Tech. Rep.}\ (\bibinfo  {institution} {LIGO},\
  \bibinfo {year} {2010})\BibitemShut {NoStop}%
\bibitem [{\citenamefont {{S. J. Waldman for the LIGO Scientific Collaboration
  and the Virgo Collaboration}}(2011)}]{WaldmanEtAl:2011}%
  \BibitemOpen
  \bibfield  {author} {\bibinfo {author} {\bibnamefont {{S. J. Waldman for the
  LIGO Scientific Collaboration and the Virgo Collaboration}}},\ }\href@noop {}
  {\enquote {\bibinfo {title} {The advanced {LIGO} gravitational wave
  detector},}\ } (\bibinfo {year} {2011}),\ \Eprint
  {http://arxiv.org/abs/1103.2728} {arXiv:1103.2728 [gr-qc]} \BibitemShut
  {NoStop}%
\bibitem [{\citenamefont {{The Virgo Collaboration}}(2012)}]{Virgo:2012}%
  \BibitemOpen
  \bibfield  {author} {\bibinfo {author} {\bibnamefont {{The Virgo
  Collaboration}}},\ }\href {\doibase 10.1088/1748-0221/7/03/P03012} {\bibfield
   {journal} {\bibinfo  {journal} {J. Instr.}\ }\textbf {\bibinfo {volume}
  {7}},\ \bibinfo {pages} {P03012} (\bibinfo {year} {2012})}\BibitemShut
  {NoStop}%
\bibitem [{\citenamefont {{Kentaro Somiya for the KAGRA
  collaboration}}(2012)}]{KAGRA:2012}%
  \BibitemOpen
  \bibfield  {author} {\bibinfo {author} {\bibnamefont {{Kentaro Somiya for the
  KAGRA collaboration}}},\ }\href {\doibase 10.1088/0264-9381/29/12/124007}
  {\bibfield  {journal} {\bibinfo  {journal} {Class. Quant. Grav.}\ }\textbf
  {\bibinfo {volume} {29}},\ \bibinfo {pages} {124007} (\bibinfo {year}
  {2012})}\BibitemShut {NoStop}%
\bibitem [{\citenamefont {{Alan J. Weinstein for the LIGO Scientific
  Collaboration and the Virgo Collaboration}}(2012)}]{WeinsteinEtAl:2012}%
  \BibitemOpen
  \bibfield  {author} {\bibinfo {author} {\bibnamefont {{Alan J. Weinstein for
  the LIGO Scientific Collaboration and the Virgo Collaboration}}},\ }\href
  {\doibase 10.1088/0264-9381/29/12/124012} {\bibfield  {journal} {\bibinfo
  {journal} {Class. Quant. Grav.}\ }\textbf {\bibinfo {volume} {29}},\ \bibinfo
  {pages} {124012} (\bibinfo {year} {2012})}\BibitemShut {NoStop}%
\bibitem [{\citenamefont {Kalogera}(2000)}]{Kalogera:2000}%
  \BibitemOpen
  \bibfield  {author} {\bibinfo {author} {\bibfnamefont {V.}~\bibnamefont
  {Kalogera}},\ }\href {\doibase 10.1086/309400} {\bibfield  {journal}
  {\bibinfo  {journal} {Astrophys. J.}\ }\textbf {\bibinfo {volume} {541}},\
  \bibinfo {pages} {319} (\bibinfo {year} {2000})}\BibitemShut {NoStop}%
\bibitem [{\citenamefont {O'{S}haughnessy}\ \emph {et~al.}(2005)\citenamefont
  {O'{S}haughnessy}, \citenamefont {Kaplan}, \citenamefont {Kalogera},\ and\
  \citenamefont {Belczynski}}]{OShaughnessyEtAl:2005}%
  \BibitemOpen
  \bibfield  {author} {\bibinfo {author} {\bibfnamefont {R.}~\bibnamefont
  {O'{S}haughnessy}}, \bibinfo {author} {\bibfnamefont {J.}~\bibnamefont
  {Kaplan}}, \bibinfo {author} {\bibfnamefont {V.}~\bibnamefont {Kalogera}}, \
  and\ \bibinfo {author} {\bibfnamefont {K.}~\bibnamefont {Belczynski}},\
  }\href {\doibase 10.1086/444346} {\bibfield  {journal} {\bibinfo  {journal}
  {Astrophys. J.}\ }\textbf {\bibinfo {volume} {632}},\ \bibinfo {pages} {1035}
  (\bibinfo {year} {2005})}\BibitemShut {NoStop}%
\bibitem [{\citenamefont {Grandcl{\'{e}}ment}\ \emph
  {et~al.}(2004)\citenamefont {Grandcl{\'{e}}ment}, \citenamefont {Ihm},
  \citenamefont {Kalogera},\ and\ \citenamefont
  {Belczynski}}]{GrandclementEtAl:2004}%
  \BibitemOpen
  \bibfield  {author} {\bibinfo {author} {\bibfnamefont {P.}~\bibnamefont
  {Grandcl{\'{e}}ment}}, \bibinfo {author} {\bibfnamefont {M.}~\bibnamefont
  {Ihm}}, \bibinfo {author} {\bibfnamefont {V.}~\bibnamefont {Kalogera}}, \
  and\ \bibinfo {author} {\bibfnamefont {K.}~\bibnamefont {Belczynski}},\
  }\href {\doibase 10.1103/PhysRevD.69.102002} {\bibfield  {journal} {\bibinfo
  {journal} {Phys. Rev. D}\ }\textbf {\bibinfo {volume} {69}},\ \bibinfo
  {pages} {102002} (\bibinfo {year} {2004})}\BibitemShut {NoStop}%
\bibitem [{\citenamefont {{LIGO Scientific Collaboration}}\ and\ \citenamefont
  {{Virgo Collaboration}}(2010)}]{LIGO:2010}%
  \BibitemOpen
  \bibfield  {author} {\bibinfo {author} {\bibnamefont {{LIGO Scientific
  Collaboration}}}\ and\ \bibinfo {author} {\bibnamefont {{Virgo
  Collaboration}}},\ }\href {\doibase doi:10.1088/0264-9381/27/17/173001}
  {\bibfield  {journal} {\bibinfo  {journal} {Class. Quant. Grav.}\ }\textbf
  {\bibinfo {volume} {27}},\ \bibinfo {pages} {173001} (\bibinfo {year}
  {2010})}\BibitemShut {NoStop}%
\bibitem [{\citenamefont {Kidder}(1995)}]{Kidder:1995}%
  \BibitemOpen
  \bibfield  {author} {\bibinfo {author} {\bibfnamefont {L.~E.}\ \bibnamefont
  {Kidder}},\ }\href {\doibase 10.1103/PhysRevD.52.821} {\bibfield  {journal}
  {\bibinfo  {journal} {Phys. Rev. D}\ }\textbf {\bibinfo {volume} {52}},\
  \bibinfo {pages} {821} (\bibinfo {year} {1995})}\BibitemShut {NoStop}%
\bibitem [{\citenamefont {Apostolatos}\ \emph {et~al.}(1994)\citenamefont
  {Apostolatos}, \citenamefont {Cutler}, \citenamefont {Sussman},\ and\
  \citenamefont {Thorne}}]{ApostolatosEtAl:1994}%
  \BibitemOpen
  \bibfield  {author} {\bibinfo {author} {\bibfnamefont {T.~A.}\ \bibnamefont
  {Apostolatos}}, \bibinfo {author} {\bibfnamefont {C.}~\bibnamefont {Cutler}},
  \bibinfo {author} {\bibfnamefont {G.~J.}\ \bibnamefont {Sussman}}, \ and\
  \bibinfo {author} {\bibfnamefont {K.~S.}\ \bibnamefont {Thorne}},\ }\href
  {\doibase 10.1103/PhysRevD.49.6274} {\bibfield  {journal} {\bibinfo
  {journal} {Phys. Rev. D}\ }\textbf {\bibinfo {volume} {49}},\ \bibinfo
  {pages} {6274} (\bibinfo {year} {1994})}\BibitemShut {NoStop}%
\bibitem [{\citenamefont {Will}\ and\ \citenamefont
  {Wiseman}(1996)}]{WillWiseman:1996}%
  \BibitemOpen
  \bibfield  {author} {\bibinfo {author} {\bibfnamefont {C.~M.}\ \bibnamefont
  {Will}}\ and\ \bibinfo {author} {\bibfnamefont {A.~G.}\ \bibnamefont
  {Wiseman}},\ }\href {\doibase 10.1103/PhysRevD.54.4813} {\bibfield  {journal}
  {\bibinfo  {journal} {Phys. Rev. D}\ }\textbf {\bibinfo {volume} {54}},\
  \bibinfo {pages} {4813} (\bibinfo {year} {1996})}\BibitemShut {NoStop}%
\bibitem [{\citenamefont {Apostolatos}(1996)}]{Apostolatos:1996}%
  \BibitemOpen
  \bibfield  {author} {\bibinfo {author} {\bibfnamefont {T.~A.}\ \bibnamefont
  {Apostolatos}},\ }\href {\doibase 10.1103/PhysRevD.54.2438} {\bibfield
  {journal} {\bibinfo  {journal} {Phys. Rev. D}\ }\textbf {\bibinfo {volume}
  {54}},\ \bibinfo {pages} {2438} (\bibinfo {year} {1996})}\BibitemShut
  {NoStop}%
\bibitem [{\citenamefont {Einstein}(2006)}]{Einstein:1918}%
  \BibitemOpen
  \bibfield  {author} {\bibinfo {author} {\bibfnamefont {A.}~\bibnamefont
  {Einstein}},\ }in\ \href
  {http://onlinelibrary.wiley.com/doi/10.1002/3527608958.ch14/summary} {\emph
  {\bibinfo {booktitle} {Albert Einstein: Akademie-Vortr\"age}}},\ \bibinfo
  {editor} {edited by\ \bibinfo {editor} {\bibfnamefont {D.}~\bibnamefont
  {Simon}}}\ (\bibinfo  {publisher} {Wiley-{VCH} Verlag {GmbH} \& Co. {KGaA}},\
  \bibinfo {address} {Weinheim, Germany},\ \bibinfo {year} {2006})\ pp.\
  \bibinfo {pages} {154--166}\BibitemShut {NoStop}%
\bibitem [{\citenamefont {Thorne}(1980)}]{Thorne:1980}%
  \BibitemOpen
  \bibfield  {author} {\bibinfo {author} {\bibfnamefont {K.~S.}\ \bibnamefont
  {Thorne}},\ }\href {\doibase 10.1103/RevModPhys.52.299} {\bibfield  {journal}
  {\bibinfo  {journal} {Rev. Mod. Phys.}\ }\textbf {\bibinfo {volume} {52}},\
  \bibinfo {pages} {299} (\bibinfo {year} {1980})}\BibitemShut {NoStop}%
\bibitem [{\citenamefont {Blanchet}\ \emph {et~al.}(2008)\citenamefont
  {Blanchet}, \citenamefont {Faye}, \citenamefont {Iyer},\ and\ \citenamefont
  {Sinha}}]{BlanchetEtAl:2008}%
  \BibitemOpen
  \bibfield  {author} {\bibinfo {author} {\bibfnamefont {L.}~\bibnamefont
  {Blanchet}}, \bibinfo {author} {\bibfnamefont {G.}~\bibnamefont {Faye}},
  \bibinfo {author} {\bibfnamefont {B.~R.}\ \bibnamefont {Iyer}}, \ and\
  \bibinfo {author} {\bibfnamefont {S.}~\bibnamefont {Sinha}},\ }\href
  {http://stacks.iop.org/0264-9381/25/i=16/a=165003} {\bibfield  {journal}
  {\bibinfo  {journal} {Class. Quant. Grav.}\ }\textbf {\bibinfo {volume}
  {25}},\ \bibinfo {pages} {165003} (\bibinfo {year} {2008})},\ \bibinfo {note}
  {note erratum}\BibitemShut {NoStop}%
\bibitem [{\citenamefont {Buonanno}\ \emph {et~al.}(2003)\citenamefont
  {Buonanno}, \citenamefont {Chen},\ and\ \citenamefont
  {Vallisneri}}]{BuonannoEtAl:2003}%
  \BibitemOpen
  \bibfield  {author} {\bibinfo {author} {\bibfnamefont {A.}~\bibnamefont
  {Buonanno}}, \bibinfo {author} {\bibfnamefont {Y.}~\bibnamefont {Chen}}, \
  and\ \bibinfo {author} {\bibfnamefont {M.}~\bibnamefont {Vallisneri}},\
  }\href {\doibase 10.1103/PhysRevD.67.104025} {\bibfield  {journal} {\bibinfo
  {journal} {Phys. Rev. D}\ }\textbf {\bibinfo {volume} {67}},\ \bibinfo
  {pages} {104025} (\bibinfo {year} {2003})}\BibitemShut {NoStop}%
\bibitem [{\citenamefont {Blanchet}\ \emph {et~al.}(2006)\citenamefont
  {Blanchet}, \citenamefont {Buonanno},\ and\ \citenamefont
  {Faye}}]{Blanchet:2006b}%
  \BibitemOpen
  \bibfield  {author} {\bibinfo {author} {\bibfnamefont {L.}~\bibnamefont
  {Blanchet}}, \bibinfo {author} {\bibfnamefont {A.}~\bibnamefont {Buonanno}},
  \ and\ \bibinfo {author} {\bibfnamefont {G.}~\bibnamefont {Faye}},\ }\href
  {\doibase 10.1103/PhysRevD.74.104034} {\bibfield  {journal} {\bibinfo
  {journal} {Phys. Rev. D}\ }\textbf {\bibinfo {volume} {74}},\ \bibinfo
  {pages} {104034} (\bibinfo {year} {2006})},\ \bibinfo {note} {note the two
  associated errata; the {arXiv} version is fully up to date},\ \Eprint
  {http://arxiv.org/abs/gr-qc/0605140v4} {arXiv:gr-qc/0605140v4} \BibitemShut
  {NoStop}%
\bibitem [{\citenamefont {Blanchet}\ \emph {et~al.}(2007)\citenamefont
  {Blanchet}, \citenamefont {Buonanno},\ and\ \citenamefont
  {Faye}}]{Blanchet:2007}%
  \BibitemOpen
  \bibfield  {author} {\bibinfo {author} {\bibfnamefont {L.}~\bibnamefont
  {Blanchet}}, \bibinfo {author} {\bibfnamefont {A.}~\bibnamefont {Buonanno}},
  \ and\ \bibinfo {author} {\bibfnamefont {G.}~\bibnamefont {Faye}},\ }\href
  {\doibase 10.1103/PhysRevD.75.049903} {\bibfield  {journal} {\bibinfo
  {journal} {Phys. Rev. D}\ }\textbf {\bibinfo {volume} {75}},\ \bibinfo
  {pages} {049903(E)} (\bibinfo {year} {2007})}\BibitemShut {NoStop}%
\bibitem [{\citenamefont {Arun}\ \emph {et~al.}(2009)\citenamefont {Arun},
  \citenamefont {Buonanno}, \citenamefont {Faye},\ and\ \citenamefont
  {Ochsner}}]{ArunEtAl:2009}%
  \BibitemOpen
  \bibfield  {author} {\bibinfo {author} {\bibfnamefont {K.~G.}\ \bibnamefont
  {Arun}}, \bibinfo {author} {\bibfnamefont {A.}~\bibnamefont {Buonanno}},
  \bibinfo {author} {\bibfnamefont {G.}~\bibnamefont {Faye}}, \ and\ \bibinfo
  {author} {\bibfnamefont {E.}~\bibnamefont {Ochsner}},\ }\href {\doibase
  10.1103/PhysRevD.79.104023} {\bibfield  {journal} {\bibinfo  {journal} {Phys.
  Rev. D}\ }\textbf {\bibinfo {volume} {79}},\ \bibinfo {pages} {104023}
  (\bibinfo {year} {2009})},\ \bibinfo {note} {note that the 1.5 and 2.5pN spin
  terms in the flux and energy expressions do not account for an erratum from
  2010}\BibitemShut {NoStop}%
\bibitem [{\citenamefont {Blanchet}\ \emph {et~al.}(2010)\citenamefont
  {Blanchet}, \citenamefont {Buonanno},\ and\ \citenamefont
  {Faye}}]{Blanchet:2010}%
  \BibitemOpen
  \bibfield  {author} {\bibinfo {author} {\bibfnamefont {L.}~\bibnamefont
  {Blanchet}}, \bibinfo {author} {\bibfnamefont {A.}~\bibnamefont {Buonanno}},
  \ and\ \bibinfo {author} {\bibfnamefont {G.}~\bibnamefont {Faye}},\ }\href
  {\doibase 10.1103/PhysRevD.81.089901} {\bibfield  {journal} {\bibinfo
  {journal} {Phys. Rev. D}\ }\textbf {\bibinfo {volume} {81}},\ \bibinfo
  {pages} {089901(E)} (\bibinfo {year} {2010})}\BibitemShut {NoStop}%
\bibitem [{\citenamefont {Marsat}\ \emph {et~al.}(2013)\citenamefont {Marsat},
  \citenamefont {Boh{\'{e}}}, \citenamefont {Faye},\ and\ \citenamefont
  {Blanchet}}]{MarsatEtAl:2012}%
  \BibitemOpen
  \bibfield  {author} {\bibinfo {author} {\bibfnamefont {S.}~\bibnamefont
  {Marsat}}, \bibinfo {author} {\bibfnamefont {A.}~\bibnamefont {Boh{\'{e}}}},
  \bibinfo {author} {\bibfnamefont {G.}~\bibnamefont {Faye}}, \ and\ \bibinfo
  {author} {\bibfnamefont {L.}~\bibnamefont {Blanchet}},\ }\href {\doibase
  10.1088/0264-9381/30/5/055007} {\bibfield  {journal} {\bibinfo  {journal}
  {Class. Quant. Grav.}\ }\textbf {\bibinfo {volume} {30}},\ \bibinfo {pages}
  {055007} (\bibinfo {year} {2013})}\BibitemShut {NoStop}%
\bibitem [{\citenamefont {Boh{\'{e}}}\ \emph {et~al.}(2013)\citenamefont
  {Boh{\'{e}}}, \citenamefont {Marsat}, \citenamefont {Faye},\ and\
  \citenamefont {Blanchet}}]{BoheEtAl:2012}%
  \BibitemOpen
  \bibfield  {author} {\bibinfo {author} {\bibfnamefont {A.}~\bibnamefont
  {Boh{\'{e}}}}, \bibinfo {author} {\bibfnamefont {S.}~\bibnamefont {Marsat}},
  \bibinfo {author} {\bibfnamefont {G.}~\bibnamefont {Faye}}, \ and\ \bibinfo
  {author} {\bibfnamefont {L.}~\bibnamefont {Blanchet}},\ }\href {\doibase
  10.1088/0264-9381/30/7/075017} {\bibfield  {journal} {\bibinfo  {journal}
  {Class. Quant. Grav.}\ }\textbf {\bibinfo {volume} {30}},\ \bibinfo {pages}
  {075017} (\bibinfo {year} {2013})}\BibitemShut {NoStop}%
\bibitem [{\citenamefont {Ajith}(2011)}]{Ajith:2011}%
  \BibitemOpen
  \bibfield  {author} {\bibinfo {author} {\bibfnamefont {P.}~\bibnamefont
  {Ajith}},\ }\href {\doibase 10.1103/PhysRevD.84.084037} {\bibfield  {journal}
  {\bibinfo  {journal} {Phys. Rev. D}\ }\textbf {\bibinfo {volume} {84}},\
  \bibinfo {pages} {084037} (\bibinfo {year} {2011})}\BibitemShut {NoStop}%
\bibitem [{\citenamefont {Schmidt}\ \emph {et~al.}(2012)\citenamefont
  {Schmidt}, \citenamefont {Hannam},\ and\ \citenamefont
  {Husa}}]{SchmidtEtAl:2012}%
  \BibitemOpen
  \bibfield  {author} {\bibinfo {author} {\bibfnamefont {P.}~\bibnamefont
  {Schmidt}}, \bibinfo {author} {\bibfnamefont {M.}~\bibnamefont {Hannam}}, \
  and\ \bibinfo {author} {\bibfnamefont {S.}~\bibnamefont {Husa}},\ }\href
  {\doibase 10.1103/PhysRevD.86.104063} {\bibfield  {journal} {\bibinfo
  {journal} {Phys. Rev. D}\ }\textbf {\bibinfo {volume} {86}},\ \bibinfo
  {pages} {104063} (\bibinfo {year} {2012})}\BibitemShut {NoStop}%
\bibitem [{\citenamefont {Brown}\ \emph
  {et~al.}(2012{\natexlab{a}})\citenamefont {Brown}, \citenamefont {Lundgren},\
  and\ \citenamefont {{O}'{S}haughnessy}}]{BrownEtAl:2012}%
  \BibitemOpen
  \bibfield  {author} {\bibinfo {author} {\bibfnamefont {D.~A.}\ \bibnamefont
  {Brown}}, \bibinfo {author} {\bibfnamefont {A.}~\bibnamefont {Lundgren}}, \
  and\ \bibinfo {author} {\bibfnamefont {R.}~\bibnamefont
  {{O}'{S}haughnessy}},\ }\href {\doibase 10.1103/PhysRevD.86.064020}
  {\bibfield  {journal} {\bibinfo  {journal} {Phys. Rev. D}\ }\textbf {\bibinfo
  {volume} {86}},\ \bibinfo {pages} {064020} (\bibinfo {year}
  {2012}{\natexlab{a}})}\BibitemShut {NoStop}%
\bibitem [{\citenamefont {Brown}\ \emph
  {et~al.}(2012{\natexlab{b}})\citenamefont {Brown}, \citenamefont {Harry},
  \citenamefont {Lundgren},\ and\ \citenamefont {Nitz}}]{BrownEtAl:2012b}%
  \BibitemOpen
  \bibfield  {author} {\bibinfo {author} {\bibfnamefont {D.~A.}\ \bibnamefont
  {Brown}}, \bibinfo {author} {\bibfnamefont {I.}~\bibnamefont {Harry}},
  \bibinfo {author} {\bibfnamefont {A.}~\bibnamefont {Lundgren}}, \ and\
  \bibinfo {author} {\bibfnamefont {A.~H.}\ \bibnamefont {Nitz}},\ }\href
  {\doibase 10.1103/PhysRevD.86.084017} {\bibfield  {journal} {\bibinfo
  {journal} {Phys. Rev. D}\ }\textbf {\bibinfo {volume} {86}},\ \bibinfo
  {pages} {084017} (\bibinfo {year} {2012}{\natexlab{b}})}\BibitemShut
  {NoStop}%
\bibitem [{\citenamefont {{O'S}haughnessy}\ \emph {et~al.}(2012)\citenamefont
  {{O'S}haughnessy}, \citenamefont {London}, \citenamefont {Healy},\ and\
  \citenamefont {Shoemaker}}]{OShaughnessyEtAl:2012}%
  \BibitemOpen
  \bibfield  {author} {\bibinfo {author} {\bibfnamefont {R.}~\bibnamefont
  {{O'S}haughnessy}}, \bibinfo {author} {\bibfnamefont {L.}~\bibnamefont
  {London}}, \bibinfo {author} {\bibfnamefont {J.}~\bibnamefont {Healy}}, \
  and\ \bibinfo {author} {\bibfnamefont {D.}~\bibnamefont {Shoemaker}},\
  }\href@noop {} {\enquote {\bibinfo {title} {Precession during merger 1:
  {S}trong polarization changes are observationally accessible features of
  strong-field gravity during binary black hole merger},}\ } (\bibinfo {year}
  {2012}),\ \Eprint {http://arxiv.org/abs/1209.3712} {arXiv:1209.3712 [gr-qc]}
  \BibitemShut {NoStop}%
\bibitem [{\citenamefont {Schmidt}\ \emph {et~al.}(2011)\citenamefont
  {Schmidt}, \citenamefont {Hannam}, \citenamefont {Husa},\ and\ \citenamefont
  {Ajith}}]{SchmidtEtAl:2011}%
  \BibitemOpen
  \bibfield  {author} {\bibinfo {author} {\bibfnamefont {P.}~\bibnamefont
  {Schmidt}}, \bibinfo {author} {\bibfnamefont {M.}~\bibnamefont {Hannam}},
  \bibinfo {author} {\bibfnamefont {S.}~\bibnamefont {Husa}}, \ and\ \bibinfo
  {author} {\bibfnamefont {P.}~\bibnamefont {Ajith}},\ }\href {\doibase
  10.1103/PhysRevD.84.024046} {\bibfield  {journal} {\bibinfo  {journal} {Phys.
  Rev. D}\ }\textbf {\bibinfo {volume} {84}},\ \bibinfo {pages} {024046}
  (\bibinfo {year} {2011})}\BibitemShut {NoStop}%
\bibitem [{\citenamefont {{O}'{S}haughnessy}\ \emph {et~al.}(2011)\citenamefont
  {{O}'{S}haughnessy}, \citenamefont {Vaishnav}, \citenamefont {Healy},
  \citenamefont {Meeks},\ and\ \citenamefont
  {Shoemaker}}]{OShaughnessyEtAl:2011}%
  \BibitemOpen
  \bibfield  {author} {\bibinfo {author} {\bibfnamefont {R.}~\bibnamefont
  {{O}'{S}haughnessy}}, \bibinfo {author} {\bibfnamefont {B.}~\bibnamefont
  {Vaishnav}}, \bibinfo {author} {\bibfnamefont {J.}~\bibnamefont {Healy}},
  \bibinfo {author} {\bibfnamefont {Z.}~\bibnamefont {Meeks}}, \ and\ \bibinfo
  {author} {\bibfnamefont {D.}~\bibnamefont {Shoemaker}},\ }\href {\doibase
  10.1103/PhysRevD.84.124002} {\bibfield  {journal} {\bibinfo  {journal} {Phys.
  Rev. D}\ }\textbf {\bibinfo {volume} {84}},\ \bibinfo {pages} {124002}
  (\bibinfo {year} {2011})}\BibitemShut {NoStop}%
\bibitem [{\citenamefont {Boyle}\ \emph {et~al.}(2011)\citenamefont {Boyle},
  \citenamefont {Owen},\ and\ \citenamefont {Pfeiffer}}]{BoyleEtAl:2011}%
  \BibitemOpen
  \bibfield  {author} {\bibinfo {author} {\bibfnamefont {M.}~\bibnamefont
  {Boyle}}, \bibinfo {author} {\bibfnamefont {R.}~\bibnamefont {Owen}}, \ and\
  \bibinfo {author} {\bibfnamefont {H.~P.}\ \bibnamefont {Pfeiffer}},\ }\href
  {\doibase 10.1103/PhysRevD.84.124011} {\bibfield  {journal} {\bibinfo
  {journal} {Phys. Rev. D}\ }\textbf {\bibinfo {volume} {84}},\ \bibinfo
  {pages} {124011} (\bibinfo {year} {2011})}\BibitemShut {NoStop}%
\bibitem [{\citenamefont {Jaranowski}\ \emph {et~al.}(1996)\citenamefont
  {Jaranowski}, \citenamefont {Kokkotas}, \citenamefont {Kr{\'{o}}lak},\ and\
  \citenamefont {Tsegas}}]{JaranowskiEtAl:1996}%
  \BibitemOpen
  \bibfield  {author} {\bibinfo {author} {\bibfnamefont {P.}~\bibnamefont
  {Jaranowski}}, \bibinfo {author} {\bibfnamefont {K.~D.}\ \bibnamefont
  {Kokkotas}}, \bibinfo {author} {\bibfnamefont {A.}~\bibnamefont
  {Kr{\'{o}}lak}}, \ and\ \bibinfo {author} {\bibfnamefont {G.}~\bibnamefont
  {Tsegas}},\ }\href {\doibase 10.1088/0264-9381/13/6/004} {\bibfield
  {journal} {\bibinfo  {journal} {Class. Quant. Grav.}\ }\textbf {\bibinfo
  {volume} {13}},\ \bibinfo {pages} {1279} (\bibinfo {year}
  {1996})}\BibitemShut {NoStop}%
\bibitem [{\citenamefont {van~der Sluys}\ \emph
  {et~al.}(2008{\natexlab{a}})\citenamefont {van~der Sluys}, \citenamefont
  {R{\"{o}}ver}, \citenamefont {Stroeer}, \citenamefont {Raymond},
  \citenamefont {Mandel}, \citenamefont {Christensen}, \citenamefont
  {Kalogera}, \citenamefont {Meyer},\ and\ \citenamefont
  {Vecchio}}]{vanderSluysEtAl:2008a}%
  \BibitemOpen
  \bibfield  {author} {\bibinfo {author} {\bibfnamefont {M.~V.}\ \bibnamefont
  {van~der Sluys}}, \bibinfo {author} {\bibfnamefont {C.}~\bibnamefont
  {R{\"{o}}ver}}, \bibinfo {author} {\bibfnamefont {A.}~\bibnamefont
  {Stroeer}}, \bibinfo {author} {\bibfnamefont {V.}~\bibnamefont {Raymond}},
  \bibinfo {author} {\bibfnamefont {I.}~\bibnamefont {Mandel}}, \bibinfo
  {author} {\bibfnamefont {N.}~\bibnamefont {Christensen}}, \bibinfo {author}
  {\bibfnamefont {V.}~\bibnamefont {Kalogera}}, \bibinfo {author}
  {\bibfnamefont {R.}~\bibnamefont {Meyer}}, \ and\ \bibinfo {author}
  {\bibfnamefont {A.}~\bibnamefont {Vecchio}},\ }\href {\doibase
  10.1086/595279} {\bibfield  {journal} {\bibinfo  {journal} {Astrophys. J.}\
  }\textbf {\bibinfo {volume} {688}},\ \bibinfo {pages} {L61} (\bibinfo {year}
  {2008}{\natexlab{a}})}\BibitemShut {NoStop}%
\bibitem [{\citenamefont {van~der Sluys}\ \emph
  {et~al.}(2008{\natexlab{b}})\citenamefont {van~der Sluys}, \citenamefont
  {Raymond}, \citenamefont {Mandel}, \citenamefont {R{\"{o}}ver}, \citenamefont
  {Christensen}, \citenamefont {Kalogera}, \citenamefont {Meyer},\ and\
  \citenamefont {Vecchio}}]{vanderSluysEtAl:2008b}%
  \BibitemOpen
  \bibfield  {author} {\bibinfo {author} {\bibfnamefont {M.}~\bibnamefont
  {van~der Sluys}}, \bibinfo {author} {\bibfnamefont {V.}~\bibnamefont
  {Raymond}}, \bibinfo {author} {\bibfnamefont {I.}~\bibnamefont {Mandel}},
  \bibinfo {author} {\bibfnamefont {C.}~\bibnamefont {R{\"{o}}ver}}, \bibinfo
  {author} {\bibfnamefont {N.}~\bibnamefont {Christensen}}, \bibinfo {author}
  {\bibfnamefont {V.}~\bibnamefont {Kalogera}}, \bibinfo {author}
  {\bibfnamefont {R.}~\bibnamefont {Meyer}}, \ and\ \bibinfo {author}
  {\bibfnamefont {A.}~\bibnamefont {Vecchio}},\ }\href {\doibase
  10.1088/0264-9381/25/18/184011} {\bibfield  {journal} {\bibinfo  {journal}
  {Class. Quant. Grav.}\ }\textbf {\bibinfo {volume} {25}},\ \bibinfo {pages}
  {184011} (\bibinfo {year} {2008}{\natexlab{b}})}\BibitemShut {NoStop}%
\bibitem [{\citenamefont {Raymond}\ \emph {et~al.}(2009)\citenamefont
  {Raymond}, \citenamefont {van~der Sluys}, \citenamefont {Mandel},
  \citenamefont {Kalogera}, \citenamefont {R{\"{o}}ver},\ and\ \citenamefont
  {Christensen}}]{RaymondEtAl:2009}%
  \BibitemOpen
  \bibfield  {author} {\bibinfo {author} {\bibfnamefont {V.}~\bibnamefont
  {Raymond}}, \bibinfo {author} {\bibfnamefont {M.~V.}\ \bibnamefont {van~der
  Sluys}}, \bibinfo {author} {\bibfnamefont {I.}~\bibnamefont {Mandel}},
  \bibinfo {author} {\bibfnamefont {V.}~\bibnamefont {Kalogera}}, \bibinfo
  {author} {\bibfnamefont {C.}~\bibnamefont {R{\"{o}}ver}}, \ and\ \bibinfo
  {author} {\bibfnamefont {N.}~\bibnamefont {Christensen}},\ }\href {\doibase
  10.1088/0264-9381/26/11/114007} {\bibfield  {journal} {\bibinfo  {journal}
  {Class. Quant. Grav.}\ }\textbf {\bibinfo {volume} {26}},\ \bibinfo {pages}
  {114007} (\bibinfo {year} {2009})}\BibitemShut {NoStop}%
\bibitem [{\citenamefont {Nissanke}\ \emph {et~al.}(2010)\citenamefont
  {Nissanke}, \citenamefont {Holz}, \citenamefont {Hughes}, \citenamefont
  {Dalal},\ and\ \citenamefont {Sievers}}]{NissankeEtAl:2010}%
  \BibitemOpen
  \bibfield  {author} {\bibinfo {author} {\bibfnamefont {S.}~\bibnamefont
  {Nissanke}}, \bibinfo {author} {\bibfnamefont {D.~E.}\ \bibnamefont {Holz}},
  \bibinfo {author} {\bibfnamefont {S.~A.}\ \bibnamefont {Hughes}}, \bibinfo
  {author} {\bibfnamefont {N.}~\bibnamefont {Dalal}}, \ and\ \bibinfo {author}
  {\bibfnamefont {J.~L.}\ \bibnamefont {Sievers}},\ }\href {\doibase
  10.1088/0004-637X/725/1/496} {\bibfield  {journal} {\bibinfo  {journal}
  {Astrophys. J.}\ }\textbf {\bibinfo {volume} {725}},\ \bibinfo {pages} {496}
  (\bibinfo {year} {2010})}\BibitemShut {NoStop}%
\bibitem [{\citenamefont {Wen}\ and\ \citenamefont
  {Chen}(2010)}]{WenChen:2010}%
  \BibitemOpen
  \bibfield  {author} {\bibinfo {author} {\bibfnamefont {L.}~\bibnamefont
  {Wen}}\ and\ \bibinfo {author} {\bibfnamefont {Y.}~\bibnamefont {Chen}},\
  }\href {\doibase 10.1103/PhysRevD.81.082001} {\bibfield  {journal} {\bibinfo
  {journal} {Phys. Rev. D}\ }\textbf {\bibinfo {volume} {81}},\ \bibinfo
  {pages} {082001} (\bibinfo {year} {2010})}\BibitemShut {NoStop}%
\bibitem [{\citenamefont {Fairhurst}(2011)}]{Fairhurst:2011}%
  \BibitemOpen
  \bibfield  {author} {\bibinfo {author} {\bibfnamefont {S.}~\bibnamefont
  {Fairhurst}},\ }\href {\doibase 10.1088/0264-9381/28/10/105021} {\bibfield
  {journal} {\bibinfo  {journal} {Class. Quant. Grav.}\ }\textbf {\bibinfo
  {volume} {28}},\ \bibinfo {pages} {105021} (\bibinfo {year}
  {2011})}\BibitemShut {NoStop}%
\bibitem [{\citenamefont {Klimenko}\ \emph {et~al.}(2011)\citenamefont
  {Klimenko}, \citenamefont {Vedovato}, \citenamefont {Drago}, \citenamefont
  {Mazzolo}, \citenamefont {Mitselmakher}, \citenamefont {Pankow},
  \citenamefont {Prodi}, \citenamefont {Re}, \citenamefont {Salemi},\ and\
  \citenamefont {Yakushin}}]{KlimenkoEtAl:2011}%
  \BibitemOpen
  \bibfield  {author} {\bibinfo {author} {\bibfnamefont {S.}~\bibnamefont
  {Klimenko}}, \bibinfo {author} {\bibfnamefont {G.}~\bibnamefont {Vedovato}},
  \bibinfo {author} {\bibfnamefont {M.}~\bibnamefont {Drago}}, \bibinfo
  {author} {\bibfnamefont {G.}~\bibnamefont {Mazzolo}}, \bibinfo {author}
  {\bibfnamefont {G.}~\bibnamefont {Mitselmakher}}, \bibinfo {author}
  {\bibfnamefont {C.}~\bibnamefont {Pankow}}, \bibinfo {author} {\bibfnamefont
  {G.}~\bibnamefont {Prodi}}, \bibinfo {author} {\bibfnamefont
  {V.}~\bibnamefont {Re}}, \bibinfo {author} {\bibfnamefont {F.}~\bibnamefont
  {Salemi}}, \ and\ \bibinfo {author} {\bibfnamefont {I.}~\bibnamefont
  {Yakushin}},\ }\href {\doibase 10.1103/PhysRevD.83.102001} {\bibfield
  {journal} {\bibinfo  {journal} {Phys. Rev. D}\ }\textbf {\bibinfo {volume}
  {83}},\ \bibinfo {pages} {102001} (\bibinfo {year} {2011})}\BibitemShut
  {NoStop}%
\bibitem [{\citenamefont {Nissanke}\ \emph {et~al.}(2011)\citenamefont
  {Nissanke}, \citenamefont {Sievers}, \citenamefont {Dalal},\ and\
  \citenamefont {Holz}}]{NissankeEtAl:2011}%
  \BibitemOpen
  \bibfield  {author} {\bibinfo {author} {\bibfnamefont {S.}~\bibnamefont
  {Nissanke}}, \bibinfo {author} {\bibfnamefont {J.}~\bibnamefont {Sievers}},
  \bibinfo {author} {\bibfnamefont {N.}~\bibnamefont {Dalal}}, \ and\ \bibinfo
  {author} {\bibfnamefont {D.}~\bibnamefont {Holz}},\ }\href {\doibase
  10.1088/0004-637X/739/2/99} {\bibfield  {journal} {\bibinfo  {journal}
  {Astrophys. J.}\ }\textbf {\bibinfo {volume} {739}},\ \bibinfo {pages} {99}
  (\bibinfo {year} {2011})}\BibitemShut {NoStop}%
\bibitem [{\citenamefont {Vallisneri}(2011)}]{Vallisneri:2011}%
  \BibitemOpen
  \bibfield  {author} {\bibinfo {author} {\bibfnamefont {M.}~\bibnamefont
  {Vallisneri}},\ }\href {\doibase 10.1103/PhysRevLett.107.191104} {\bibfield
  {journal} {\bibinfo  {journal} {Phys. Rev. Lett.}\ }\textbf {\bibinfo
  {volume} {107}},\ \bibinfo {pages} {191104} (\bibinfo {year}
  {2011})}\BibitemShut {NoStop}%
\bibitem [{\citenamefont {Vitale}\ and\ \citenamefont
  {Zanolin}(2011)}]{VitaleZanolin:2011}%
  \BibitemOpen
  \bibfield  {author} {\bibinfo {author} {\bibfnamefont {S.}~\bibnamefont
  {Vitale}}\ and\ \bibinfo {author} {\bibfnamefont {M.}~\bibnamefont
  {Zanolin}},\ }\href {\doibase 10.1103/PhysRevD.84.104020} {\bibfield
  {journal} {\bibinfo  {journal} {Phys. Rev. D}\ }\textbf {\bibinfo {volume}
  {84}},\ \bibinfo {pages} {104020} (\bibinfo {year} {2011})}\BibitemShut
  {NoStop}%
\bibitem [{\citenamefont {Veitch}\ \emph {et~al.}(2012)\citenamefont {Veitch},
  \citenamefont {Mandel}, \citenamefont {Aylott}, \citenamefont {Farr},
  \citenamefont {Raymond}, \citenamefont {Rodriguez}, \citenamefont {van~der
  Sluys}, \citenamefont {Kalogera},\ and\ \citenamefont
  {Vecchio}}]{VeitchEtAl:2012}%
  \BibitemOpen
  \bibfield  {author} {\bibinfo {author} {\bibfnamefont {J.}~\bibnamefont
  {Veitch}}, \bibinfo {author} {\bibfnamefont {I.}~\bibnamefont {Mandel}},
  \bibinfo {author} {\bibfnamefont {B.}~\bibnamefont {Aylott}}, \bibinfo
  {author} {\bibfnamefont {B.}~\bibnamefont {Farr}}, \bibinfo {author}
  {\bibfnamefont {V.}~\bibnamefont {Raymond}}, \bibinfo {author} {\bibfnamefont
  {C.}~\bibnamefont {Rodriguez}}, \bibinfo {author} {\bibfnamefont
  {M.}~\bibnamefont {van~der Sluys}}, \bibinfo {author} {\bibfnamefont
  {V.}~\bibnamefont {Kalogera}}, \ and\ \bibinfo {author} {\bibfnamefont
  {A.}~\bibnamefont {Vecchio}},\ }\href {\doibase 10.1103/PhysRevD.85.104045}
  {\bibfield  {journal} {\bibinfo  {journal} {Phys. Rev. D}\ }\textbf {\bibinfo
  {volume} {85}},\ \bibinfo {pages} {104045} (\bibinfo {year}
  {2012})}\BibitemShut {NoStop}%
\bibitem [{\citenamefont {{P. Ajith \emph{et~al.}}}(2012)}]{NINJA2:2012}%
  \BibitemOpen
  \bibfield  {author} {\bibinfo {author} {\bibnamefont {{P. Ajith
  \emph{et~al.}}}},\ }\href {\doibase 10.1088/0264-9381/29/12/124001}
  {\bibfield  {journal} {\bibinfo  {journal} {Class. Quant. Grav.}\ }\textbf
  {\bibinfo {volume} {29}},\ \bibinfo {pages} {124001} (\bibinfo {year}
  {2012})}\BibitemShut {NoStop}%
\bibitem [{\citenamefont {Kozameh}\ \emph {et~al.}(2008)\citenamefont
  {Kozameh}, \citenamefont {Newman},\ and\ \citenamefont
  {Silva-Ortigoza}}]{KozamehEtAl:2008}%
  \BibitemOpen
  \bibfield  {author} {\bibinfo {author} {\bibfnamefont {C.}~\bibnamefont
  {Kozameh}}, \bibinfo {author} {\bibfnamefont {E.~T.}\ \bibnamefont {Newman}},
  \ and\ \bibinfo {author} {\bibfnamefont {G.}~\bibnamefont {Silva-Ortigoza}},\
  }\href {\doibase 10.1088/0264-9381/25/14/145001} {\bibfield  {journal}
  {\bibinfo  {journal} {Class. Quant. Grav.}\ }\textbf {\bibinfo {volume}
  {25}},\ \bibinfo {pages} {145001} (\bibinfo {year} {2008})}\BibitemShut
  {NoStop}%
\bibitem [{\citenamefont {Moreschi}(2004)}]{Moreschi:2004}%
  \BibitemOpen
  \bibfield  {author} {\bibinfo {author} {\bibfnamefont {O.~M.}\ \bibnamefont
  {Moreschi}},\ }\href {\doibase 10.1088/0264-9381/21/23/008} {\bibfield
  {journal} {\bibinfo  {journal} {Class. Quant. Grav.}\ }\textbf {\bibinfo
  {volume} {21}},\ \bibinfo {pages} {5409} (\bibinfo {year}
  {2004})}\BibitemShut {NoStop}%
\bibitem [{\citenamefont {Helfer}(2010)}]{Helfer:2010}%
  \BibitemOpen
  \bibfield  {author} {\bibinfo {author} {\bibfnamefont {A.~D.}\ \bibnamefont
  {Helfer}},\ }\href {\doibase 10.1103/PhysRevD.81.084001} {\bibfield
  {journal} {\bibinfo  {journal} {Phys. Rev. D}\ }\textbf {\bibinfo {volume}
  {81}},\ \bibinfo {pages} {084001} (\bibinfo {year} {2010})}\BibitemShut
  {NoStop}%
\bibitem [{\citenamefont {Goldberg}\ \emph {et~al.}(1967)\citenamefont
  {Goldberg}, \citenamefont {Macfarlane}, \citenamefont {Newman}, \citenamefont
  {Rohrlich},\ and\ \citenamefont {Sudarshan}}]{GoldbergEtAl:1967}%
  \BibitemOpen
  \bibfield  {author} {\bibinfo {author} {\bibfnamefont {J.~N.}\ \bibnamefont
  {Goldberg}}, \bibinfo {author} {\bibfnamefont {A.~J.}\ \bibnamefont
  {Macfarlane}}, \bibinfo {author} {\bibfnamefont {E.~T.}\ \bibnamefont
  {Newman}}, \bibinfo {author} {\bibfnamefont {F.}~\bibnamefont {Rohrlich}}, \
  and\ \bibinfo {author} {\bibfnamefont {E.~C.~G.}\ \bibnamefont {Sudarshan}},\
  }\href {\doibase 10.1063/1.1705135} {\bibfield  {journal} {\bibinfo
  {journal} {J. Math. Phys.}\ }\textbf {\bibinfo {volume} {8}},\ \bibinfo
  {pages} {2155} (\bibinfo {year} {1967})}\BibitemShut {NoStop}%
\bibitem [{\citenamefont {Gualtieri}\ \emph {et~al.}(2008)\citenamefont
  {Gualtieri}, \citenamefont {Berti}, \citenamefont {Cardoso},\ and\
  \citenamefont {Sperhake}}]{GualtieriEtAl:2008}%
  \BibitemOpen
  \bibfield  {author} {\bibinfo {author} {\bibfnamefont {L.}~\bibnamefont
  {Gualtieri}}, \bibinfo {author} {\bibfnamefont {E.}~\bibnamefont {Berti}},
  \bibinfo {author} {\bibfnamefont {V.}~\bibnamefont {Cardoso}}, \ and\
  \bibinfo {author} {\bibfnamefont {U.}~\bibnamefont {Sperhake}},\ }\href
  {\doibase 10.1103/PhysRevD.78.044024} {\bibfield  {journal} {\bibinfo
  {journal} {Phys. Rev. D}\ }\textbf {\bibinfo {volume} {78}},\ \bibinfo
  {pages} {044024} (\bibinfo {year} {2008})}\BibitemShut {NoStop}%
\bibitem [{Cpl(2012)}]{Cplusplus}%
  \BibitemOpen
  \href {http://www.stroustrup.com/C++.html} {\enquote {\bibinfo {title} {The
  {C}++ programming language},}\ } (\bibinfo {year} {2012})\BibitemShut
  {NoStop}%
\bibitem [{\citenamefont {Galassi}(2009)}]{GSL:2009}%
  \BibitemOpen
  \bibfield  {author} {\bibinfo {author} {\bibfnamefont {M.}~\bibnamefont
  {Galassi}},\ }\href {http://www.gnu.org/software/gsl/} {\emph {\bibinfo
  {title} {{GNU} {S}cientific {L}ibrary Reference Manual}}},\ \bibinfo
  {edition} {3rd}\ ed. (\bibinfo {year} {2009})\BibitemShut {NoStop}%
\bibitem [{Pyt(2012)}]{Python}%
  \BibitemOpen
  \href {http://www.python.org/} {\enquote {\bibinfo {title} {Python
  programming language},}\ } (\bibinfo {year} {2012})\BibitemShut {NoStop}%
\bibitem [{\citenamefont {P\'erez}\ and\ \citenamefont
  {Granger}(2007)}]{IPython:2007}%
  \BibitemOpen
  \bibfield  {author} {\bibinfo {author} {\bibfnamefont {F.}~\bibnamefont
  {P\'erez}}\ and\ \bibinfo {author} {\bibfnamefont {B.~E.}\ \bibnamefont
  {Granger}},\ }\href {http://ipython.org/} {\bibfield  {journal} {\bibinfo
  {journal} {{C}omput. {S}ci. {E}ng.}\ }\textbf {\bibinfo {volume} {9}},\
  \bibinfo {pages} {21} (\bibinfo {year} {2007})}\BibitemShut {NoStop}%
\bibitem [{\citenamefont {Doran}\ and\ \citenamefont
  {Lasenby}(2003)}]{DoranLasenby:2003}%
  \BibitemOpen
  \bibfield  {author} {\bibinfo {author} {\bibfnamefont {C.}~\bibnamefont
  {Doran}}\ and\ \bibinfo {author} {\bibfnamefont {A.}~\bibnamefont
  {Lasenby}},\ }\href@noop {} {\emph {\bibinfo {title} {Geometric algebra for
  physicists}}},\ \bibinfo {edition} {3rd}\ ed.\ (\bibinfo  {publisher}
  {Cambridge University Press},\ \bibinfo {year} {2003})\BibitemShut {NoStop}%
\bibitem [{\citenamefont {Szekeres}(2006)}]{Szekeres:2006}%
  \BibitemOpen
  \bibfield  {author} {\bibinfo {author} {\bibfnamefont {P.}~\bibnamefont
  {Szekeres}},\ }\href@noop {} {\emph {\bibinfo {title} {A course in modern
  mathematical physics: {G}roups, {H}ilbert space, and differential
  geometry}}}\ (\bibinfo  {publisher} {Cambridge University Press},\ \bibinfo
  {year} {2006})\BibitemShut {NoStop}%
\bibitem [{\citenamefont {Faye}\ \emph {et~al.}(2006)\citenamefont {Faye},
  \citenamefont {Blanchet},\ and\ \citenamefont {Buonanno}}]{FayeEtAl:2006}%
  \BibitemOpen
  \bibfield  {author} {\bibinfo {author} {\bibfnamefont {G.}~\bibnamefont
  {Faye}}, \bibinfo {author} {\bibfnamefont {L.}~\bibnamefont {Blanchet}}, \
  and\ \bibinfo {author} {\bibfnamefont {A.}~\bibnamefont {Buonanno}},\ }\href
  {\doibase 10.1103/PhysRevD.74.104033} {\bibfield  {journal} {\bibinfo
  {journal} {Phys. Rev. D}\ }\textbf {\bibinfo {volume} {74}},\ \bibinfo
  {pages} {104033} (\bibinfo {year} {2006})}\BibitemShut {NoStop}%
\bibitem [{\citenamefont {Owen}(1996)}]{Owen:1996}%
  \BibitemOpen
  \bibfield  {author} {\bibinfo {author} {\bibfnamefont {B.~J.}\ \bibnamefont
  {Owen}},\ }\href {\doibase 10.1103/PhysRevD.53.6749} {\bibfield  {journal}
  {\bibinfo  {journal} {Phys. Rev. D}\ }\textbf {\bibinfo {volume} {53}},\
  \bibinfo {pages} {6749} (\bibinfo {year} {1996})}\BibitemShut {NoStop}%
\bibitem [{\citenamefont {Pan}\ \emph {et~al.}(2004)\citenamefont {Pan},
  \citenamefont {Buonanno}, \citenamefont {Chen},\ and\ \citenamefont
  {Vallisneri}}]{PanEtAl:2004}%
  \BibitemOpen
  \bibfield  {author} {\bibinfo {author} {\bibfnamefont {Y.}~\bibnamefont
  {Pan}}, \bibinfo {author} {\bibfnamefont {A.}~\bibnamefont {Buonanno}},
  \bibinfo {author} {\bibfnamefont {Y.}~\bibnamefont {Chen}}, \ and\ \bibinfo
  {author} {\bibfnamefont {M.}~\bibnamefont {Vallisneri}},\ }\href {\doibase
  10.1103/PhysRevD.69.104017} {\bibfield  {journal} {\bibinfo  {journal} {Phys.
  Rev. D}\ }\textbf {\bibinfo {volume} {69}},\ \bibinfo {pages} {104017}
  (\bibinfo {year} {2004})}\BibitemShut {NoStop}%
\bibitem [{\citenamefont {Owen}(2007)}]{Owen:2007}%
  \BibitemOpen
  \bibfield  {author} {\bibinfo {author} {\bibfnamefont {R.}~\bibnamefont
  {Owen}},\ }\emph {\bibinfo {title} {Topics in numerical relativity: {T}he
  periodic standing-wave approximation, the stability of constraints in free
  evolution, and the spin of dynamical black holes}},\ \href
  {http://etd.caltech.edu/etd/available/etd-05252007-143511/} {Ph.D. thesis},\
  \bibinfo  {school} {California Institute of Technology} (\bibinfo {year}
  {2007})\BibitemShut {NoStop}%
\bibitem [{\citenamefont {Cook}\ and\ \citenamefont
  {Whiting}(2007)}]{CookWhiting:2007}%
  \BibitemOpen
  \bibfield  {author} {\bibinfo {author} {\bibfnamefont {G.~B.}\ \bibnamefont
  {Cook}}\ and\ \bibinfo {author} {\bibfnamefont {B.~F.}\ \bibnamefont
  {Whiting}},\ }\href {\doibase 10.1103/PhysRevD.76.041501} {\bibfield
  {journal} {\bibinfo  {journal} {Phys. Rev. D}\ }\textbf {\bibinfo {volume}
  {76}},\ \bibinfo {pages} {041501} (\bibinfo {year} {2007})}\BibitemShut
  {NoStop}%
\bibitem [{\citenamefont {Lovelace}\ \emph {et~al.}(2008)\citenamefont
  {Lovelace}, \citenamefont {Owen}, \citenamefont {Pfeiffer},\ and\
  \citenamefont {Chu}}]{LovelaceEtAl:2008}%
  \BibitemOpen
  \bibfield  {author} {\bibinfo {author} {\bibfnamefont {G.}~\bibnamefont
  {Lovelace}}, \bibinfo {author} {\bibfnamefont {R.}~\bibnamefont {Owen}},
  \bibinfo {author} {\bibfnamefont {H.~P.}\ \bibnamefont {Pfeiffer}}, \ and\
  \bibinfo {author} {\bibfnamefont {T.}~\bibnamefont {Chu}},\ }\href {\doibase
  10.1103/PhysRevD.78.084017} {\bibfield  {journal} {\bibinfo  {journal} {Phys.
  Rev. D}\ }\textbf {\bibinfo {volume} {78}},\ \bibinfo {pages} {084017}
  (\bibinfo {year} {2008})}\BibitemShut {NoStop}%
\bibitem [{\citenamefont {Alvi}(2001)}]{Alvi:2001}%
  \BibitemOpen
  \bibfield  {author} {\bibinfo {author} {\bibfnamefont {K.}~\bibnamefont
  {Alvi}},\ }\href {http://link.aps.org/doi/10.1103/PhysRevD.64.104020}
  {\bibfield  {journal} {\bibinfo  {journal} {Phys. Rev. D}\ }\textbf {\bibinfo
  {volume} {64}},\ \bibinfo {pages} {104020} (\bibinfo {year}
  {2001})}\BibitemShut {NoStop}%
\bibitem [{\citenamefont {Lovelace}\ \emph {et~al.}(2012)\citenamefont
  {Lovelace}, \citenamefont {Boyle}, \citenamefont {Scheel},\ and\
  \citenamefont {Szil{\'{a}}gyi}}]{LovelaceEtAl:2012}%
  \BibitemOpen
  \bibfield  {author} {\bibinfo {author} {\bibfnamefont {G.}~\bibnamefont
  {Lovelace}}, \bibinfo {author} {\bibfnamefont {M.}~\bibnamefont {Boyle}},
  \bibinfo {author} {\bibfnamefont {M.~A.}\ \bibnamefont {Scheel}}, \ and\
  \bibinfo {author} {\bibfnamefont {B.}~\bibnamefont {Szil{\'{a}}gyi}},\ }\href
  {\doibase 10.1088/0264-9381/29/4/045003} {\bibfield  {journal} {\bibinfo
  {journal} {Class. Quant. Grav.}\ }\textbf {\bibinfo {volume} {29}},\ \bibinfo
  {pages} {045003} (\bibinfo {year} {2012})}\BibitemShut {NoStop}%
\bibitem [{\citenamefont {Ochsner}\ and\ \citenamefont
  {{O}'{S}haughnessy}(2012)}]{OchsnerOShaughnessy:2012}%
  \BibitemOpen
  \bibfield  {author} {\bibinfo {author} {\bibfnamefont {E.}~\bibnamefont
  {Ochsner}}\ and\ \bibinfo {author} {\bibfnamefont {R.}~\bibnamefont
  {{O}'{S}haughnessy}},\ }\href {\doibase 10.1103/PhysRevD.86.104037}
  {\bibfield  {journal} {\bibinfo  {journal} {Phys. Rev. D}\ }\textbf {\bibinfo
  {volume} {86}},\ \bibinfo {pages} {104037} (\bibinfo {year}
  {2012})}\BibitemShut {NoStop}%
\bibitem [{\citenamefont {Moreschi}(1986)}]{Moreschi:1986}%
  \BibitemOpen
  \bibfield  {author} {\bibinfo {author} {\bibfnamefont {O.~M.}\ \bibnamefont
  {Moreschi}},\ }\href {http://stacks.iop.org/0264-9381/3/i=4/a=006} {\bibfield
   {journal} {\bibinfo  {journal} {Class. Quant. Grav.}\ }\textbf {\bibinfo
  {volume} {3}},\ \bibinfo {pages} {503} (\bibinfo {year} {1986})}\BibitemShut
  {NoStop}%
\bibitem [{\citenamefont {Adamo}\ \emph {et~al.}(2009)\citenamefont {Adamo},
  \citenamefont {Kozameh},\ and\ \citenamefont {Newman}}]{AdamoEtAl:2009}%
  \BibitemOpen
  \bibfield  {author} {\bibinfo {author} {\bibfnamefont {T.~M.}\ \bibnamefont
  {Adamo}}, \bibinfo {author} {\bibfnamefont {C.}~\bibnamefont {Kozameh}}, \
  and\ \bibinfo {author} {\bibfnamefont {E.~T.}\ \bibnamefont {Newman}},\
  }\href {http://www.livingreviews.org/lrr-2009-6} {\bibfield  {journal}
  {\bibinfo  {journal} {Living Rev. Relativity}\ }\textbf {\bibinfo {volume}
  {12}} (\bibinfo {year} {2009})}\BibitemShut {NoStop}%
\bibitem [{\citenamefont {Buonanno}\ \emph {et~al.}(2013)\citenamefont
  {Buonanno}, \citenamefont {Faye},\ and\ \citenamefont
  {Hinderer}}]{BuonannoEtAl:2012}%
  \BibitemOpen
  \bibfield  {author} {\bibinfo {author} {\bibfnamefont {A.}~\bibnamefont
  {Buonanno}}, \bibinfo {author} {\bibfnamefont {G.}~\bibnamefont {Faye}}, \
  and\ \bibinfo {author} {\bibfnamefont {T.}~\bibnamefont {Hinderer}},\ }\href
  {\doibase 10.1103/PhysRevD.87.044009} {\bibfield  {journal} {\bibinfo
  {journal} {Phys. Rev. D}\ }\textbf {\bibinfo {volume} {87}},\ \bibinfo
  {pages} {044009} (\bibinfo {year} {2013})}\BibitemShut {NoStop}%
\bibitem [{\citenamefont {Cohen}\ and\ \citenamefont
  {Giacomo}(1987)}]{IUPAPRedBookSymbols:1987}%
  \BibitemOpen
  \bibfield  {author} {\bibinfo {author} {\bibfnamefont {E.~R.}\ \bibnamefont
  {Cohen}}\ and\ \bibinfo {author} {\bibfnamefont {P.}~\bibnamefont
  {Giacomo}},\ }\href {\doibase 10.1016/0378-4371(87)90220-2} {\bibfield
  {journal} {\bibinfo  {journal} {Physica {A}}\ }\textbf {\bibinfo {volume}
  {146}},\ \bibinfo {pages} {47} (\bibinfo {year} {1987})}\BibitemShut
  {NoStop}%
\bibitem [{\citenamefont {{ISO Technical Committee
  12}}(2009)}]{ISO80000-2:2009}%
  \BibitemOpen
  \bibfield  {author} {\bibinfo {author} {\bibnamefont {{ISO Technical
  Committee 12}}},\ }\href
  {http://www.iso.org/iso/catalogue_detail?csnumber=31887} {\emph {\bibinfo
  {title} {Quantities and Units: Mathematical signs and symbols to be used in
  the natural sciences and technology}}},\ \bibinfo {type} {Tech. Rep.}\
  \bibinfo {number} {80000-2}\ (\bibinfo  {institution} {International
  Organization for Standardization},\ \bibinfo {year} {2009})\BibitemShut
  {NoStop}%
\bibitem [{\citenamefont {Ajith}\ \emph {et~al.}(2011)\citenamefont {Ajith},
  \citenamefont {Boyle}, \citenamefont {Brown}, \citenamefont {Fairhurst},
  \citenamefont {Hannam}, \citenamefont {Hinder}, \citenamefont {Husa},
  \citenamefont {Krishnan}, \citenamefont {Mercer}, \citenamefont {Ohme},
  \citenamefont {Ott}, \citenamefont {Read}, \citenamefont {Santamar{\'{i}}a},\
  and\ \citenamefont {Whelan}}]{NINJA2:2011}%
  \BibitemOpen
  \bibfield  {author} {\bibinfo {author} {\bibfnamefont {P.}~\bibnamefont
  {Ajith}}, \bibinfo {author} {\bibfnamefont {M.}~\bibnamefont {Boyle}},
  \bibinfo {author} {\bibfnamefont {D.~A.}\ \bibnamefont {Brown}}, \bibinfo
  {author} {\bibfnamefont {S.}~\bibnamefont {Fairhurst}}, \bibinfo {author}
  {\bibfnamefont {M.}~\bibnamefont {Hannam}}, \bibinfo {author} {\bibfnamefont
  {I.}~\bibnamefont {Hinder}}, \bibinfo {author} {\bibfnamefont
  {S.}~\bibnamefont {Husa}}, \bibinfo {author} {\bibfnamefont {B.}~\bibnamefont
  {Krishnan}}, \bibinfo {author} {\bibfnamefont {R.~A.}\ \bibnamefont
  {Mercer}}, \bibinfo {author} {\bibfnamefont {F.}~\bibnamefont {Ohme}},
  \bibinfo {author} {\bibfnamefont {C.~D.}\ \bibnamefont {Ott}}, \bibinfo
  {author} {\bibfnamefont {J.~S.}\ \bibnamefont {Read}}, \bibinfo {author}
  {\bibfnamefont {L.}~\bibnamefont {Santamar{\'{i}}a}}, \ and\ \bibinfo
  {author} {\bibfnamefont {J.~T.}\ \bibnamefont {Whelan}},\ }\href@noop {}
  {\enquote {\bibinfo {title} {Data formats for numerical relativity},}\ }
  (\bibinfo {year} {2011}),\ \Eprint {http://arxiv.org/abs/0709.0093v3}
  {arXiv:0709.0093v3 [gr-qc]} \BibitemShut {NoStop}%
\bibitem [{\citenamefont {Campanelli}\ \emph {et~al.}(2009)\citenamefont
  {Campanelli}, \citenamefont {Lousto}, \citenamefont {Nakano},\ and\
  \citenamefont {Zlochower}}]{CampanelliEtAl:2009}%
  \BibitemOpen
  \bibfield  {author} {\bibinfo {author} {\bibfnamefont {M.}~\bibnamefont
  {Campanelli}}, \bibinfo {author} {\bibfnamefont {C.~O.}\ \bibnamefont
  {Lousto}}, \bibinfo {author} {\bibfnamefont {H.}~\bibnamefont {Nakano}}, \
  and\ \bibinfo {author} {\bibfnamefont {Y.}~\bibnamefont {Zlochower}},\ }\href
  {\doibase 10.1103/PhysRevD.79.084010} {\bibfield  {journal} {\bibinfo
  {journal} {Phys. Rev. D}\ }\textbf {\bibinfo {volume} {79}},\ \bibinfo
  {pages} {084010} (\bibinfo {year} {2009})}\BibitemShut {NoStop}%
\bibitem [{\citenamefont {Boyle}\ \emph {et~al.}(2007)\citenamefont {Boyle},
  \citenamefont {Brown}, \citenamefont {Kidder}, \citenamefont {Mroue},
  \citenamefont {Pfeiffer}, \citenamefont {Scheel}, \citenamefont {Cook},\ and\
  \citenamefont {Teukolsky}}]{BoyleEtAl:2007}%
  \BibitemOpen
  \bibfield  {author} {\bibinfo {author} {\bibfnamefont {M.}~\bibnamefont
  {Boyle}}, \bibinfo {author} {\bibfnamefont {D.~A.}\ \bibnamefont {Brown}},
  \bibinfo {author} {\bibfnamefont {L.~E.}\ \bibnamefont {Kidder}}, \bibinfo
  {author} {\bibfnamefont {A.~H.}\ \bibnamefont {Mroue}}, \bibinfo {author}
  {\bibfnamefont {H.~P.}\ \bibnamefont {Pfeiffer}}, \bibinfo {author}
  {\bibfnamefont {M.~A.}\ \bibnamefont {Scheel}}, \bibinfo {author}
  {\bibfnamefont {G.~B.}\ \bibnamefont {Cook}}, \ and\ \bibinfo {author}
  {\bibfnamefont {S.~A.}\ \bibnamefont {Teukolsky}},\ }\href {\doibase
  10.1103/PhysRevD.76.124038} {\bibfield  {journal} {\bibinfo  {journal} {Phys.
  Rev. D}\ }\textbf {\bibinfo {volume} {76}},\ \bibinfo {pages} {124038}
  (\bibinfo {year} {2007})}\BibitemShut {NoStop}%
\bibitem [{\citenamefont {Boyle}(2008)}]{Boyle:2008}%
  \BibitemOpen
  \bibfield  {author} {\bibinfo {author} {\bibfnamefont {M.}~\bibnamefont
  {Boyle}},\ }\emph {\bibinfo {title} {Accurate gravitational waveforms from
  binary black-hole systems}},\ \href
  {http://etd.caltech.edu/etd/available/etd-01122009-143851/} {Ph.D. thesis},\
  \bibinfo  {school} {California Institute of Technology} (\bibinfo {year}
  {2008})\BibitemShut {NoStop}%
\bibitem [{\citenamefont {Boyle}\ and\ \citenamefont
  {Mrou{\'{e}}}(2009)}]{BoyleMroue:2009}%
  \BibitemOpen
  \bibfield  {author} {\bibinfo {author} {\bibfnamefont {M.}~\bibnamefont
  {Boyle}}\ and\ \bibinfo {author} {\bibfnamefont {A.~H.}\ \bibnamefont
  {Mrou{\'{e}}}},\ }\href {\doibase 10.1103/PhysRevD.80.124045} {\bibfield
  {journal} {\bibinfo  {journal} {Phys. Rev. D}\ }\textbf {\bibinfo {volume}
  {80}},\ \bibinfo {pages} {124045} (\bibinfo {year} {2009})}\BibitemShut
  {NoStop}%
\bibitem [{\citenamefont {Bishop}\ \emph {et~al.}(1996)\citenamefont {Bishop},
  \citenamefont {Gomez}, \citenamefont {Holvorcem}, \citenamefont {Matzner},
  \citenamefont {Papadopoulos},\ and\ \citenamefont
  {Winicour}}]{BishopEtAl:1996}%
  \BibitemOpen
  \bibfield  {author} {\bibinfo {author} {\bibfnamefont {N.~T.}\ \bibnamefont
  {Bishop}}, \bibinfo {author} {\bibfnamefont {R.}~\bibnamefont {Gomez}},
  \bibinfo {author} {\bibfnamefont {P.~R.}\ \bibnamefont {Holvorcem}}, \bibinfo
  {author} {\bibfnamefont {R.~A.}\ \bibnamefont {Matzner}}, \bibinfo {author}
  {\bibfnamefont {P.}~\bibnamefont {Papadopoulos}}, \ and\ \bibinfo {author}
  {\bibfnamefont {J.}~\bibnamefont {Winicour}},\ }\href {\doibase
  10.1103/PhysRevLett.76.4303} {\bibfield  {journal} {\bibinfo  {journal}
  {Phys. Rev. Lett.}\ }\textbf {\bibinfo {volume} {76}},\ \bibinfo {pages}
  {4303} (\bibinfo {year} {1996})}\BibitemShut {NoStop}%
\bibitem [{\citenamefont {Winicour}(2012)}]{Winicour:2012}%
  \BibitemOpen
  \bibfield  {author} {\bibinfo {author} {\bibfnamefont {J.}~\bibnamefont
  {Winicour}},\ }\href
  {http://relativity.livingreviews.org/Articles/lrr-2012-2/} {\bibfield
  {journal} {\bibinfo  {journal} {Living Rev. Relativity}\ }\textbf {\bibinfo
  {volume} {15}} (\bibinfo {year} {2012})}\BibitemShut {NoStop}%
\bibitem [{\citenamefont {Reisswig}\ \emph {et~al.}(2012)\citenamefont
  {Reisswig}, \citenamefont {Bishop},\ and\ \citenamefont
  {Pollney}}]{ReisswigEtAl:2012}%
  \BibitemOpen
  \bibfield  {author} {\bibinfo {author} {\bibfnamefont {C.}~\bibnamefont
  {Reisswig}}, \bibinfo {author} {\bibfnamefont {N.~T.}\ \bibnamefont
  {Bishop}}, \ and\ \bibinfo {author} {\bibfnamefont {D.}~\bibnamefont
  {Pollney}},\ }\href@noop {} {\enquote {\bibinfo {title} {General relativistic
  null-cone evolutions with a high-order scheme},}\ } (\bibinfo {year}
  {2012}),\ \Eprint {http://arxiv.org/abs/1208.3891} {arXiv:1208.3891 [gr-qc]}
  \BibitemShut {NoStop}%
\bibitem [{\citenamefont {Boyle}\ \emph {et~al.}(2009)\citenamefont {Boyle},
  \citenamefont {Brown},\ and\ \citenamefont {Pekowsky}}]{BoyleEtAl:2009}%
  \BibitemOpen
  \bibfield  {author} {\bibinfo {author} {\bibfnamefont {M.}~\bibnamefont
  {Boyle}}, \bibinfo {author} {\bibfnamefont {D.~A.}\ \bibnamefont {Brown}}, \
  and\ \bibinfo {author} {\bibfnamefont {L.}~\bibnamefont {Pekowsky}},\ }\href
  {http://www.iop.org/EJ/abstract/0264-9381/26/11/114006} {\bibfield  {journal}
  {\bibinfo  {journal} {Class. Quant. Grav.}\ }\textbf {\bibinfo {volume}
  {26}},\ \bibinfo {pages} {114006} (\bibinfo {year} {2009})}\BibitemShut
  {NoStop}%
\bibitem [{\citenamefont {Boyle}(2011)}]{Boyle:2011}%
  \BibitemOpen
  \bibfield  {author} {\bibinfo {author} {\bibfnamefont {M.}~\bibnamefont
  {Boyle}},\ }\href {\doibase 10.1103/PhysRevD.84.064013} {\bibfield  {journal}
  {\bibinfo  {journal} {Phys. Rev. D}\ }\textbf {\bibinfo {volume} {84}},\
  \bibinfo {pages} {064013} (\bibinfo {year} {2011})}\BibitemShut {NoStop}%
\bibitem [{\citenamefont {{MacDonald}}\ \emph {et~al.}(2011)\citenamefont
  {{MacDonald}}, \citenamefont {Nissanke},\ and\ \citenamefont
  {Pfeiffer}}]{MacDonaldEtAl:2011}%
  \BibitemOpen
  \bibfield  {author} {\bibinfo {author} {\bibfnamefont {I.}~\bibnamefont
  {{MacDonald}}}, \bibinfo {author} {\bibfnamefont {S.}~\bibnamefont
  {Nissanke}}, \ and\ \bibinfo {author} {\bibfnamefont {H.~P.}\ \bibnamefont
  {Pfeiffer}},\ }\href {\doibase 10.1088/0264-9381/28/13/134002} {\bibfield
  {journal} {\bibinfo  {journal} {Class. Quant. Grav.}\ }\textbf {\bibinfo
  {volume} {28}},\ \bibinfo {pages} {134002} (\bibinfo {year}
  {2011})}\BibitemShut {NoStop}%
\bibitem [{\citenamefont {{MacDonald}}\ \emph {et~al.}(2013)\citenamefont
  {{MacDonald}}, \citenamefont {Mrou{\'{e}}}, \citenamefont {Pfeiffer},
  \citenamefont {Boyle}, \citenamefont {Kidder}, \citenamefont {Scheel},
  \citenamefont {Szil{\'{a}}gyi},\ and\ \citenamefont
  {Taylor}}]{MacDonaldEtAl:2013}%
  \BibitemOpen
  \bibfield  {author} {\bibinfo {author} {\bibfnamefont {I.}~\bibnamefont
  {{MacDonald}}}, \bibinfo {author} {\bibfnamefont {A.~H.}\ \bibnamefont
  {Mrou{\'{e}}}}, \bibinfo {author} {\bibfnamefont {H.~P.}\ \bibnamefont
  {Pfeiffer}}, \bibinfo {author} {\bibfnamefont {M.}~\bibnamefont {Boyle}},
  \bibinfo {author} {\bibfnamefont {L.~E.}\ \bibnamefont {Kidder}}, \bibinfo
  {author} {\bibfnamefont {M.~A.}\ \bibnamefont {Scheel}}, \bibinfo {author}
  {\bibfnamefont {B.}~\bibnamefont {Szil{\'{a}}gyi}}, \ and\ \bibinfo {author}
  {\bibfnamefont {N.~W.}\ \bibnamefont {Taylor}},\ }\href {\doibase
  10.1103/PhysRevD.87.024009} {\bibfield  {journal} {\bibinfo  {journal} {Phys.
  Rev. D}\ }\textbf {\bibinfo {volume} {87}},\ \bibinfo {pages} {024009}
  (\bibinfo {year} {2013})}\BibitemShut {NoStop}%
\bibitem [{\citenamefont {Boyle}\ \emph {et~al.}(2008)\citenamefont {Boyle},
  \citenamefont {Buonanno}, \citenamefont {Kidder}, \citenamefont
  {Mrou{\'{e}}}, \citenamefont {Pan}, \citenamefont {Pfeiffer},\ and\
  \citenamefont {Scheel}}]{BoyleEtAl:2008}%
  \BibitemOpen
  \bibfield  {author} {\bibinfo {author} {\bibfnamefont {M.}~\bibnamefont
  {Boyle}}, \bibinfo {author} {\bibfnamefont {A.}~\bibnamefont {Buonanno}},
  \bibinfo {author} {\bibfnamefont {L.~E.}\ \bibnamefont {Kidder}}, \bibinfo
  {author} {\bibfnamefont {A.~H.}\ \bibnamefont {Mrou{\'{e}}}}, \bibinfo
  {author} {\bibfnamefont {Y.}~\bibnamefont {Pan}}, \bibinfo {author}
  {\bibfnamefont {H.~P.}\ \bibnamefont {Pfeiffer}}, \ and\ \bibinfo {author}
  {\bibfnamefont {M.~A.}\ \bibnamefont {Scheel}},\ }\href {\doibase
  10.1103/PhysRevD.78.104020} {\bibfield  {journal} {\bibinfo  {journal} {Phys.
  Rev. D}\ }\textbf {\bibinfo {volume} {78}},\ \bibinfo {pages} {104020}
  (\bibinfo {year} {2008})}\BibitemShut {NoStop}%
\bibitem [{\citenamefont {Scheel}\ \emph {et~al.}(2006)\citenamefont {Scheel},
  \citenamefont {Pfeiffer}, \citenamefont {Lindblom}, \citenamefont {Kidder},
  \citenamefont {Rinne},\ and\ \citenamefont {Teukolsky}}]{ScheelEtAl:2006}%
  \BibitemOpen
  \bibfield  {author} {\bibinfo {author} {\bibfnamefont {M.~A.}\ \bibnamefont
  {Scheel}}, \bibinfo {author} {\bibfnamefont {H.~P.}\ \bibnamefont
  {Pfeiffer}}, \bibinfo {author} {\bibfnamefont {L.}~\bibnamefont {Lindblom}},
  \bibinfo {author} {\bibfnamefont {L.~E.}\ \bibnamefont {Kidder}}, \bibinfo
  {author} {\bibfnamefont {O.}~\bibnamefont {Rinne}}, \ and\ \bibinfo {author}
  {\bibfnamefont {S.~A.}\ \bibnamefont {Teukolsky}},\ }\href {\doibase
  10.1103/PhysRevD.74.104006} {\bibfield  {journal} {\bibinfo  {journal} {Phys.
  Rev. D}\ }\textbf {\bibinfo {volume} {74}},\ \bibinfo {pages} {104006}
  (\bibinfo {year} {2006})}\BibitemShut {NoStop}%
\bibitem [{\citenamefont {Hemberger}\ \emph {et~al.}(2012)\citenamefont
  {Hemberger}, \citenamefont {Scheel}, \citenamefont {Kidder}, \citenamefont
  {Szil{\'{a}}gyi},\ and\ \citenamefont {Teukolsky}}]{HembergerEtAl:2012}%
  \BibitemOpen
  \bibfield  {author} {\bibinfo {author} {\bibfnamefont {D.~A.}\ \bibnamefont
  {Hemberger}}, \bibinfo {author} {\bibfnamefont {M.~A.}\ \bibnamefont
  {Scheel}}, \bibinfo {author} {\bibfnamefont {L.~E.}\ \bibnamefont {Kidder}},
  \bibinfo {author} {\bibfnamefont {B.}~\bibnamefont {Szil{\'{a}}gyi}}, \ and\
  \bibinfo {author} {\bibfnamefont {S.~A.}\ \bibnamefont {Teukolsky}},\
  }\href@noop {} {\enquote {\bibinfo {title} {Dynamical excision boundaries in
  spectral evolutions of binary black hole spacetimes},}\ } (\bibinfo {year}
  {2012}),\ \Eprint {http://arxiv.org/abs/1211.6079} {arXiv:1211.6079 [gr-qc]}
  \BibitemShut {NoStop}%
\bibitem [{\citenamefont {Pan}\ \emph {et~al.}(2011)\citenamefont {Pan},
  \citenamefont {Buonanno}, \citenamefont {Boyle}, \citenamefont {Buchman},
  \citenamefont {Kidder}, \citenamefont {Pfeiffer},\ and\ \citenamefont
  {Scheel}}]{PanEtAl:2011}%
  \BibitemOpen
  \bibfield  {author} {\bibinfo {author} {\bibfnamefont {Y.}~\bibnamefont
  {Pan}}, \bibinfo {author} {\bibfnamefont {A.}~\bibnamefont {Buonanno}},
  \bibinfo {author} {\bibfnamefont {M.}~\bibnamefont {Boyle}}, \bibinfo
  {author} {\bibfnamefont {L.~T.}\ \bibnamefont {Buchman}}, \bibinfo {author}
  {\bibfnamefont {L.~E.}\ \bibnamefont {Kidder}}, \bibinfo {author}
  {\bibfnamefont {H.~P.}\ \bibnamefont {Pfeiffer}}, \ and\ \bibinfo {author}
  {\bibfnamefont {M.~A.}\ \bibnamefont {Scheel}},\ }\href {\doibase
  10.1103/PhysRevD.84.124052} {\bibfield  {journal} {\bibinfo  {journal} {Phys.
  Rev. D}\ }\textbf {\bibinfo {volume} {84}},\ \bibinfo {pages} {124052}
  (\bibinfo {year} {2011})}\BibitemShut {NoStop}%
\bibitem [{\citenamefont {Kelly}\ and\ \citenamefont
  {Baker}(2012)}]{KellyBaker:2012}%
  \BibitemOpen
  \bibfield  {author} {\bibinfo {author} {\bibfnamefont {B.~J.}\ \bibnamefont
  {Kelly}}\ and\ \bibinfo {author} {\bibfnamefont {J.~G.}\ \bibnamefont
  {Baker}},\ }\href@noop {} {\enquote {\bibinfo {title} {Decoding mode-mixing
  in black-hole merger ringdown},}\ } (\bibinfo {year} {2012}),\ \Eprint
  {http://arxiv.org/abs/1212.5553} {arXiv:1212.5553 [gr-qc]} \BibitemShut
  {NoStop}%
\bibitem [{\citenamefont {Yang}\ \emph {et~al.}(2012)\citenamefont {Yang},
  \citenamefont {Nichols}, \citenamefont {Zhang}, \citenamefont {Zimmerman},
  \citenamefont {Zhang},\ and\ \citenamefont {Chen}}]{YangEtAl:2012}%
  \BibitemOpen
  \bibfield  {author} {\bibinfo {author} {\bibfnamefont {H.}~\bibnamefont
  {Yang}}, \bibinfo {author} {\bibfnamefont {D.~A.}\ \bibnamefont {Nichols}},
  \bibinfo {author} {\bibfnamefont {F.}~\bibnamefont {Zhang}}, \bibinfo
  {author} {\bibfnamefont {A.}~\bibnamefont {Zimmerman}}, \bibinfo {author}
  {\bibfnamefont {Z.}~\bibnamefont {Zhang}}, \ and\ \bibinfo {author}
  {\bibfnamefont {Y.}~\bibnamefont {Chen}},\ }\href {\doibase
  10.1103/PhysRevD.86.104006} {\bibfield  {journal} {\bibinfo  {journal} {Phys.
  Rev. D}\ }\textbf {\bibinfo {volume} {86}},\ \bibinfo {pages} {104006}
  (\bibinfo {year} {2012})}\BibitemShut {NoStop}%
\bibitem [{\citenamefont {{Wolfram Software, Inc.}}(2010)}]{Mathematica8}%
  \BibitemOpen
  \bibfield  {author} {\bibinfo {author} {\bibnamefont {{Wolfram Software,
  Inc.}}},\ }\href {http://www.wolfram.com/mathematica/} {\enquote {\bibinfo
  {title} {Mathematica, {V}ersion 8.0},}\ } (\bibinfo {year}
  {2010})\BibitemShut {NoStop}%
\bibitem [{\citenamefont {Newman}\ and\ \citenamefont
  {Penrose}(1965)}]{NewmanPenrose:1965}%
  \BibitemOpen
  \bibfield  {author} {\bibinfo {author} {\bibfnamefont {E.~T.}\ \bibnamefont
  {Newman}}\ and\ \bibinfo {author} {\bibfnamefont {R.}~\bibnamefont
  {Penrose}},\ }\href {\doibase 10.1103/PhysRevLett.15.231} {\bibfield
  {journal} {\bibinfo  {journal} {Phys. Rev. Lett.}\ }\textbf {\bibinfo
  {volume} {15}},\ \bibinfo {pages} {231} (\bibinfo {year} {1965})}\BibitemShut
  {NoStop}%
\bibitem [{\citenamefont {Janis}\ and\ \citenamefont
  {Newman}(1965)}]{JanisNewman:1965}%
  \BibitemOpen
  \bibfield  {author} {\bibinfo {author} {\bibfnamefont {A.~I.}\ \bibnamefont
  {Janis}}\ and\ \bibinfo {author} {\bibfnamefont {E.~T.}\ \bibnamefont
  {Newman}},\ }\href {\doibase doi:10.1063/1.1704349} {\bibfield  {journal}
  {\bibinfo  {journal} {J. Math. Phys.}\ }\textbf {\bibinfo {volume} {6}},\
  \bibinfo {pages} {902} (\bibinfo {year} {1965})}\BibitemShut {NoStop}%
\bibitem [{\citenamefont {Newman}\ and\ \citenamefont
  {Penrose}(1966)}]{NewmanPenrose:1966}%
  \BibitemOpen
  \bibfield  {author} {\bibinfo {author} {\bibfnamefont {E.~T.}\ \bibnamefont
  {Newman}}\ and\ \bibinfo {author} {\bibfnamefont {R.}~\bibnamefont
  {Penrose}},\ }\href {\doibase 10.1063/1.1931221} {\bibfield  {journal}
  {\bibinfo  {journal} {J. Math. Phys.}\ }\textbf {\bibinfo {volume} {7}},\
  \bibinfo {pages} {863} (\bibinfo {year} {1966})}\BibitemShut {NoStop}%
\bibitem [{\citenamefont {Wigner}(1959)}]{Wigner:1959}%
  \BibitemOpen
  \bibfield  {author} {\bibinfo {author} {\bibfnamefont {E.~P.}\ \bibnamefont
  {Wigner}},\ }\href@noop {} {\emph {\bibinfo {title} {Group Theory and Its
  Applications to the Quantum Mechanics of Atomic Spectra}}}\ (\bibinfo
  {publisher} {Academic Press},\ \bibinfo {address} {New York, NY},\ \bibinfo
  {year} {1959})\BibitemShut {NoStop}%
\bibitem [{\citenamefont {Ajith}\ \emph {et~al.}(2007)\citenamefont {Ajith},
  \citenamefont {Babak}, \citenamefont {Chen}, \citenamefont {Hewitson},
  \citenamefont {Krishnan}, \citenamefont {Whelan}, \citenamefont
  {Br{\"{u}}gmann}, \citenamefont {Diener}, \citenamefont {Gonzalez},
  \citenamefont {Hannam}, \citenamefont {Husa}, \citenamefont {Koppitz},
  \citenamefont {Pollney}, \citenamefont {Rezzolla}, \citenamefont
  {Santamar{\'{i}}a}, \citenamefont {Sintes}, \citenamefont {Sperhake},\ and\
  \citenamefont {Thornburg}}]{AjithEtAl:2007}%
  \BibitemOpen
  \bibfield  {author} {\bibinfo {author} {\bibfnamefont {P.}~\bibnamefont
  {Ajith}}, \bibinfo {author} {\bibfnamefont {S.}~\bibnamefont {Babak}},
  \bibinfo {author} {\bibfnamefont {Y.}~\bibnamefont {Chen}}, \bibinfo {author}
  {\bibfnamefont {M.}~\bibnamefont {Hewitson}}, \bibinfo {author}
  {\bibfnamefont {B.}~\bibnamefont {Krishnan}}, \bibinfo {author}
  {\bibfnamefont {J.~T.}\ \bibnamefont {Whelan}}, \bibinfo {author}
  {\bibfnamefont {B.}~\bibnamefont {Br{\"{u}}gmann}}, \bibinfo {author}
  {\bibfnamefont {P.}~\bibnamefont {Diener}}, \bibinfo {author} {\bibfnamefont
  {J.}~\bibnamefont {Gonzalez}}, \bibinfo {author} {\bibfnamefont
  {M.}~\bibnamefont {Hannam}}, \bibinfo {author} {\bibfnamefont
  {S.}~\bibnamefont {Husa}}, \bibinfo {author} {\bibfnamefont {M.}~\bibnamefont
  {Koppitz}}, \bibinfo {author} {\bibfnamefont {D.}~\bibnamefont {Pollney}},
  \bibinfo {author} {\bibfnamefont {L.}~\bibnamefont {Rezzolla}}, \bibinfo
  {author} {\bibfnamefont {L.}~\bibnamefont {Santamar{\'{i}}a}}, \bibinfo
  {author} {\bibfnamefont {A.~M.}\ \bibnamefont {Sintes}}, \bibinfo {author}
  {\bibfnamefont {U.}~\bibnamefont {Sperhake}}, \ and\ \bibinfo {author}
  {\bibfnamefont {J.}~\bibnamefont {Thornburg}},\ }\href {\doibase
  10.1088/0264-9381/24/19/S31} {\bibfield  {journal} {\bibinfo  {journal}
  {Class. Quant. Grav.}\ }\textbf {\bibinfo {volume} {24}},\ \bibinfo {pages}
  {S689} (\bibinfo {year} {2007})}\BibitemShut {NoStop}%
\bibitem [{\citenamefont {Miller}(1972)}]{Miller:1972}%
  \BibitemOpen
  \bibfield  {author} {\bibinfo {author} {\bibfnamefont {W.}~\bibnamefont
  {Miller}},\ }\href {http://www.ima.umn.edu/~miller/symmetrygroups.html}
  {\emph {\bibinfo {title} {Symmetry Groups and Their Applications}}},\ Pure
  and Applied Mathematics\ (\bibinfo  {publisher} {Academic Press},\ \bibinfo
  {address} {New York},\ \bibinfo {year} {1972})\BibitemShut {NoStop}%
\bibitem [{\citenamefont {Duistermaat}\ and\ \citenamefont
  {Kolk}(1999)}]{DuistermaatKolk:1999}%
  \BibitemOpen
  \bibfield  {author} {\bibinfo {author} {\bibfnamefont {J.~J.}\ \bibnamefont
  {Duistermaat}}\ and\ \bibinfo {author} {\bibfnamefont {J.~A.~C.}\
  \bibnamefont {Kolk}},\ }\href@noop {} {\emph {\bibinfo {title} {Lie
  Groups}}}\ (\bibinfo  {publisher} {Springer-Verlag},\ \bibinfo {address} {New
  York, NY},\ \bibinfo {year} {1999})\BibitemShut {NoStop}%
\bibitem [{\citenamefont {Grassia}(1998)}]{Grassia:1998}%
  \BibitemOpen
  \bibfield  {author} {\bibinfo {author} {\bibfnamefont {F.~S.}\ \bibnamefont
  {Grassia}},\ }\href {\doibase 10.1080/10867651.1998.10487493} {\bibfield
  {journal} {\bibinfo  {journal} {Journal of Graphics Tools}\ }\textbf
  {\bibinfo {volume} {3}},\ \bibinfo {pages} {29} (\bibinfo {year}
  {1998})}\BibitemShut {NoStop}%
\bibitem [{\citenamefont {Shoemake}(1985)}]{Shoemake:1985}%
  \BibitemOpen
  \bibfield  {author} {\bibinfo {author} {\bibfnamefont {K.}~\bibnamefont
  {Shoemake}},\ }\href {\doibase 10.1145/325165.325242} {\bibfield  {journal}
  {\bibinfo  {journal} {{SIGGRAPH} Comput. Graph.}\ }\textbf {\bibinfo {volume}
  {19}},\ \bibinfo {pages} {245} (\bibinfo {year} {1985})}\BibitemShut
  {NoStop}%
\bibitem [{\citenamefont {Kim}\ \emph {et~al.}(1995)\citenamefont {Kim},
  \citenamefont {Kim},\ and\ \citenamefont {Shin}}]{KimEtAl:1995}%
  \BibitemOpen
  \bibfield  {author} {\bibinfo {author} {\bibfnamefont {M.}~\bibnamefont
  {Kim}}, \bibinfo {author} {\bibfnamefont {M.}~\bibnamefont {Kim}}, \ and\
  \bibinfo {author} {\bibfnamefont {S.~Y.}\ \bibnamefont {Shin}},\ }in\ \href
  {\doibase 10.1145/218380.218486} {\emph {\bibinfo {booktitle} {Proceedings of
  the 22nd annual conference on computer graphics and interactive
  techniques}}},\ \bibinfo {series and number} {{SIGGRAPH} '95}\ (\bibinfo
  {publisher} {{ACM}},\ \bibinfo {address} {New York, {NY}, {USA}},\ \bibinfo
  {year} {1995})\ pp.\ \bibinfo {pages} {369--376}\BibitemShut {NoStop}%
\end{thebibliography}%
